\documentclass[sigconf]{aamas} 

\usepackage{balance} %
\usepackage{hyperref}
\usepackage{url}

\usepackage{todonotes}
\setuptodonotes{inline}

\usepackage{microtype}
\usepackage{graphicx}
\usepackage{subcaption}
\usepackage{booktabs} %

\usepackage{amsmath}
\usepackage{amsthm}
\usepackage{xparse}

\usepackage[capitalize,noabbrev]{cleveref}
\usepackage{amsfonts}       %
\usepackage{nicefrac}       %
\usepackage{enumerate}
\usepackage{tabularx}       %
\usepackage{array,multirow}
\usepackage{algorithm}
\usepackage{algpseudocode}
\usepackage{mathtools}
\crefname{figure}{figure}{figures}
\crefname{equation}{equation}{equations}
\crefname{table}{table}{tables}
\crefname{section}{section}{sections}
\usepackage[
    separate-uncertainty=true,
    detect-weight,
    table-number-alignment = center,
	table-text-alignment = center,
	table-format=3.2(4),
    table-sign-mantissa,
    group-digits=integer,
    group-separator={,},
]{siunitx}
\usepackage{etoolbox}
\DeclareMathOperator*{\argmax}{arg\,max}

\newcommand{\exarg}[2]{\mathbb{E}_{#1}\hspace{-2pt}\left[ #2 \right]}
\newcommand{\exnoarg}[1]{\mathbb{E}_{#1}}
\NewDocumentCommand\ex{ m g }{
	\IfNoValueTF{#2}{\exnoarg{#1}}{\exarg{#1}{#2}}
}

\newcommand{\Ag}{I}
\newcommand{\St}{S}

\newcommand{\st}{s}
\newcommand{\Ob}{O}
\newcommand{\jOb}{\mathbf{O}}
\newcommand{\ob}{o}
\newcommand{\job}{\mathbf{o}}
\newcommand{\Ac}{A}
\newcommand{\jAc}{\mathbf{A}}
\newcommand{\ac}{a}
\newcommand{\jac}{\mathbf{a}}
\newcommand{\rew}{r}
\newcommand{\jrew}{\mathbf{r}}

\newcommand{\instdist}{\mu}
\newcommand{\Stf}{\mathcal{T}}
\newcommand{\Obf}{\mathcal{O}}
\newcommand{\Rew}{\mathcal{R}}

\newcommand{\pol}{\pi}
\newcommand{\jpol}{\mathbf{\pi}}

\newcommand{\his}{h}
\newcommand{\jhis}{\mathbf{h}}
\newcommand{\dsc}{\gamma}

\newcommand{\loss}{\mathcal{L}}
\newcommand{\data}{\mathcal{D}}
\newcommand{\batch}{B}

\newcolumntype{C}{D{,}{,}{-1}} %

\newcommand{\h}[1]{\emph{#1}}
\newcommand{\seehere}[1]{(\Cref{#1})}

\makeatletter
\gdef\@copyrightpermission{
  \begin{minipage}{0.2\columnwidth}
   \href{https://creativecommons.org/licenses/by/4.0/}{\includegraphics[width=0.90\textwidth]{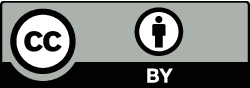}}
  \end{minipage}\hfill
  \begin{minipage}{0.8\columnwidth}
   \href{https://creativecommons.org/licenses/by/4.0/}{This work is licensed under a Creative Commons Attribution International 4.0 License.}
  \end{minipage}
  \vspace{5pt}
}
\makeatother

\setcopyright{ifaamas}
\acmConference[AAMAS '25]{Proc.\@ of the 24th International Conference
on Autonomous Agents and Multiagent Systems (AAMAS 2025)}{May 19 -- 23, 2025}
{Detroit, Michigan, USA}{Y.~Vorobeychik, S.~Das, A.~Nowé  (eds.)}
\copyrightyear{2025}
\acmYear{2025}
\acmDOI{}
\acmPrice{}
\acmISBN{}

\acmSubmissionID{550}

\title{Ensemble Value Functions for Efficient Exploration in Multi-Agent Reinforcement Learning}

\author{Lukas Sch\"afer}
\affiliation{
    \institution{University of Edinburgh}
    \city{Edinburgh}
    \country{United Kingdom}
}
\email{l.schaefer@ed.ac.uk}
\authornote{Now at Microsoft Research Cambridge, contact at lukas.schaefer@microsoft.com.}

\author{Oliver Slumbers}
\affiliation{
    \institution{University College London}
    \city{London}
    \country{United Kingdom}
}
\email{o.slumbers@cs.ucl.ac.uk}

\author{Stephen McAleer}
\affiliation{
    \institution{Carnegie Mellon University}
    \city{Pittsburgh}
    \country{USA}
}
\email{mcaleer.stephen@gmail.com}

\author{Yali Du}
\affiliation{
    \institution{King's College London}
    \city{London}
    \country{United Kingdom}
}
\email{yali.du@kcl.ac.uk}

\author{Stefano V. Albrecht}
\affiliation{
    \institution{University of Edinburgh}
    \city{Edinburgh}
    \country{United Kingdom}
}
\email{s.albrecht@ed.ac.uk}

\author{David Mguni}
\affiliation{
    \institution{Queen Mary University London}
    \city{London}
    \country{United Kingdom}
}
\email{d.mguni@qmul.ac.uk}

\begin{abstract}
    Multi-agent reinforcement learning (MARL) requires agents to explore within a vast joint action space to find joint actions that lead to coordination. Existing value-based MARL algorithms commonly rely on random exploration, such as $\epsilon$-greedy, to explore the environment which is not systematic and inefficient at identifying effective actions in multi-agent problems. 
    Additionally, the concurrent training of the policies of multiple agents during training can render the optimisation non-stationary. This can lead to unstable value estimates, highly variant gradients, and ultimately hinder coordination between agents.
    To address these challenges, we propose ensemble value functions for multi-agent exploration (EMAX). EMAX is a framework to seamlessly extend value-based MARL algorithms. EMAX leverages an ensemble of value functions for each agent to guide their exploration, reduce the variance of their optimisation, and makes their policies more robust to miscoordination. EMAX achieves these benefits by (1) systematically guiding the exploration of agents with a UCB policy towards parts of the environment that require multiple agents to coordinate. (2) EMAX computes average value estimates across the ensemble as target values to reduce the variance of gradients and make optimisation more stable. (3) During evaluation, EMAX selects actions following a majority vote across the ensemble to reduce the likelihood of miscoordination. %
    We first instantiate independent DQN with EMAX and evaluate it in 11 general-sum tasks with sparse rewards. We show that EMAX improves final evaluation returns by 185\% across all tasks. We then evaluate EMAX on top of IDQN, VDN and QMIX in 21 common-reward tasks, and show that EMAX improves sample efficiency and final evaluation returns across all tasks over all three vanilla algorithms by 60\%, 47\%, and 538\%, respectively. %
\end{abstract}


\begin{document}

\pagestyle{fancy}
\fancyhead{}

\maketitle 

\section{Introduction}
\label{sec:introduction}

\begin{figure}[t]
    \centering
    \includegraphics[width=.9\linewidth]{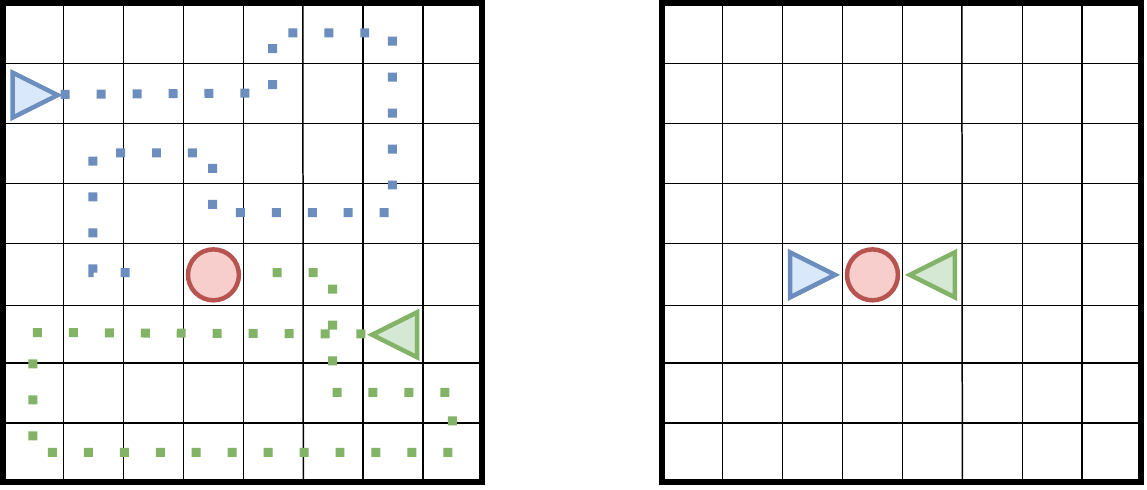}
    \caption[Motivational example for guiding exploration towards states that require agent interactions.]{Motivational example: Two agents (triangles) can independently explore (left), but they must coordinate to pick up the object (circle) and complete the task (right).}
    \label{fig:emax:motivational_example}
\end{figure}

Multi-agent reinforcement learning (MARL)~\citep{marl-book} is a common paradigm to concurrently train autonomous agents in decision-making tasks that require coordination between agents. To discover such coordinated actions, agents need to explore the state space and vast joint action space of the task. However, many value-based MARL algorithms still rely on random exploration, such as an $\epsilon$-greedy policy \citep{sunehag2018value,rashid2020monotonic} that is not systematic and so can take many iterations to discover the optimal behaviour, in particular in states in which multiple agents need to coordinate their actions~\citep{papoudakis2021benchmarking}. To better understand this inefficiency, we consider the following example (\Cref{fig:emax:motivational_example}) in which two agents have to navigate a grid-world to jointly pick up a heavy object. To learn to pick up the goal object, agents need to both select the pick-up action in a state where both agents are next to the object (\Cref{fig:emax:motivational_example}, right). Such coordination is highly unlikely when following a random exploration policy. In contrast, the exploration is not required to be coordinated if agents are not next to the object (\Cref{fig:emax:motivational_example}, left) so random exploration might not be highly inefficient in this case. This example illustrates that random exploration is particularly inefficient in states in which multiple agents are required to coordinate their actions, and demonstrates the need for more systematic exploration of such states. Furthermore, the concurrent training of multiple agents in MARL and the resulting non-stationarity of the policies of other agents makes it challenging for agents to robustly learn to solve the task, and can result in miscoordination between agents~\citep{papoudakis2019nonstationarity}.

Motivated by these challenges in MARL, we propose \h{ensemble value functions for multi-agent exploration} (EMAX), a general framework to seamlessly extend any value-based MARL algorithms by training ensembles of value functions for each agent. EMAX systematically explores states and actions that may require multiple agents to coordinate by following an upper-confidence bound (UCB) policy~\citep{auer2002using} over the average and disagreement of value estimates across the ensemble. This exploration strategy prioritises the exploration of actions that appear promising (as measured by high average value estimates) and that might require coordination between agents (as measured by high disagreement in value estimates). Beyond its exploration policy, EMAX computes average target values across the ensemble to reduce the variance of gradients and eliminate the need for additional target networks. Lastly, EMAX selects actions during evaluation by following a majority vote across the greedy actions of all value functions in the ensemble to reduce the likelihood of selecting sub-optimal actions.

To evaluate the efficacy of EMAX, we first extend independent DQN~\citep{mnih2015human,tan1993multi} with EMAX and evaluate it in 11 mixed-objective tasks, in which agents receive individual rewards but must coordinate their actions in some states. 
In this setting, EMAX improves the final evaluation returns of IDQN by 185\% across all tasks~\seehere{sec:emax:experiments_results}. Afterwards, we focus on the cooperative MARL setting and extend IDQN as well as VDN~\citep{sunehag2018value} and QMIX~\citep{rashid2020monotonic} with EMAX, and conduct an extensive evaluation of EMAX in 21 common-reward tasks across four diverse environments. Across all common-reward tasks, EMAX improves sample efficiency and final achieved returns over all three vanilla algorithms (IDQN, VDN, QMIX) by 60\%, 47\%, and 538\%, respectively \seehere{sec:emax:experiments_results}. 
Lastly, we empirically validate that all three major components of EMAX are essential for its performance in an ablation study, and further demonstrate the effects of the proposed exploration policy, target computation, and evaluation policy~\seehere{sec:emax:analysis}.

\section{Background}
\label{sec:background}

\textbf{Partially observable stochastic games:} We formalise multi-agent environments as partially observable stochastic games (POSG) \citep{littman1994markov,hansen2004dynamic} defined by $(\Ag, \St, \jOb, \jAc, \Stf, \Obf, \{\Rew_i\}_{i\in \Ag}, \dsc)$. Agents are indexed by $i\in\Ag = \{1,\ldots,N\}$, $\St$ denotes the state space of the environment and $\jAc = \Ac_1\times\ldots\times \Ac_N$ denotes the joint action space of all agents. Each agent has access to its local observations $\ob_i \in \Ob_i$. The joint observation space is denoted $\jOb = \Ob_i \times \ldots \times \Ob_N$. $\Stf: \St \times \jAc \mapsto \Delta(\St)$ denotes the transition function of the environment and defines a distribution of successor states given the current state and the applied joint action. The observation transition function $\Obf: \St \times \jAc \times \Delta(\jOb)$ defines a distribution of next joint observations received by agents given the current state and joint action of all agents. $\Rew_i: \St \times \jAc \times \St \mapsto \mathbb{R}$ denotes the reward function for each agent $i$. Each agent learns a policy $\pol_i(\ac^t_i \mid \his_i^t)$ conditioned on its history of observations until time step $t$, i.e.\ $\his_i^t = (\ob_i^0, \ob_i^1, \ldots, \ob_i^t)$. The objective of a POSG is for all agents to learn a joint policy $\jpol = (\pol_1, \ldots, \pol_N)$ such that the expected discounted returns of each agent are maximised with respect to the policies of all other agents. The discounted returns for agent $i$ can be written as
\begin{equation}
    \ex{\ac^t_i \sim \pol_i(\his^t_i); \jac^t_{-i} \sim \pol_{-i}(\jhis^t_{-i})}{\sum_{t=0}^\infty \dsc^t \Rew_i(\st^t, \jac^t, \st^{t+1})}
    \label{eq:background:marl-return}
\end{equation}
where $\dsc \in [0; 1)$ denotes the discount factor, $\jac^t = (\ac_1^t, \ldots, \ac_N^t)$ and $\jhis^t = (\his_1^t, \ldots, \his_N^t)$ denote the joint action and observation history, respectively, and subscript $-i$ denotes all agents but agent $i$. We also consider the special case of common-reward environments, often formalised as a Dec-POMDP~\citep{bernstein2002complexity, oliehoek2016concise}, in which agents collectively maximise the cumulative discounted sum of shared rewards. 

\begin{figure*}[t]
    \centering
    \includegraphics[width=\linewidth]{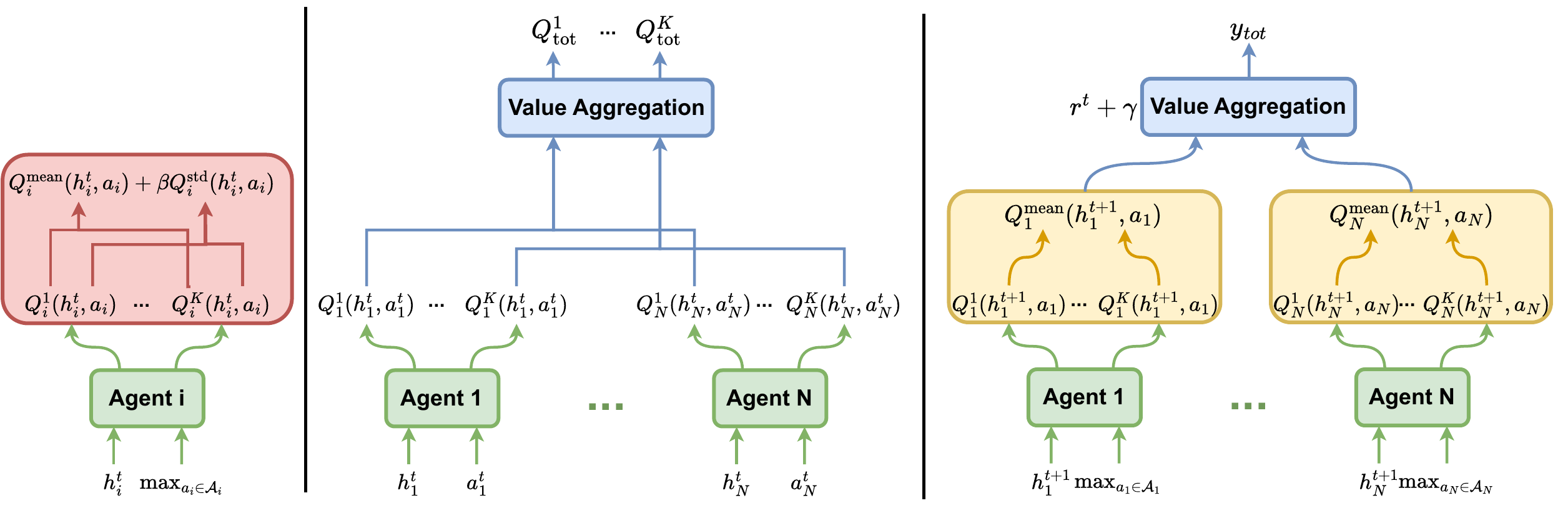}
    \caption{Illustration of the components of the EMAX algorithm. Left: UCB exploration strategy for agent $i$. Middle: value estimation with value decomposition. Right: target computation with value decomposition. The value functions of individual agents are highlighted in green, the exploration policy in red, value decomposition in blue, and target computation in orange.}
    \label{fig:emax:emax}
\end{figure*}

\textbf{Independent deep Q-networks:}
Independent deep Q-networks (IDQN) extends DQN~\citep{mnih2015human} for MARL and independently learns a value function $Q_i$~\citep{tan1993multi}, parameterised by $\theta_i$, for each agent $i$. Agents store tuples $(s^t, h^t, a^t, r^t, s^{t+1}, h^{t+1})$ of experience consisting of state $s^t$, joint observation history $h^t$, applied joint action $a^t$, received reward $r^t$, next state $s^{t+1}$, and next joint observation history $h^{t+1}$, respectively, in a replay buffer. The value function of agent $i$ is then optimised by minimising the average loss across sampled batches of experience:
\begin{equation}
    \mathcal{L}(\theta_i) = \left[r_i^t + \gamma \max_{a_i \in \Ac_i} \bar{Q}_i(h^{t+1}_i, a_i; \bar{\theta}_i) - Q_i(h^t_i, a^t_i; \theta_i)\right]^2
    \label{eq:idqn_loss}
\end{equation}
with $\bar{\theta}_i$ denoting the parameters of the target network $\bar{Q}$ which are periodically copied from $\theta_i$. 

\textbf{Value decomposition:}
Independent learning serves as an effective baseline in many cooperative MARL tasks~\citep{papoudakis2021benchmarking} but suffers from the multi-agent credit assignment problem, i.e.\ agents need to identify their individual contribution to received rewards~\citep{du2019liir,rashid2020monotonic}. Value decomposition algorithms extend IDQN by learning a decomposed centralised state-action value function $Q_{\text{tot}}$, conditioned on the state and joint action of all agents.\footnote{In environments, where the state is not available during training, we use the joint observation as a proxy for the state.} Directly learning such a value function is often computationally infeasible due to the exponential growth of the joint action space with the number of agents, so the centralised value function is approximated with an aggregation of individual utility functions of all agents conditioned on the local observation history. These individual utility functions of agents estimate their contribution to the centralised state-action value function and, thus, address the multi-agent credit assignment problem. All functions are jointly optimised by minimising the following loss with $y_\text{tot}$ denoting centralised target values:
\begin{equation}
    \mathcal{L}(\theta) = \left[ Q_{\text{tot}}(s^t, a^t; \theta) - y_{\text{tot}}\right]^2
    \label{eq:vd_loss}
\end{equation}
Two common value decomposition algorithms are VDN~\citep{sunehag2018value} and QMIX~\citep{rashid2020monotonic} that assume a linear and monotonic decomposition of the centralised value function, respectively.

\section{Ensemble Value Functions for Multi-Agent Exploration}
\label{sec:emax:methodology}

In this section, we present \h{ensemble value functions for multi-agent exploration} (EMAX), a general framework that trains an ensemble of value functions for each agent in value-based MARL. Formally, each agent $i$ trains an ensemble of $K$ value functions $\{Q^k_i\}_{k=1}^K$ with $Q^k_i$ being parameterised by $\theta^k_i$ and conditioned on agent $i$'s local observation history. EMAX leverages these ensembles of value functions to guide the exploration of agents and stabilise their optimisation. %
\Cref{fig:emax:emax} illustrates the exploration policy as well as the value and target estimation of EMAX. We provide pseudocode for EMAX in \Cref{app:sec:emax:pseudocode}.\footnote{Appendices are available at \url{https://arxiv.org/abs/2302.03439}.} In the following, we denote the average and standard deviation across the ensemble of value functions of agent $i$ with parameters $\theta_i = \{\theta^k_i\}_{k=1}^K$ as follows:
\vspace{-.75em}
\begin{align}
    Q_i^{\text{mean}}(\his^t_i, \ac^t_i; \theta_i) &= \frac{1}{K} \sum_{k=1}^K Q_i^{k}(\his_i^t, \ac_i^t; \theta^k_i)\\
    Q_i^{\text{std}}(\his^t_i, \ac^t_i; \theta_i) &= \sqrt{\frac{\sum_{k=1}^K \left(Q_i^{k}(\his^t_i, \ac^t_i; \theta^k_i) - Q_i^{\text{mean}}(\his^t_i, \ac^t_i; \theta_i)\right)^2}{K}}
    \label{eq:emax:ensemble_mean_std}
\end{align}
\vspace{-1em}

\textbf{Exploration policy:} EMAX follow a UCB policy using the average and standard deviation of value estimates across the ensemble:
\begin{equation}
    \pol_i^{\text{expl}}(\his^t_i; \theta_i) \in \argmax_{\ac_i\in\Ac_i} Q_i^{\text{mean}}(\his^t_i, \ac_i; \theta_i) + \beta Q_i^{\text{std}}(\his^t_i, \ac_i; \theta_i)
    \label{eq:emax:expl_policy}
\end{equation}
with $\beta > 0$ denoting a weighting hyperparameter. As measured by the mean value estimate, this policy guides agents to explore actions that are deemed promising. Prior work in single-agent RL already established that the disagreement across an ensemble of value functions can indicate epistemic uncertainty and the need for exploration~\citep{auer2002using,lee2021sunrise,liang2022reducing}. In this work, we argue that in MARL this disagreement of value estimates can additionally indicate whether state-action pairs require agents to coordinate their actions. To see why, consider states in which multiple agents have to coordinate, i.e.\ multiple agents need to select specific actions, to receive a large reward (as in our motivational example in \Cref{fig:emax:motivational_example}, right). If any agent deviates from this joint action, the agents receive no reward. In such states, received rewards for a given action of agent $i$ will vary significantly whenever other agents follow stochastic policies, since the reward depends on the stochastic actions of other agents. In contrast, in states where agent $i$ receives identical rewards independent of the actions of other agents (as in \Cref{fig:emax:motivational_example}, left), no such variability of rewards is experienced. Due to this variability of rewards (or lack thereof), value estimates across the ensemble will exhibit high disagreement in states that might require coordination, and little disagreement in states that might require no or limited coordination. Therefore, the EMAX exploration policy systematically focuses the exploration of agents on state-action pairs that might require coordination in contrast to common random exploration for value-based MARL such as $\epsilon$-greedy policies. 
We note that the disagreement diminishes throughout training as value functions and policies converge. Once agents always succeed at coordinating in a state with such potential, returns will no longer be variable, and the disagreement of value estimates will reduce. Furthermore, the disagreement of value estimates also incentivises the exploration of states with potential for future rather than just immediate coordination since value functions estimate expected returns over entire episodes. %
We empirically validate these effects and benefits of the EMAX exploration policy in \Cref{sec:emax:analysis}.

\textbf{Optimisation:} To extend IDQN with EMAX, we optimise the $k$-th value function of agent $i$ by minimising the following loss:
\vspace{-.75em}
\begin{multline}
    \loss(\theta_i^k) = \mathbb{E}_{(\his_i^t, \ac_i^t, \rew_i^t, \his_i^{t+1}) \sim \data} \left[\left(\rew_i^t + \dsc \max_{\ac_i \in \Ac_i} Q_i^{\text{mean}}(\his^{t+1}_i, \ac_i; \theta_i) -\right.\right.\\
    \left.\left.Q_i^k(\his^t_i, \ac^t_i; \theta^k_i)\right)^2\right]
    \label{eq:emax:imeanq_loss}
\end{multline}
\vspace{-.75em}

\noindent
Computing target values as the average across all value estimates of the ensemble~\citep{liang2022reducing} reduces the computational and memory cost of training ensemble networks by eliminating the need for target networks. Additionally, as we empirically show in \Cref{sec:emax:analysis}, these target values reduce the variability of gradients and improve the stability of training. Such improved stability is particularly valuable in MARL where the non-stationarity of the policies of other agents can otherwise result in unstable or inefficient training~\citep{papoudakis2019nonstationarity,papoudakis2021benchmarking}.

\textbf{Evaluation policy:} Value-based MARL algorithms typically exploit using the greedy policy with respect to their value function. In EMAX, agent $i$ selects its action during evaluation using a majority vote across the greedy actions of all models in its ensemble~\citep{osband2016deep}:

\vspace{-1.5em}
\begin{multline}
    \pi_i^{\text{eval}}(\his^t_i; \theta_i) \in \argmax_{\ac_i \in \Ac_i} \sum_{k=1}^K [1]_{\mathcal{A}_{\text{opt},i}^k}(\ac_i)\\
    \mathcal{A}_{\text{opt},i}^k = \{\ac_i \in \Ac_i \mid \ac_i \in \argmax_{\ac_i'} Q^k_i(\his^t_i, \ac_i'; \theta^k_i)\}
    \label{eq:emax:eval_policy}
\end{multline}
\vspace{-1.25em}

\noindent
with indicator function $[1]_{\mathcal{A}_{\text{opt},i}^k}(a) = 1$ for the greedy action(s) of the $k$-th value function of agent $i$ and $0$ otherwise. Such a policy decreases the likelihood of taking poor actions because any individual value function preferring a poor action due to errors in value estimates does not impact the action selection as long as the majority of models agree on the optimal action. We empirically demonstrate the benefits of such an evaluation policy in \Cref{sec:emax:analysis}.

\textbf{Ensemble value functions:} All aforementioned techniques rely on value functions within the ensemble to be sufficiently diverse early in training. To ensure such diversity, we employ three techniques: (1) Ensemble models share no parameters and are randomly initialised. (2) Each model in the ensemble is trained on separately sampled batches of experiences~\citep{liang2022reducing}. (3) Each model is trained on bootstrapped samples of the entire experience collected~\citep{osband2016deep}. We provide more details on the bootstrapping procedure in \Cref{app:sec:emax:bootstrapped_sampling}.

\textbf{Value decomposition:} So far, we presented EMAX as an extension of IDQN. We now discuss the application of EMAX to value decomposition algorithms for common-reward tasks in which agents suffer from the multi-agent credit assignment problem. In EMAX, this problem has the additional implication that the exploration policy defined in \Cref{eq:emax:expl_policy} does not distinguish which agents need to coordinate their actions in a particular state. To make sure that each agent explores states and actions in which that particular agent's coordination, rather than any agents' coordination, is required, we integrate EMAX into value decomposition algorithms such as VDN~\citep{sunehag2018value} and QMIX~\citep{rashid2020monotonic}. These algorithms enable agents to learn individual value functions that identify their contribution to received common rewards. In EMAX, we train an ensemble of these utility functions for each agent. The parameters of the $k$-th utility function of all agents $\theta^k = \{\theta^k_i\}_{i\in\Ag}$ will be optimised to minimise \Cref{eq:vd_loss}. For VDN, the decomposition and target value are defined as follows:

\vspace{-1.25em}
\begin{align}
    \label{eq:emax:ensemble_vdn_estimate}
    Q_{\text{tot}}(s^t, a^t; \theta^k) &= \sum_{i\in\Ag} Q_i^k(\his_i^t, \ac_i^t; \theta_i^k)\\
    y_{\text{tot}} &= \rew^t + \dsc \sum_{i\in\Ag} \max_{\ac_i \in \Ac_i} Q_i^{\text{mean}}(\his_i^{t+1}, \ac_i; \theta_i)
    \label{eq:emax:ensemble_vdn_target}
\end{align}
\vspace{-.75em}

\noindent
and for QMIX is defined as follows:
\vspace{-.5em}
\begin{multline}
    Q_{\text{tot}}(s^t, a^t; \theta^k) = f_{\text{mix}}\left(Q_1^k(\his_1^t, \ac_1^t; \theta_1^k), \ldots, Q_N^k(\his^t_N, \ac^t_N; \theta_N^k); \theta_{\text{mix}}\right)\\
    y_{\text{tot}} = \rew^t + \dsc f_{\text{mix}} \begin{pmatrix}\max_{\ac_1\in\Ac_1}Q_1^{\text{mean}}(\his_1^{t+1}, \ac_1; \theta_1), & \\ \ldots & ; \widebar{\theta}_{\text{mix}}\\ \max_{\ac_N\in\Ac_N}Q_N^{\text{mean}}(\his_N^{t+1}, \ac_N; \theta_N) & \end{pmatrix}
    \label{eq:emax:ensemble-qmix-estimate}
\end{multline}

\noindent
For QMIX, we use a single mixing network and target mixing network with parameters $\theta_{\text{mix}}$ and $\widebar{\theta}_{\text{mix}}$, respectively, to aggregate the utility estimates for all utility functions in the ensemble.

\section{Evaluation}
\label{sec:emax:deep_experiments}

\begin{figure}[t]
    \centering
    \begin{subfigure}{.24\linewidth}
        \centering
        \includegraphics[width=\linewidth]{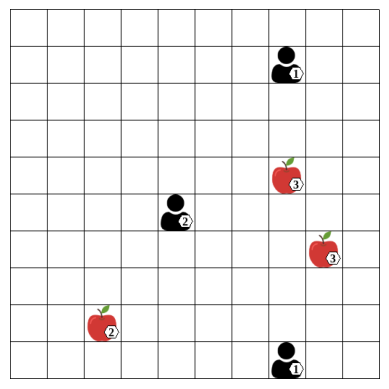}
        \caption{LBF}
        \label{fig:emax:lbf_env}
    \end{subfigure}
    \begin{subfigure}{.24\linewidth}
        \centering
        \includegraphics[width=\linewidth]{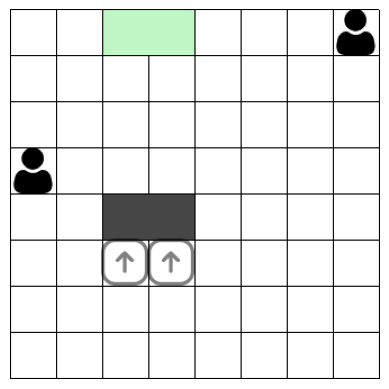}
        \caption{BPUSH}
        \label{fig:emax:bpush_env}
    \end{subfigure}
    \begin{subfigure}{.22\linewidth}
        \centering
        \includegraphics[width=\linewidth]{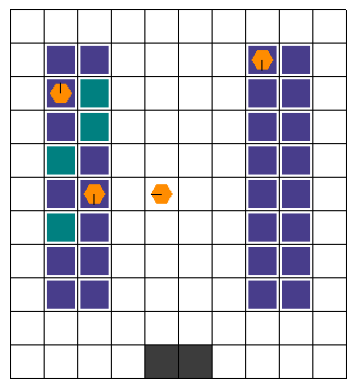}
        \caption{RWARE}
        \label{fig:emax:rware_env}
    \end{subfigure}
    \begin{subfigure}{.21\linewidth}
        \centering
        \includegraphics[width=\linewidth]{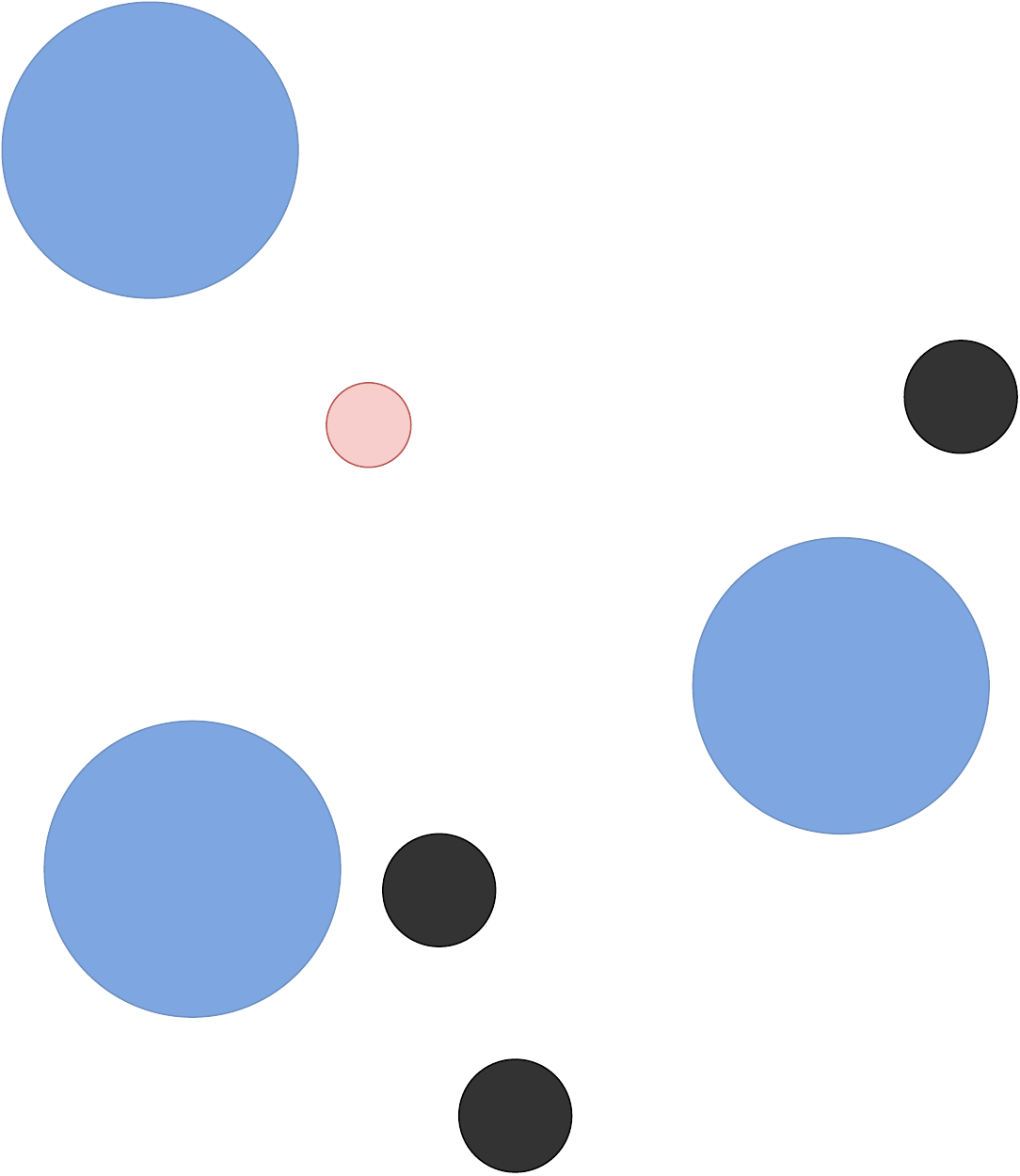}
        \caption{MPE}
        \label{fig:emax:mpe_env}
    \end{subfigure}
    \caption{Visualisations of four multi-agent environments.}
    \label{fig:emax:env_renders}
\end{figure}

We evaluate EMAX in 11 mixed-objective tasks, in which all agents receive individual rewards, and in 21 common-reward tasks across four multi-agent environments shown in \Cref{fig:emax:env_renders}: level-based foraging (LBF)~\citep{albrecht2013game,papoudakis2021benchmarking}, multi-robot warehouse (RWARE)~\citep{christianos2020shared,papoudakis2021benchmarking}, boulder-push (BPUSH)~\citep{christianos2022pareto}, and multi-agent particle environment (MPE)~\citep{mordatch2018emergence,lowe2017multi}.
In mixed-objective tasks, we evaluate IDQN with and without EMAX in LBF and RWARE tasks that require a mixture of cooperation in competition, represented by agents picking up food either by themselves or collectively in LBF and by agents delivering shelves and avoiding to block each others' path in RWARE. In common-reward tasks, we evaluate EMAX as an extension of IDQN, VDN and QMIX. All considered common-reward tasks require agents to cooperate to achieve high rewards. Additionally, many of these tasks feature sparse rewards and, thus, are challenging for exploration, making them particularly suited to evaluate the benefits of the systematic exploration of EMAX. We provide detailed descriptions of all environments in \Cref{app:sec:environments}. In the common-reward setting, we compare EMAX to additional baselines in three value-based exploration algorithms with MAVEN~\citep{mahajan2019maven}, CDS~\citep{li2021celebrating}, and EMC~\citep{zheng2021episodic}, as well as independent and multi-agent PPO (IPPO and MAPPO) that have been shown to exhibit strong MARL performance~\citep{papoudakis2021benchmarking,yu2022surprising}. Lastly, we provide an analysis of each component of EMAX to investigate our hypotheses on their benefits and effects, and provide an ablation study~\seehere{sec:emax:analysis}.

\begin{figure*}[t]
    \centering

    \begin{minipage}{.76\linewidth}
        \centering
        \includegraphics[width=\linewidth]{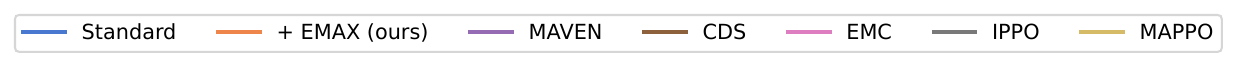}
    
        \begin{subfigure}{.49\linewidth}
            \includegraphics[width=\linewidth]{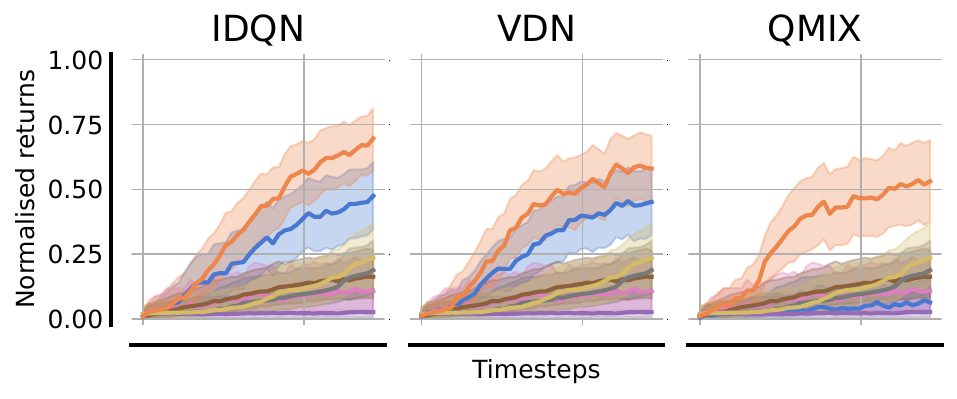}
            \caption{Normalised evaluation returns}
            \label{fig:emax:norm_returns_all}
        \end{subfigure}
        \begin{subfigure}{.49\linewidth}
            \includegraphics[width=\linewidth]{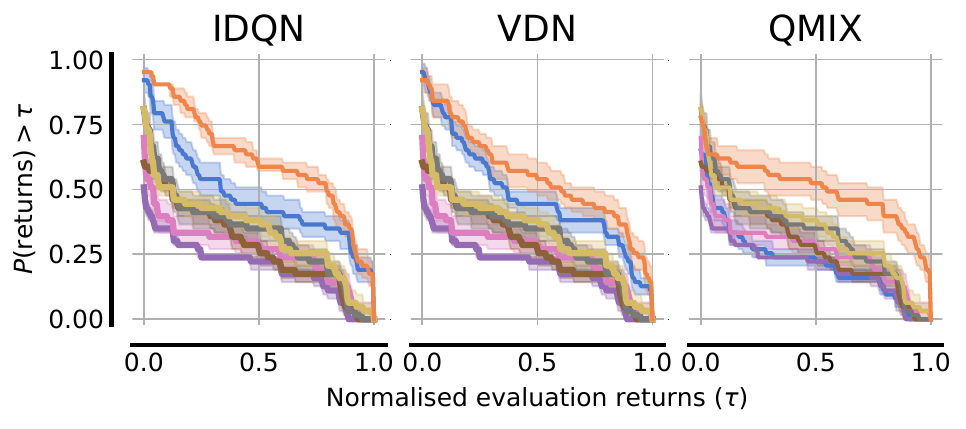}
            \caption{Performance profile}
            \label{fig:emax:deep_profiles}
        \end{subfigure}
    \end{minipage}
    \begin{minipage}{.23\linewidth}
        \centering
        \begin{subfigure}{\linewidth}
            \includegraphics[width=\textwidth]{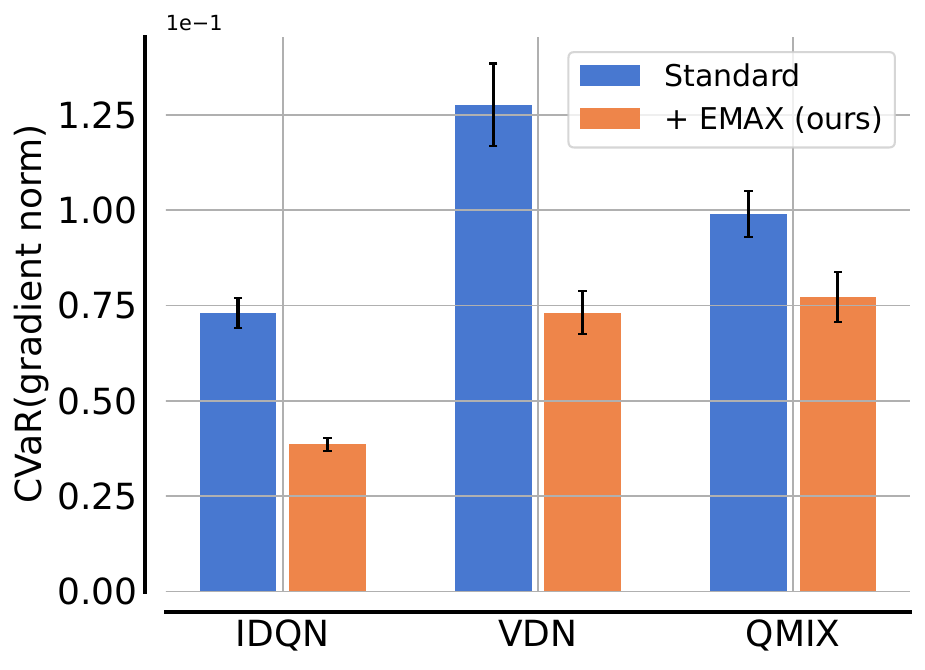}
            \caption{Gradient CVaR}
            \label{fig:emax:coop_gradcvar}
        \end{subfigure}
    \end{minipage}

    \caption{Common-reward evaluation across 21 tasks. (a) Normalised evaluation returns and (b) performance profile of all algorithms aggregated across all tasks. (c) Average and standard error of gradient stability~\seehere{eq:emax:grad_cvar}.}
    \label{fig:emax:coop_deep_results}
\end{figure*}

\begin{figure}[t]
    \centering
    \includegraphics[width=.6\linewidth]{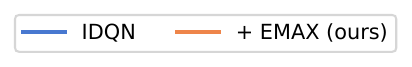}
    
    \begin{subfigure}{.49\linewidth}
        \centering
        \includegraphics[width=\linewidth]{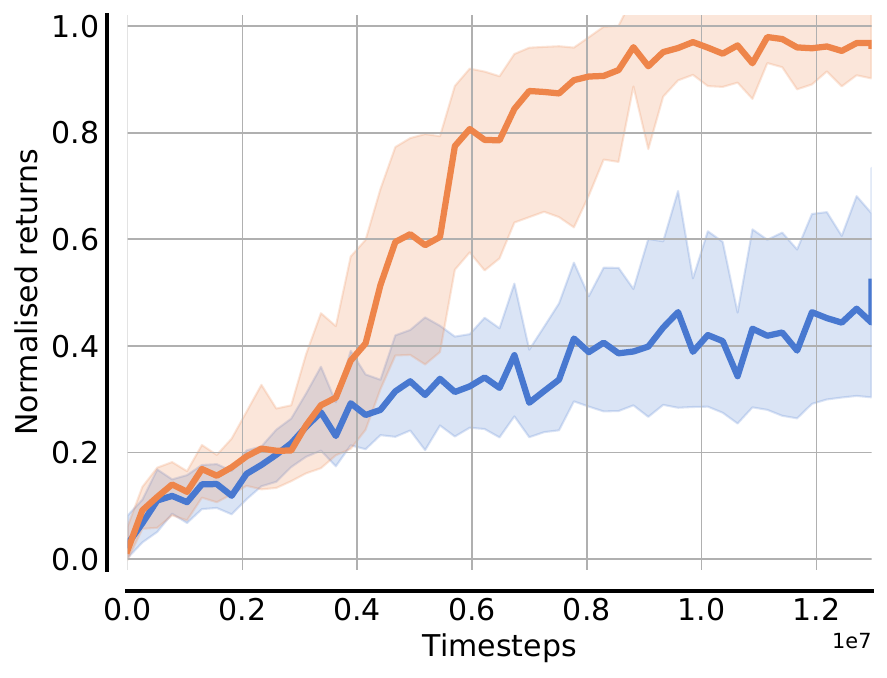}
        \caption{LBF -- Normalised}
        \label{fig:emax:lbf_ind_norm_returns}
    \end{subfigure}
    \begin{subfigure}{.49\linewidth}
        \centering
        \includegraphics[width=\linewidth]{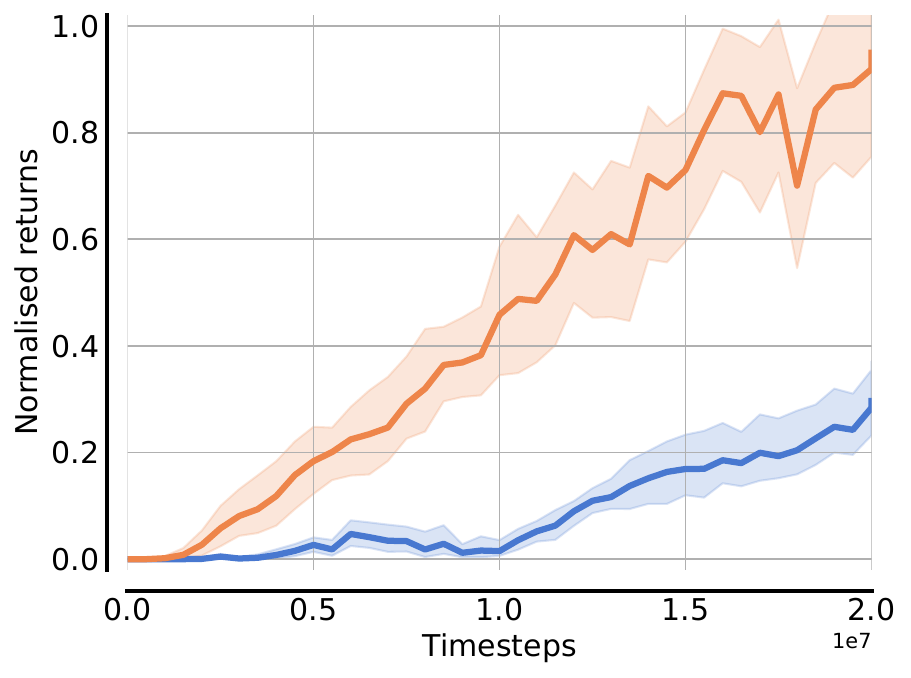}
        \caption{RWARE -- Normalised}
        \label{fig:emax:rware_ind_norm_returns}
    \end{subfigure}

    \begin{subfigure}{.49\linewidth}
        \centering
        \includegraphics[width=\linewidth]{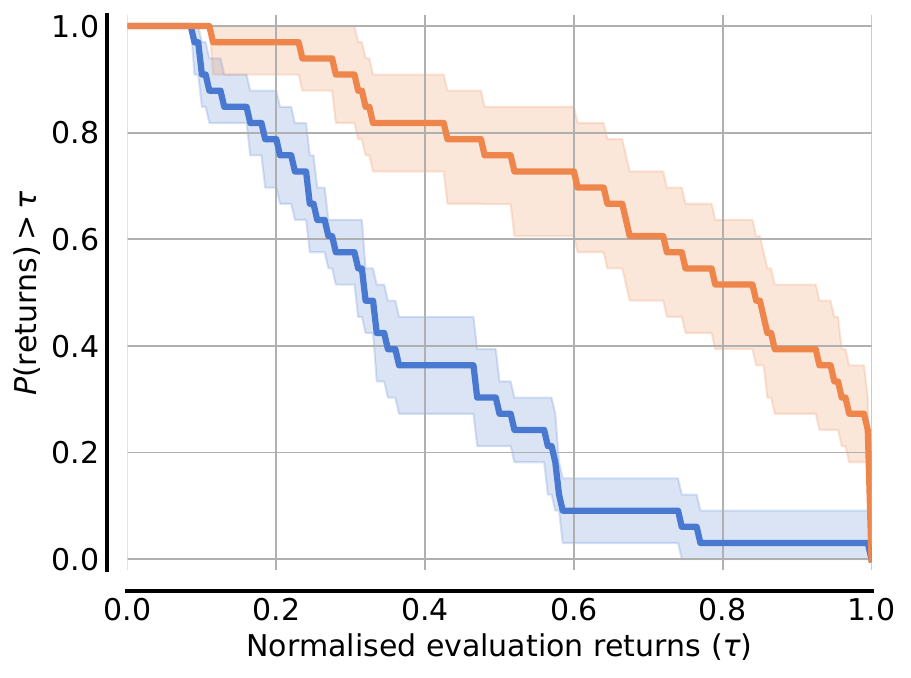}
        \caption{Performance profile}
        \label{fig:emax:ind_profiles}
    \end{subfigure}
    \begin{subfigure}{.49\linewidth}
        \centering
        \includegraphics[width=\textwidth]{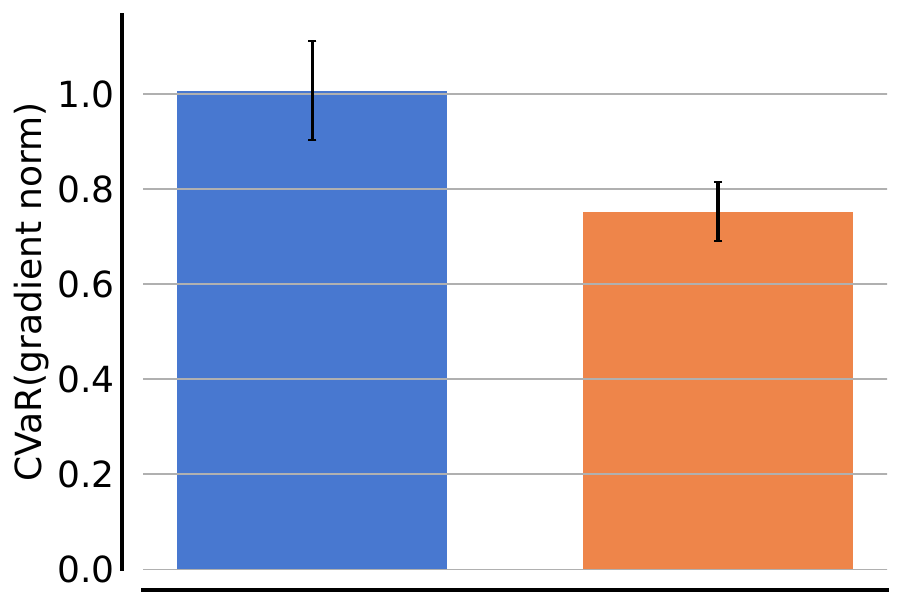}
        \caption{Gradient CVaR}
        \label{fig:emax:ind_gradcvar}
    \end{subfigure}

    \caption{Mixed-objective evaluation across 11 tasks. Evaluation returns in (a) LBF and (b) RWARE, (c) performance profiles across all tasks, and (d) average and standard error of gradient stability~\seehere{eq:emax:grad_cvar}.}
    \label{fig:emax:ind_deep_results}
\end{figure}

\textbf{Evaluation metrics:} We report the mean evaluation returns as well as 95\% confidence intervals computed over five runs in all individual tasks across both settings. In common-reward tasks, we report the returns computed over the common rewards, and in mixed-objective tasks we report the sum of all agents' evaluation returns. Following the methodology of \citet{agarwal2021deep}, we report aggregated normalised evaluation returns\footnote{We follow the task-based normalisation procedure of \citet{papoudakis2021benchmarking}.} and performance profiles with the interquartile mean (IQM) and 95\% confidence intervals computed over all tasks in each setting. The learning curves indicate the sample efficiency of agents, and performance profiles allow to compare the distribution of final evaluation returns indicating the robustness of the final policies learned by each algorithm. 

To evaluate the training stability of algorithms, we would like to capture how variable and noisy gradients are during training. To measure this variability, we detrend gradient norms by deducting each gradient norm from its subsequent norm, and compute the conditional value at risk (CVaR) of detrended gradient norms:
\begin{equation}
    \text{CVaR}(g') = \mathbb{E}\left[g' \mid g' \geq \text{VaR}_{95\%}(g')\right]\ \text{and} \ g_t' = |\nabla_{t+1}| - |\nabla_t|
    \label{eq:emax:grad_cvar}
\end{equation}
where the value at risk (VaR) corresponds to the value at the 95\% quantile of all detrended gradient norm values. This metric corresponds to the short-term risk across time suggested by \citet{chan2020measuring}. A larger CVaR value indicates more variability in gradients which can indicate unstable training, while a smaller CVaR value indicates less variability in gradients and more stable training.

\textbf{Implementation details:} In all experiments, agents share parameters with each other to improve sample efficiency~\citep{christianos2021scaling,papoudakis2021benchmarking}. To allow for agent specialisation, shared networks receive one-hot vectors that indicate agents' identity as additional inputs. Unless stated otherwise, EMAX trains an ensemble of $K=5$ value functions. For more details on chosen hyperparameters, see \Cref{app:sec:emax:deep_hyperparams}.

\subsection{Evaluation Results}
\label{sec:emax:experiments_results}

\Cref{fig:emax:ind_deep_results} shows the learning curves of IDQN with and without EMAX in 11 mixed-objective tasks in LBF and RWARE with normalised evaluation returns, and a performance profile at the end of training. Across all 11 tasks, EMAX improves final evaluation returns of IDQN by 189\%, with 105\% and 275\% improvement in LBF and RWARE, respectively. The performance profile also shows that EMAX significantly improves the robustness of IDQN. \Cref{app:sec:emax:ind_deep_results} provides learning curves in each individual task.

\begin{figure*}[t]
    \centering
    \includegraphics[height=2.5em]{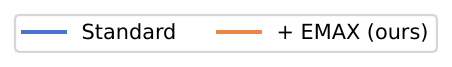}

    \begin{subfigure}{.49\linewidth}
        \centering
        \includegraphics[width=\linewidth]{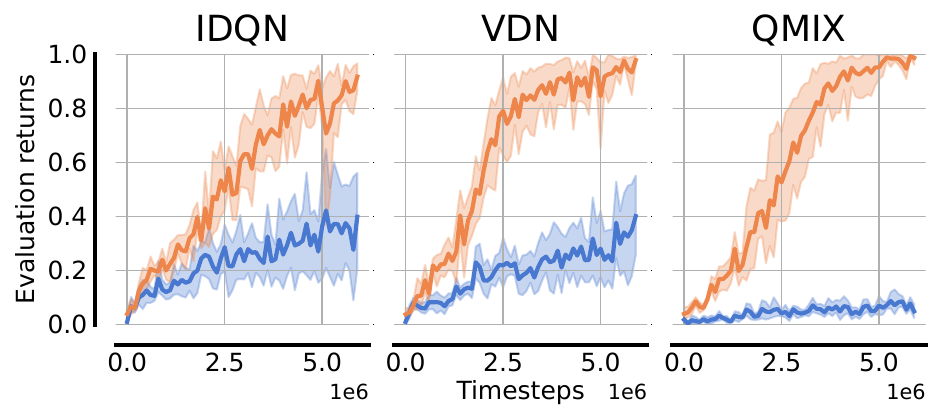}
        \caption{Evaluation returns}
        \label{fig:emax:lbf-10x10-3p-5f_returns}
    \end{subfigure}
    \begin{subfigure}{.49\linewidth}
        \centering
        \includegraphics[width=\linewidth]{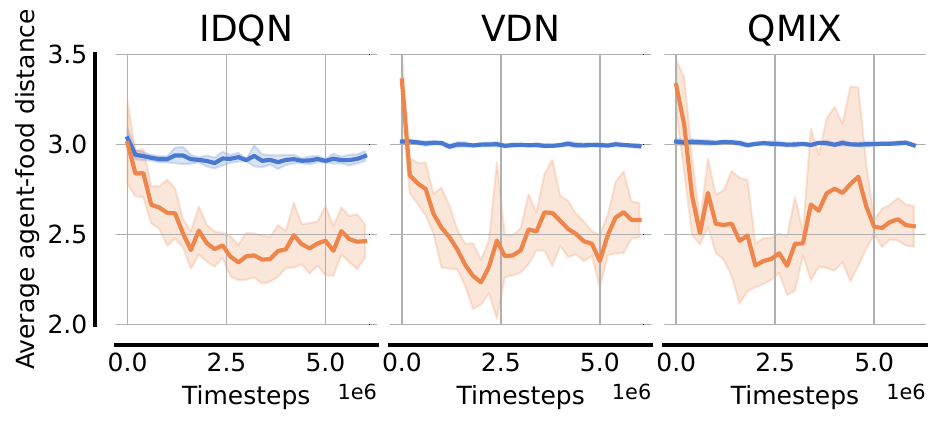}
        \caption{Food distances}
        \label{fig:emax:lbf-10x10-3p-5f_food_dist}
    \end{subfigure}
    \begin{subfigure}{.49\linewidth}
        \centering
        \includegraphics[height=2em]{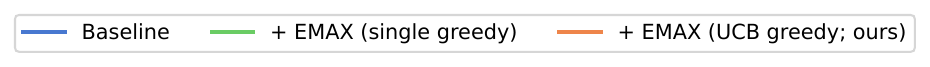}
        \includegraphics[width=\linewidth]{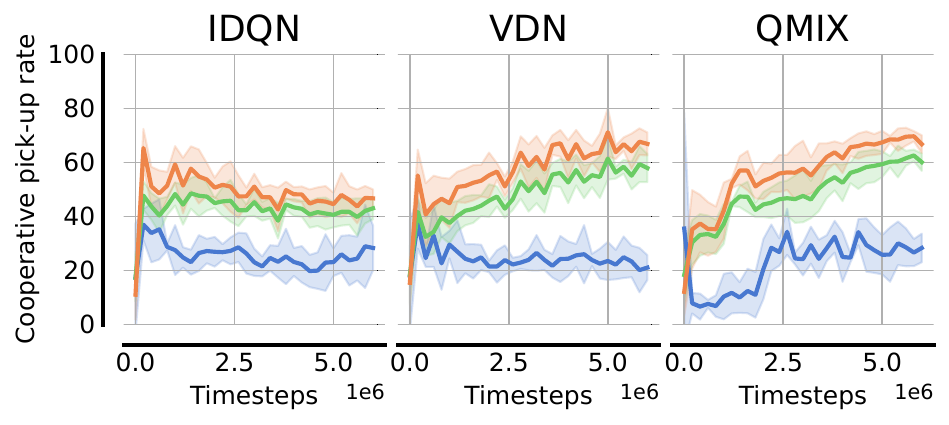}
        \caption{Cooperation percentage}
        \label{fig:emax:lbf-10x10-3p-5f_coop_percentage}
    \end{subfigure}
    \begin{subfigure}{.49\linewidth}
        \centering
        \includegraphics[height=2em]{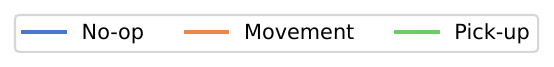}
        \includegraphics[width=\linewidth]{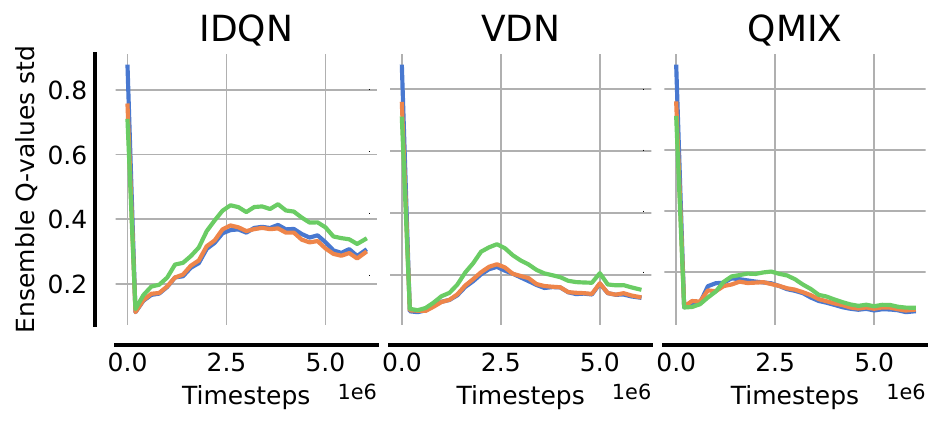}
        \caption{Ensemble Q-value std}
        \label{fig:emax:lbf-10x10-3p-5f_coop_q-values_std}
    \end{subfigure}
    \caption{Exploration policy analysis in LBF 10x10-3p-5f with common rewards. (a) Mean and 95\% confidence intervals of evaluation returns, mean and standard deviation of (b) average food distances across rollouts, (c) percentages of agents selecting the pick-up action in states that require coordination, and (d) the standard deviation of value estimates for the no-op, movement, and pick-up actions in states that require coordination.}
    \label{fig:emax:exploration_policy_performance}
\end{figure*}

Following the same evaluation protocol, we evaluate EMAX on top of IDQN, VDN, and QMIX across 21 common-reward tasks. \Cref{fig:emax:coop_deep_results} visualises the learning curve and performance profile of evaluation returns of all algorithms. Similar to mixed-objective tasks, EMAX substantially improves final evaluation returns of IDQN, VDN, and QMIX in common-reward tasks, shown in \Cref{fig:emax:norm_returns_all}, by 60\%, 47\%, and 538\%, leading to higher final returns compared to their vanilla baselines in 18, 16, and 20 out of 21 tasks, respectively. These results arise from EMAX improving the sample efficiency and learning stability of the vanilla algorithms, as we will show in \Cref{sec:emax:analysis}. Additionally, QMIX-EMAX is able to learn effective policies in several hard exploration tasks where QMIX fails to achieve any reward. From the performance profile in \Cref{fig:emax:deep_profiles} we also see that algorithms with EMAX achieve higher returns with a higher probability at the end of training. We provide normalised evaluation returns for each environment and learning curves in individual tasks in \Cref{app:sec:emax:deep_norm_eval_returns} and \Cref{app:sec:emax:deep_ind_eval_returns}, respectively.

In LBF, EMAX significantly improves the performance of QMIX whereas minor improvements can be seen for IDQN and VDN in the common-reward setting~\seehere{fig:emax:lbf_norm}, and significant gains are observed in the mixed-objective setting~\seehere{fig:emax:lbf_ind_norm_returns}. Learning curves of individual tasks~\seehere{app:sec:emax:ind_deep_results} show that QMIX, MAVEN, CDS and EMC fail to achieve any rewards in several LBF tasks with particularly sparse rewards. A similar trend can be observed in BPUSH where, most notably, VDN-EMAX and QMIX-EMAX learn to solve a BPUSH task in which four agents need to cooperate and no baseline demonstrates any positive rewards (see \Cref{app:fig:emax:bpush_tiny-4ag-easy}). 

In RWARE, prior work found that no value-based algorithm was able to achieve notable rewards within four million time steps of training due to the significant sparsity of rewards, and on-policy algorithms like IPPO and MAPPO vastly outperformed value-based algorithms~\citep{papoudakis2021benchmarking}. In contrast, IDQN-EMAX is able to achieve notable rewards in all RWARE tasks and outperforms all baselines, including both IPPO and MAPPO, in four out of six RWARE tasks. To the best of our knowledge, IDQN-EMAX is the first value-based algorithm that outperforms on-policy algorithms such as IPPO and MAPPO in RWARE tasks. IDQN and VDN achieve 330\% and 252\% higher final evaluation returns with EMAX than their vanilla algorithms, respectively, whereas QMIX with and without EMAX fail to learn~\seehere{fig:emax:rware_norm}. Similarly significant improvements can be seen for IDQN in the mixed-objective setting~\seehere{fig:emax:rware_ind_norm_returns}. %

\looseness=-1
Lastly, we evaluate in three common-reward tasks of the MPE environment. In contrast to other environments, MPE features continuous observations and dense rewards, and the adversary and predator-prey tasks contain stochastic transitions due to the adversarial agent being controlled by a pre-trained policy. In all three MPE tasks, we see improvements in sample efficiency and final performance for algorithms with EMAX compared to vanilla algorithms, even if the improvements are less severe than in the other environments that feature sparse rewards. These improvements are particularly notable in the predator-prey and adversary tasks that feature stochastic transitions due to pretrained policy of adversary agents, indicating that EMAX is able to effectively guide the exploration even in environments with such stochasticity.

\subsection{Analysis}
\label{sec:emax:analysis}

\begin{figure*}[t]
    \centering

    \begin{subfigure}{.49\linewidth}
        \centering
        \includegraphics[height=1.8em]{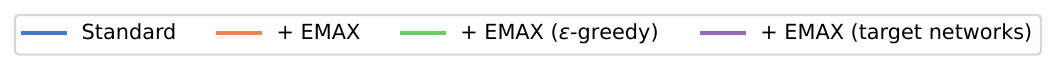}
        \includegraphics[width=\linewidth]{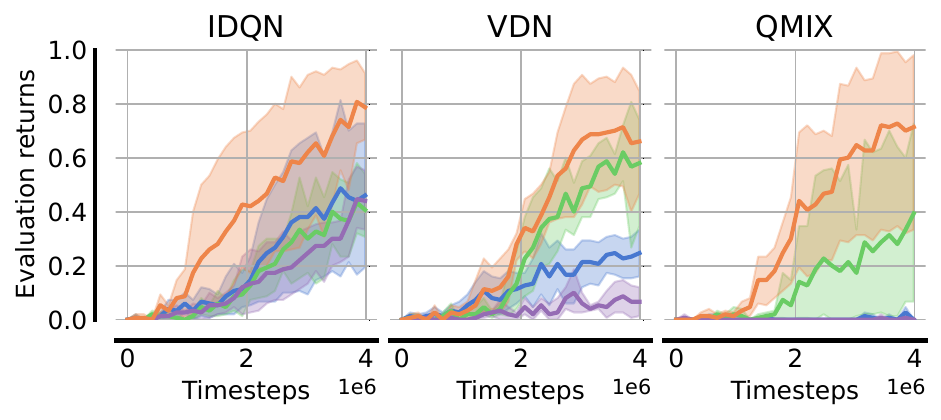}

        \caption{Exploration policy and target ablations}
        \label{fig:emax:lbf-10x10-4p-3f-coop-ablation}
    \end{subfigure}
    \begin{subfigure}{.49\linewidth}
        \centering
        \includegraphics[height=1.8em]{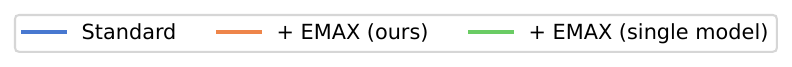}
        \includegraphics[width=.99\linewidth]{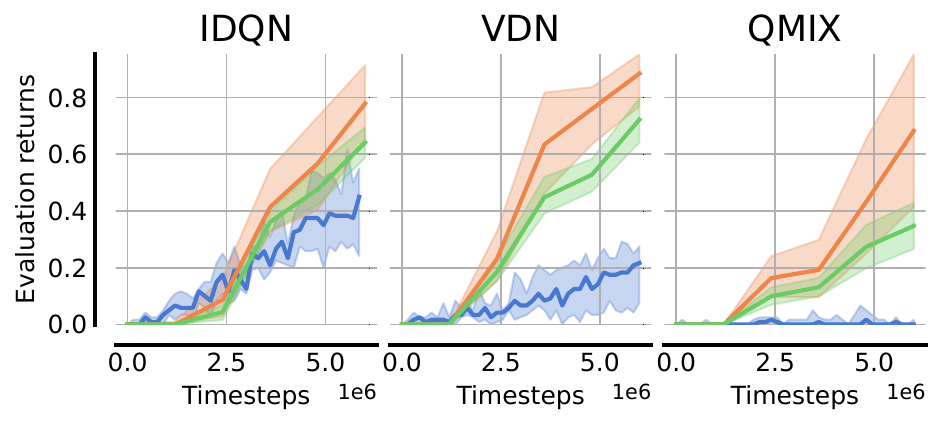}
        \caption{Evaluation policy ablation}
        \label{fig:emax:eval_policy_ablation}
    \end{subfigure}
    \caption{Evaluation returns for all vanilla and EMAX algorithms with ablations of (a) the exploration policy and target computation in LBF 10x10-4p-3f-coop, and (b) of the evaluation policy in LBF 10x10-4p-4f-coop task. We ablate the EMAX exploration policy with an $\epsilon$-greedy policy (green in a), the EMAX target computation with target networks (purple in a), and the EMAX evaluation policy by greedily following any single value function within the ensemble (b).}
\end{figure*}

We now further investigate the efficacy of all components of EMAX to study our hypotheses that (1) EMAX targets reduce the variability of gradients during training, (2) the EMAX exploration policy leads to more exploration of states and actions with the potential for coordination, and (3) the EMAX evaluation policy reduces the likelihood of selecting sub-optimal actions.

\textbf{Training stability:} To demonstrate that EMAX target computation reduces the variability of gradients during training, and, thus, improves stability of the optimisation, we visualise the average and standard error of the stability of gradients, as defined in \Cref{eq:emax:grad_cvar}. We observe that for IDQN with and without EMAX across 11 mixed-objective tasks \seehere{fig:emax:ind_gradcvar} as well as for IDQN, VDN, QMIX with and without EMAX across all 21 common-reward tasks \seehere{fig:emax:coop_gradcvar}, EMAX significantly reduces the CVaR of gradient norms for all algorithms. These results indicate more stable optimisation and confirm our hypothesis. The difference for QMIX in the common-reward setting is less pronounced since the base algorithm fails to learn in several tasks, leading to training with low gradient variability independent of target values. 

\textbf{Exploration policy:} To validate our hypothesis that the EMAX exploration policy leads to more exploration of states and actions with the potential for coordination (\Cref{sec:emax:methodology}), we train IDQN, VDN, and QMIX with and without EMAX in the LBF 10x10-3p-5f task with common rewards where agents need to cooperate to pick up some of the food items. \Cref{fig:emax:exploration_policy_performance} shows the evaluation returns throughout training, the average distances of agents to the closest food, and the percentage of agents selecting the pick-up action in states where multiple agents need to coordinate their actions to pick up food.\footnote{Average distances to food and cooperation rates are determined over 50 rollout episodes of the exploration policy of baseline algorithms and EMAX every \num{200000} time steps of training.} These results validate our hypotheses about the EMAX exploration policy in the tested task. We observe that agents following the EMAX exploration policy (1) seek out states with the potential for coordination more often compared to the baseline following a random exploration policy, as indicated by the lower average distance of EMAX agents to food items compared to the baseline in \Cref{fig:emax:lbf-10x10-3p-5f_food_dist}, and (2) are more likely to select the cooperative pick-up action in states with potential for coordination, as shown in \Cref{fig:emax:lbf-10x10-3p-5f_coop_percentage}. Together, these effects lead to EMAX agents learning significantly more efficiently and achieving higher evaluation returns compared to the baseline~\seehere{fig:emax:lbf-10x10-3p-5f_returns}.

To separate of the exploration policy and other components of EMAX, we also compare to the percentage of choosing the pick-up action in states that require coordination by greedily following any individual value functions in the ensemble instead of following the UCB policy. While this ablation leads to a significant improvement over the vanilla algorithms, it still exhibits a lower rate of coordinating compared to the EMAX exploration policy \seehere{fig:emax:lbf-10x10-3p-5f_coop_percentage}.

Lastly, \Cref{fig:emax:lbf-10x10-3p-5f_coop_q-values_std} visualises the standard deviation of action-value estimates across the ensemble for the no-op action, movement actions, and the pick-up action in states with the potential for cooperation between agents. This plot shows that the value estimate deviations across the ensemble are similar for all actions early in training but once agents sometimes cooperate successfully and sometimes fail to cooperate, the deviation for the pick-up action with potential for cooperation rises higher than the deviation for other actions in states with the potential for cooperation. Furthermore, alongside \Cref{fig:emax:lbf-10x10-3p-5f_returns} we can see that once agents successfully cooperate most of the time (indicated by high returns), the standard deviation for the pick-up action starts to reduce. For QMIX with EMAX, we can see that this reduction ends in the standard deviation of action values for the cooperative pick-up action and non-cooperative actions reaching similar levels once close-to-optimal performance is reached since now agents almost always cooperative successfully. This further indicates that, as desired, EMAX incentivises exploration of cooperative actions as long as such cooperation is not reliably achieved yet, but this bias towards cooperative actions diminishes as the policy starts to reliably cooperate successfully.

\textbf{Ablations:} To demonstrate the importance of all components of the EMAX algorithm to its performance, we provide ablations of its main components. First, we ablate the exploration policy and target computation and evaluate them in the LBF 10x10-4p-3f-coop task with common rewards \seehere{fig:emax:lbf-10x10-4p-3f-coop-ablation}. In these ablations, we replace the EMAX exploration policy with an $\epsilon$-greedy policy, and substitute the target computation with target networks in which each value function in the ensemble has its own target network. We observe that both components significantly improve the performance of all algorithms. Second, we ablate the EMAX evaluation policy in the LBF 10x10-4p-4f-coop task \seehere{fig:emax:eval_policy_ablation} by following the greedy policy with respect to any of the individual value functions within the ensemble. We highlight that no value functions were trained for this ablation so the only difference between the ablation and EMAX is the followed policy, not the underlying value functions. This experiment indicates the improved robustness in performance resulting from the EMAX evaluation policy.

\begin{figure}[t]
    \centering
    \includegraphics[width=\linewidth]{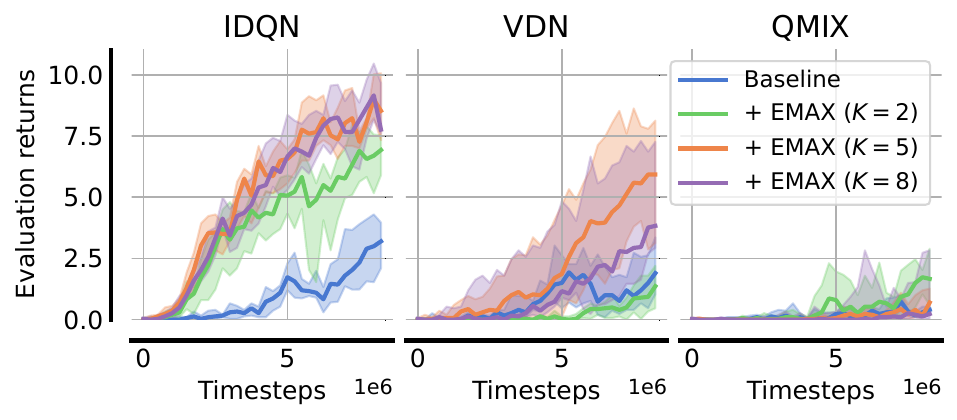}
    \label{fig:emax:rware-tiny-4ag_ensemble}
    \caption{Evaluation returns for varying ensemble sizes $K\in\{2,5,8\}$ in the RWARE $11\times10$ 4ag task.}
\end{figure}

\textbf{Ensemble size:} Training ensemble models is expensive and its cost scales with the ensemble size $K$. We report the training cost for all algorithms and varying $K$ in \Cref{app:sec:emax:ensemble_size_cost}. To identify how many models are needed for the benefits of EMAX, we train all algorithms with varying $K$ in the RWARE 11x10 task with four agents and common rewards (\Cref{fig:emax:rware-tiny-4ag_ensemble}), in which EMAX led to substantial improvements for IDQN and VDN. We observe that the benefits of EMAX saturate at $K=5$, and larger ensemble sizes ($K=8$) can result in worse performance. We hypothesise that larger ensembles might require more data to train, thus leading to diminishing benefits for ensembles of many value functions. Lastly, we compare EMAX to the baselines with larger non-ensemble networks and find that even at comparable or larger computational budget, the baselines perform significantly worse than EMAX~\seehere{app:sec:emax:large_network_baselines}.

\section{Related Work}
\label{sec:emax:related_work}

\textbf{Uncertainty for exploration in RL:}
Using uncertainty to guide exploration is a well-established idea in RL . One family of algorithms that leverages this idea are randomised value functions~\citep{osband2019deep} that build on the idea of Thompson sampling~\citep{thompson1933likelihood} from the multi-armed bandits literature~\citep{scott2010modern,chapelle2011empirical}. Posterior sampling RL extends Thompson sampling by maintaining a distribution of plausible tasks, computes optimal policies for sampled tasks, and continually updates its distribution of tasks from the collected experience~\citep{osband2013more}. This approach has extensive theoretical guarantees~\citep{osband2017posterior} but is difficult to apply to complex tasks~\citep{osband2016generalization}. This limitation has been addressed in subsequent works~\citep{osband2016generalization,osband2016deep,janz2019successor}, most notably in bootstrapped DQN~\citep{osband2016deep} which approximates randomised value functions with an ensemble of value functions. SUNRISE~\citep{lee2021sunrise} and MeanQ~\citep{liang2022reducing} also leverage an ensemble of value functions but instead of sampling value functions to explore, they follow a UCB policy using the average and standard deviation of value estimates across the ensemble to explore. Additionally, SUNRISE computes a weighting of values loss terms based on the variance of target values, and MeanQ stabilises the optimisation by computing lower variance target values as the average value estimate across the ensemble~\citep{anschel2017averaged}. Separately, \citet{fu2022model} extend posterior sampling to model-based RL by learning a probabilistic model of the environment, and \citet{dearden1998bayesian} applied these ideas to tabular Q-learning to learn distributions over Q-values and approximate the value of information of actions. Related to all these ideas, optimistic value estimates in the face of uncertainty can be used to promote exploration for actor-critic~\citep{ciosek2019better} and model-based RL~\citep{sessa2022efficient}. All these approaches leverage uncertainty to guide their exploration, similar to EMAX. However, in contrast to discussed approaches, EMAX focuses on exploration of agent coordination in environments with multiple concurrently learning agents.

\looseness=-1
\textbf{Multi-agent exploration:} There exist a plethora of exploration methods in the multi-agent setting. Several approaches provide agents with intrinsic incentives to explore, e.g.\ by rewarding them for interacting with each other as measured by influences on their transitions or value estimates~\citep{wang2019influence} or for reaching identified goal states~\citep{liu2021cooperative}. However, intrinsic rewards for exploration have to be carefully balanced for each task due to the modified optimisation objective~\citep{schaefer2022derl}. To address this challenge, LIGS~\citep{mguni2022ligs} trains an agent to determine when and which intrinsic reward should be given to each agents. Orthogonally to this line of work, experience and parameter sharing have been leveraged to greatly improve sample efficiency for MARL by synchronising agents' learning and make use of more data~\citep{christianos2020shared,christianos2021scaling}. However, there is little research using distributional and ensemble-based techniques for MARL exploration. \citet{zhou2019posterior} extend posterior sampling~\citep{osband2013more} for MARL, but are limited to two-player zero-sum extensive games. We aim to close this gap by proposing EMAX, an ensemble-based technique for efficient exploration in MARL. We further highlight that EMAX is a plug-and-play algorithm that can enhance any value-based MARL algorithm, including most existing MARL exploration techniques described in this paragraph.

\section{Conclusion}
\label{sec:emax:conclusion}
\looseness=-1
In this paper, we proposed EMAX, a general framework to seamlessly extend any value-based MARL algorithm using ensembles of value functions. EMAX leverages the disagreement of value estimates across the ensemble to systematically guide the exploration of agents towards parts of the environment that might require them to coordinate with other agents. Additionally, EMAX computes low-variance target values across the ensemble to stabilise MARL training that is otherwise prone to be unstable due to the non-stationarity of the policies of other agents, and reduces the risk of miscoordination by computing optimal actions through a majority vote across the ensemble. We empirically demonstrated the benefits of EMAX for sample efficiency, final performance, and training stability as an extension of IDQN, VDN, and QMIX across 11 mixed-objective tasks and 21 common-reward tasks across four environments. Further analysis and ablations established the efficacy of the EMAX exploration policy, target computation, and evaluation policy of EMAX, and we discussed the computational cost introduced by EMAX with experiments indicating that comparably small ensemble models are sufficient to achieve the demonstrated improvements. We believe that EMAX is a promising approach to improve exploration in MARL due to its plug-and-play nature and demonstrated efficacy in complex tasks. In this work, we design EMAX as an extension of value-based MARL algorithms but future work could investigate the application of EMAX to actor-critic MARL algorithms such as IPPO and MAPPO by training an ensemble of value functions and policies. A further limitation of the current EMAX approach is the considerable computation cost of training ensembles of value functions. Future work could consider the application of hypernetworks~\citep{dwaracherla2020hypermodels} or latent-conditioned models~\citep{shen2021implicit} to approximate the ensemble and, thereby, reduce the computational cost of EMAX.

\begin{acks}
    We would like to thank Jun Wang, Matthieu Zimmer, Alexandre Maraval, and Juliusz Ziomek for providing helpful feedback during the ideation of this project. Their comments have helped to focus and strengthen this work.
\end{acks}

\bibliographystyle{ACM-Reference-Format} 
\bibliography{references}


\begin{thebibliography}{49}


\ifx \showCODEN    \undefined \def \showCODEN     #1{\unskip}     \fi
\ifx \showDOI      \undefined \def \showDOI       #1{#1}\fi
\ifx \showISBNx    \undefined \def \showISBNx     #1{\unskip}     \fi
\ifx \showISBNxiii \undefined \def \showISBNxiii  #1{\unskip}     \fi
\ifx \showISSN     \undefined \def \showISSN      #1{\unskip}     \fi
\ifx \showLCCN     \undefined \def \showLCCN      #1{\unskip}     \fi
\ifx \shownote     \undefined \def \shownote      #1{#1}          \fi
\ifx \showarticletitle \undefined \def \showarticletitle #1{#1}   \fi
\ifx \showURL      \undefined \def \showURL       {\relax}        \fi
\providecommand\bibfield[2]{#2}
\providecommand\bibinfo[2]{#2}
\providecommand\natexlab[1]{#1}
\providecommand\showeprint[2][]{arXiv:#2}

\bibitem[\protect\citeauthoryear{Agarwal, Schwarzer, Castro, Courville, and Bellemare}{Agarwal et~al\mbox{.}}{2021}]%
        {agarwal2021deep}
\bibfield{author}{\bibinfo{person}{Rishabh Agarwal}, \bibinfo{person}{Max Schwarzer}, \bibinfo{person}{Pablo~Samuel Castro}, \bibinfo{person}{Aaron~C. Courville}, {and} \bibinfo{person}{Marc Bellemare}.} \bibinfo{year}{2021}\natexlab{}.
\newblock \showarticletitle{Deep reinforcement learning at the edge of the statistical precipice}. In \bibinfo{booktitle}{\emph{Advances in Neural Information Processing Systems}}.
\newblock


\bibitem[\protect\citeauthoryear{Albrecht, Christianos, and Sch\"afer}{Albrecht et~al\mbox{.}}{2024}]%
        {marl-book}
\bibfield{author}{\bibinfo{person}{Stefano~V. Albrecht}, \bibinfo{person}{Filippos Christianos}, {and} \bibinfo{person}{Lukas Sch\"afer}.} \bibinfo{year}{2024}\natexlab{}.
\newblock \bibinfo{booktitle}{\emph{Multi-Agent Reinforcement Learning: Foundations and Modern Approaches}}.
\newblock \bibinfo{publisher}{MIT Press}.
\newblock
\urldef\tempurl%
\url{https://www.marl-book.com}
\showURL{%
\tempurl}


\bibitem[\protect\citeauthoryear{Albrecht and Ramamoorthy}{Albrecht and Ramamoorthy}{2013}]%
        {albrecht2013game}
\bibfield{author}{\bibinfo{person}{Stefano~V. Albrecht} {and} \bibinfo{person}{Subramanian Ramamoorthy}.} \bibinfo{year}{2013}\natexlab{}.
\newblock \showarticletitle{A Game-Theoretic Model and Best-Response Learning Method for Ad Hoc Coordination in Multiagent Systems}. In \bibinfo{booktitle}{\emph{International Conference on Autonomous Agents and Multi-Agent Systems}}.
\newblock


\bibitem[\protect\citeauthoryear{Anschel, Baram, and Shimkin}{Anschel et~al\mbox{.}}{2017}]%
        {anschel2017averaged}
\bibfield{author}{\bibinfo{person}{Oron Anschel}, \bibinfo{person}{Nir Baram}, {and} \bibinfo{person}{Nahum Shimkin}.} \bibinfo{year}{2017}\natexlab{}.
\newblock \showarticletitle{Averaged-{DQN}: Variance reduction and stabilization for deep reinforcement learning}. In \bibinfo{booktitle}{\emph{International Conference on Machine Learning}}.
\newblock


\bibitem[\protect\citeauthoryear{Auer}{Auer}{2002}]%
        {auer2002using}
\bibfield{author}{\bibinfo{person}{Peter Auer}.} \bibinfo{year}{2002}\natexlab{}.
\newblock \showarticletitle{Using confidence bounds for exploitation-exploration trade-offs}.
\newblock \bibinfo{journal}{\emph{Journal of Machine Learning Research}}  \bibinfo{volume}{3} (\bibinfo{year}{2002}).
\newblock


\bibitem[\protect\citeauthoryear{Bernstein, Givan, Immerman, and Zilberstein}{Bernstein et~al\mbox{.}}{2002}]%
        {bernstein2002complexity}
\bibfield{author}{\bibinfo{person}{Daniel~S. Bernstein}, \bibinfo{person}{Robert Givan}, \bibinfo{person}{Neil Immerman}, {and} \bibinfo{person}{Shlomo Zilberstein}.} \bibinfo{year}{2002}\natexlab{}.
\newblock \showarticletitle{The complexity of decentralized control of {Markov} decision processes}.
\newblock \bibinfo{journal}{\emph{Mathematics of Operations Research}} (\bibinfo{year}{2002}).
\newblock


\bibitem[\protect\citeauthoryear{Chan, Fishman, Canny, Korattikara, and Guadarrama}{Chan et~al\mbox{.}}{2020}]%
        {chan2020measuring}
\bibfield{author}{\bibinfo{person}{Stephanie~C.Y. Chan}, \bibinfo{person}{Samuel Fishman}, \bibinfo{person}{John Canny}, \bibinfo{person}{Anoop Korattikara}, {and} \bibinfo{person}{Sergio Guadarrama}.} \bibinfo{year}{2020}\natexlab{}.
\newblock \showarticletitle{Measuring the reliability of reinforcement learning algorithms}. In \bibinfo{booktitle}{\emph{International Conference on Learning Representations}}.
\newblock


\bibitem[\protect\citeauthoryear{Chapelle and Li}{Chapelle and Li}{2011}]%
        {chapelle2011empirical}
\bibfield{author}{\bibinfo{person}{Olivier Chapelle} {and} \bibinfo{person}{Lihong Li}.} \bibinfo{year}{2011}\natexlab{}.
\newblock \showarticletitle{An empirical evaluation of {Thompson} sampling}. In \bibinfo{booktitle}{\emph{Advances in Neural Information Processing Systems}}.
\newblock


\bibitem[\protect\citeauthoryear{Cho, Van~Merri{\"e}nboer, G{\"u}l{\c{c}}ehre, Bahdanau, Bougares, Schwenk, and Bengio}{Cho et~al\mbox{.}}{2014}]%
        {cho2014learning}
\bibfield{author}{\bibinfo{person}{Kyunghyun Cho}, \bibinfo{person}{Bart Van~Merri{\"e}nboer}, \bibinfo{person}{{\c{C}}aglar G{\"u}l{\c{c}}ehre}, \bibinfo{person}{Dzmitry Bahdanau}, \bibinfo{person}{Fethi Bougares}, \bibinfo{person}{Holger Schwenk}, {and} \bibinfo{person}{Yoshua Bengio}.} \bibinfo{year}{2014}\natexlab{}.
\newblock \showarticletitle{Learning phrase representations using {RNN} encoder-decoder for statistical machine translation}. In \bibinfo{booktitle}{\emph{Conference on Empirical Methods in Natural Language Processing}}.
\newblock


\bibitem[\protect\citeauthoryear{Christianos, Papoudakis, and Albrecht}{Christianos et~al\mbox{.}}{2023}]%
        {christianos2022pareto}
\bibfield{author}{\bibinfo{person}{Filippos Christianos}, \bibinfo{person}{Georgios Papoudakis}, {and} \bibinfo{person}{Stefano~V. Albrecht}.} \bibinfo{year}{2023}\natexlab{}.
\newblock \showarticletitle{Pareto Actor-Critic for Equilibrium Selection in Multi-Agent Reinforcement Learning}.
\newblock \bibinfo{journal}{\emph{Transactions on Machine Learning Research}} (\bibinfo{year}{2023}).
\newblock


\bibitem[\protect\citeauthoryear{Christianos, Papoudakis, Rahman, and Albrecht}{Christianos et~al\mbox{.}}{2021}]%
        {christianos2021scaling}
\bibfield{author}{\bibinfo{person}{Filippos Christianos}, \bibinfo{person}{Georgios Papoudakis}, \bibinfo{person}{Muhammad~A. Rahman}, {and} \bibinfo{person}{Stefano~V. Albrecht}.} \bibinfo{year}{2021}\natexlab{}.
\newblock \showarticletitle{Scaling multi-agent reinforcement learning with selective parameter sharing}. In \bibinfo{booktitle}{\emph{International Conference on Machine Learning}}.
\newblock


\bibitem[\protect\citeauthoryear{Christianos, Sch\"afer, and Albrecht}{Christianos et~al\mbox{.}}{2020}]%
        {christianos2020shared}
\bibfield{author}{\bibinfo{person}{Filippos Christianos}, \bibinfo{person}{Lukas Sch\"afer}, {and} \bibinfo{person}{Stefano~V. Albrecht}.} \bibinfo{year}{2020}\natexlab{}.
\newblock \showarticletitle{Shared Experience Actor-Critic for Multi-Agent Reinforcement Learning}. In \bibinfo{booktitle}{\emph{Advances in Neural Information Processing Systems}}.
\newblock


\bibitem[\protect\citeauthoryear{Ciosek, Vuong, Loftin, and Hofmann}{Ciosek et~al\mbox{.}}{2019}]%
        {ciosek2019better}
\bibfield{author}{\bibinfo{person}{Kamil Ciosek}, \bibinfo{person}{Quan Vuong}, \bibinfo{person}{Robert Loftin}, {and} \bibinfo{person}{Katja Hofmann}.} \bibinfo{year}{2019}\natexlab{}.
\newblock \showarticletitle{Better Exploration with Optimistic Actor Critic}. In \bibinfo{booktitle}{\emph{Advances in Neural Information Processing Systems}}.
\newblock


\bibitem[\protect\citeauthoryear{Dearden, Friedman, and Russell}{Dearden et~al\mbox{.}}{1998}]%
        {dearden1998bayesian}
\bibfield{author}{\bibinfo{person}{Richard Dearden}, \bibinfo{person}{Nir Friedman}, {and} \bibinfo{person}{Stuart Russell}.} \bibinfo{year}{1998}\natexlab{}.
\newblock \showarticletitle{{Bayesian} {Q}-learning}.
\newblock \bibinfo{journal}{\emph{AAAI}} (\bibinfo{year}{1998}).
\newblock


\bibitem[\protect\citeauthoryear{Du, Han, Fang, Liu, Dai, and Tao}{Du et~al\mbox{.}}{2019}]%
        {du2019liir}
\bibfield{author}{\bibinfo{person}{Yali Du}, \bibinfo{person}{Lei Han}, \bibinfo{person}{Meng Fang}, \bibinfo{person}{Ji Liu}, \bibinfo{person}{Tianhong Dai}, {and} \bibinfo{person}{Dacheng Tao}.} \bibinfo{year}{2019}\natexlab{}.
\newblock \showarticletitle{{LIIR}: Learning Individual Intrinsic Reward in Multi-Agent Reinforcement Learning}. In \bibinfo{booktitle}{\emph{Advances in Neural Information Processing Systems}}.
\newblock


\bibitem[\protect\citeauthoryear{Dwaracherla, Lu, Ibrahimi, Osband, Wen, and van Roy}{Dwaracherla et~al\mbox{.}}{2020}]%
        {dwaracherla2020hypermodels}
\bibfield{author}{\bibinfo{person}{Vikranth Dwaracherla}, \bibinfo{person}{Xiuyuan Lu}, \bibinfo{person}{Morteza Ibrahimi}, \bibinfo{person}{Ian Osband}, \bibinfo{person}{Zheng Wen}, {and} \bibinfo{person}{Benjamin van Roy}.} \bibinfo{year}{2020}\natexlab{}.
\newblock \showarticletitle{Hypermodels for exploration}. In \bibinfo{booktitle}{\emph{International Conference on Learning Representations}}.
\newblock


\bibitem[\protect\citeauthoryear{Fu, Yu, Littman, and Konidaris}{Fu et~al\mbox{.}}{2022}]%
        {fu2022model}
\bibfield{author}{\bibinfo{person}{Haotian Fu}, \bibinfo{person}{Shangqun Yu}, \bibinfo{person}{Michael Littman}, {and} \bibinfo{person}{George Konidaris}.} \bibinfo{year}{2022}\natexlab{}.
\newblock \showarticletitle{Model-based Lifelong Reinforcement Learning with {Bayesian} Exploration}. In \bibinfo{booktitle}{\emph{Advances in Neural Information Processing Systems}}.
\newblock


\bibitem[\protect\citeauthoryear{Hansen, Bernstein, and Zilberstein}{Hansen et~al\mbox{.}}{2004}]%
        {hansen2004dynamic}
\bibfield{author}{\bibinfo{person}{Eric~A. Hansen}, \bibinfo{person}{Daniel~S. Bernstein}, {and} \bibinfo{person}{Shlomo Zilberstein}.} \bibinfo{year}{2004}\natexlab{}.
\newblock \showarticletitle{Dynamic programming for partially observable stochastic games}. In \bibinfo{booktitle}{\emph{AAAI Conference on Artificial Intelligence}}.
\newblock


\bibitem[\protect\citeauthoryear{Janz, Hron, Mazur, Hofmann, Hern{\'a}ndez-Lobato, and Tschiatschek}{Janz et~al\mbox{.}}{2019}]%
        {janz2019successor}
\bibfield{author}{\bibinfo{person}{David Janz}, \bibinfo{person}{Jiri Hron}, \bibinfo{person}{Przemys{\l}aw Mazur}, \bibinfo{person}{Katja Hofmann}, \bibinfo{person}{Jos{\'e}~Miguel Hern{\'a}ndez-Lobato}, {and} \bibinfo{person}{Sebastian Tschiatschek}.} \bibinfo{year}{2019}\natexlab{}.
\newblock \showarticletitle{Successor uncertainties: Exploration and uncertainty in temporal difference learning}. In \bibinfo{booktitle}{\emph{Advances in Neural Information Processing Systems}}.
\newblock


\bibitem[\protect\citeauthoryear{Lee, Laskin, Srinivas, and Abbeel}{Lee et~al\mbox{.}}{2021}]%
        {lee2021sunrise}
\bibfield{author}{\bibinfo{person}{Kimin Lee}, \bibinfo{person}{Michael Laskin}, \bibinfo{person}{Aravind Srinivas}, {and} \bibinfo{person}{Pieter Abbeel}.} \bibinfo{year}{2021}\natexlab{}.
\newblock \showarticletitle{{SUNRISE}: A simple unified framework for ensemble learning in deep reinforcement learning}. In \bibinfo{booktitle}{\emph{International Conference on Machine Learning}}.
\newblock


\bibitem[\protect\citeauthoryear{Li, Wang, Wu, Zhao, Yang, and Zhang}{Li et~al\mbox{.}}{2021}]%
        {li2021celebrating}
\bibfield{author}{\bibinfo{person}{Chenghao Li}, \bibinfo{person}{Tonghan Wang}, \bibinfo{person}{Chengjie Wu}, \bibinfo{person}{Qianchuan Zhao}, \bibinfo{person}{Jun Yang}, {and} \bibinfo{person}{Chongjie Zhang}.} \bibinfo{year}{2021}\natexlab{}.
\newblock \showarticletitle{Celebrating diversity in shared multi-agent reinforcement learning}. In \bibinfo{booktitle}{\emph{Advances in Neural Information Processing Systems}}.
\newblock


\bibitem[\protect\citeauthoryear{Liang, Xu, McAleer, Hu, Ihler, Abbeel, and Fox}{Liang et~al\mbox{.}}{2022}]%
        {liang2022reducing}
\bibfield{author}{\bibinfo{person}{Litian Liang}, \bibinfo{person}{Yaosheng Xu}, \bibinfo{person}{Stephen McAleer}, \bibinfo{person}{Dailin Hu}, \bibinfo{person}{Alexander Ihler}, \bibinfo{person}{Pieter Abbeel}, {and} \bibinfo{person}{Roy Fox}.} \bibinfo{year}{2022}\natexlab{}.
\newblock \showarticletitle{Reducing Variance in Temporal-Difference Value Estimation via Ensemble of Deep Networks}. In \bibinfo{booktitle}{\emph{International Conference on Machine Learning}}.
\newblock


\bibitem[\protect\citeauthoryear{Littman}{Littman}{1994}]%
        {littman1994markov}
\bibfield{author}{\bibinfo{person}{Michael~L. Littman}.} \bibinfo{year}{1994}\natexlab{}.
\newblock \showarticletitle{Markov games as a framework for multi-agent reinforcement learning}.
\newblock In \bibinfo{booktitle}{\emph{Machine Learning}}. \bibinfo{publisher}{Elsevier}, \bibinfo{pages}{157--163}.
\newblock


\bibitem[\protect\citeauthoryear{Liu, Jain, Yeh, and Schwing}{Liu et~al\mbox{.}}{2021}]%
        {liu2021cooperative}
\bibfield{author}{\bibinfo{person}{Iou-Jen Liu}, \bibinfo{person}{Unnat Jain}, \bibinfo{person}{Raymond~A. Yeh}, {and} \bibinfo{person}{Alexander Schwing}.} \bibinfo{year}{2021}\natexlab{}.
\newblock \showarticletitle{Cooperative exploration for multi-agent deep reinforcement learning}. In \bibinfo{booktitle}{\emph{International Conference on Machine Learning}}.
\newblock


\bibitem[\protect\citeauthoryear{Lowe, Wu, Tamar, Harb, Abbeel, and Mordatch}{Lowe et~al\mbox{.}}{2017}]%
        {lowe2017multi}
\bibfield{author}{\bibinfo{person}{Ryan Lowe}, \bibinfo{person}{Yi Wu}, \bibinfo{person}{Aviv Tamar}, \bibinfo{person}{Jean Harb}, \bibinfo{person}{Pieter Abbeel}, {and} \bibinfo{person}{Igor Mordatch}.} \bibinfo{year}{2017}\natexlab{}.
\newblock \showarticletitle{Multi-Agent Actor-Critic for Mixed Cooperative-Competitive Environments}. In \bibinfo{booktitle}{\emph{Advances in Neural Information Processing Systems}}.
\newblock


\bibitem[\protect\citeauthoryear{Mahajan, Rashid, Samvelyan, and Whiteson}{Mahajan et~al\mbox{.}}{2019}]%
        {mahajan2019maven}
\bibfield{author}{\bibinfo{person}{Anuj Mahajan}, \bibinfo{person}{Tabish Rashid}, \bibinfo{person}{Mikayel Samvelyan}, {and} \bibinfo{person}{Shimon Whiteson}.} \bibinfo{year}{2019}\natexlab{}.
\newblock \showarticletitle{{MAVEN}: Multi-Agent Variational Exploration}. In \bibinfo{booktitle}{\emph{Advances in Neural Information Processing Systems}}.
\newblock


\bibitem[\protect\citeauthoryear{Mguni, Jafferjee, Wang, Perez-Nieves, Slumbers, Tong, Li, Zhu, Yang, and Wang}{Mguni et~al\mbox{.}}{2022}]%
        {mguni2022ligs}
\bibfield{author}{\bibinfo{person}{David~Henry Mguni}, \bibinfo{person}{Taher Jafferjee}, \bibinfo{person}{Jianhong Wang}, \bibinfo{person}{Nicolas Perez-Nieves}, \bibinfo{person}{Oliver Slumbers}, \bibinfo{person}{Feifei Tong}, \bibinfo{person}{Yang Li}, \bibinfo{person}{Jiangcheng Zhu}, \bibinfo{person}{Yaodong Yang}, {and} \bibinfo{person}{Jun Wang}.} \bibinfo{year}{2022}\natexlab{}.
\newblock \showarticletitle{{LIGS}: Learnable Intrinsic-Reward Generation Selection for Multi-Agent Learning}. In \bibinfo{booktitle}{\emph{International Conference on Learning Representations}}.
\newblock


\bibitem[\protect\citeauthoryear{Mnih, Kavukcuoglu, Silver, Rusu, Veness, Bellemare, Graves, Riedmiller, Fidjeland, Ostrovski, Petersen, Beattie, Sadik, Antonoglou, King, Kumaran, Wierstra, Legg, and Hassabis}{Mnih et~al\mbox{.}}{2015}]%
        {mnih2015human}
\bibfield{author}{\bibinfo{person}{Volodymyr Mnih}, \bibinfo{person}{Koray Kavukcuoglu}, \bibinfo{person}{David Silver}, \bibinfo{person}{Andrei~A. Rusu}, \bibinfo{person}{Joel Veness}, \bibinfo{person}{Marc~G. Bellemare}, \bibinfo{person}{Alex Graves}, \bibinfo{person}{Martin~A. Riedmiller}, \bibinfo{person}{Andreas~K. Fidjeland}, \bibinfo{person}{Georg Ostrovski}, \bibinfo{person}{Stig Petersen}, \bibinfo{person}{Charles Beattie}, \bibinfo{person}{Amir Sadik}, \bibinfo{person}{Ioannis Antonoglou}, \bibinfo{person}{Helen King}, \bibinfo{person}{Dharshan Kumaran}, \bibinfo{person}{Daan Wierstra}, \bibinfo{person}{Shane Legg}, {and} \bibinfo{person}{Demis Hassabis}.} \bibinfo{year}{2015}\natexlab{}.
\newblock \showarticletitle{Human-level control through deep reinforcement learning}.
\newblock \bibinfo{journal}{\emph{Nature}} \bibinfo{volume}{518}, \bibinfo{number}{7540} (\bibinfo{year}{2015}), \bibinfo{pages}{529--533}.
\newblock


\bibitem[\protect\citeauthoryear{Mordatch and Abbeel}{Mordatch and Abbeel}{2018}]%
        {mordatch2018emergence}
\bibfield{author}{\bibinfo{person}{Igor Mordatch} {and} \bibinfo{person}{Pieter Abbeel}.} \bibinfo{year}{2018}\natexlab{}.
\newblock \showarticletitle{Emergence of grounded compositional language in multi-agent populations}. In \bibinfo{booktitle}{\emph{AAAI Conference on Artificial Intelligence}}.
\newblock


\bibitem[\protect\citeauthoryear{Oliehoek and Amato}{Oliehoek and Amato}{2016}]%
        {oliehoek2016concise}
\bibfield{author}{\bibinfo{person}{Frans~A. Oliehoek} {and} \bibinfo{person}{Christopher Amato}.} \bibinfo{year}{2016}\natexlab{}.
\newblock \bibinfo{booktitle}{\emph{A Concise Introduction to Decentralized {POMDPs}}}. Vol.~\bibinfo{volume}{1}.
\newblock \bibinfo{publisher}{Springer}.
\newblock


\bibitem[\protect\citeauthoryear{Osband, Blundell, Pritzel, and van Roy}{Osband et~al\mbox{.}}{2016a}]%
        {osband2016deep}
\bibfield{author}{\bibinfo{person}{Ian Osband}, \bibinfo{person}{Charles Blundell}, \bibinfo{person}{Alexander Pritzel}, {and} \bibinfo{person}{Benjamin van Roy}.} \bibinfo{year}{2016}\natexlab{a}.
\newblock \showarticletitle{Deep exploration via bootstrapped {DQN}}. In \bibinfo{booktitle}{\emph{Advances in Neural Information Processing Systems}}.
\newblock


\bibitem[\protect\citeauthoryear{Osband, Russo, and van Roy}{Osband et~al\mbox{.}}{2013}]%
        {osband2013more}
\bibfield{author}{\bibinfo{person}{Ian Osband}, \bibinfo{person}{Daniel Russo}, {and} \bibinfo{person}{Benjamin van Roy}.} \bibinfo{year}{2013}\natexlab{}.
\newblock \showarticletitle{({M}ore) efficient reinforcement learning via posterior sampling}. In \bibinfo{booktitle}{\emph{Advances in Neural Information Processing Systems}}.
\newblock


\bibitem[\protect\citeauthoryear{Osband and van Roy}{Osband and van Roy}{2017}]%
        {osband2017posterior}
\bibfield{author}{\bibinfo{person}{Ian Osband} {and} \bibinfo{person}{Benjamin van Roy}.} \bibinfo{year}{2017}\natexlab{}.
\newblock \showarticletitle{Why is posterior sampling better than optimism for reinforcement learning?}. In \bibinfo{booktitle}{\emph{International Conference on Machine Learning}}.
\newblock


\bibitem[\protect\citeauthoryear{Osband, van Roy, Russo, and Wen}{Osband et~al\mbox{.}}{2019}]%
        {osband2019deep}
\bibfield{author}{\bibinfo{person}{Ian Osband}, \bibinfo{person}{Benjamin van Roy}, \bibinfo{person}{Daniel~J. Russo}, {and} \bibinfo{person}{Zheng Wen}.} \bibinfo{year}{2019}\natexlab{}.
\newblock \showarticletitle{Deep Exploration via Randomized Value Functions}.
\newblock \bibinfo{journal}{\emph{Journal of Machine Learning Research}} \bibinfo{volume}{20}, \bibinfo{number}{124} (\bibinfo{year}{2019}).
\newblock


\bibitem[\protect\citeauthoryear{Osband, van Roy, and Wen}{Osband et~al\mbox{.}}{2016b}]%
        {osband2016generalization}
\bibfield{author}{\bibinfo{person}{Ian Osband}, \bibinfo{person}{Benjamin van Roy}, {and} \bibinfo{person}{Zheng Wen}.} \bibinfo{year}{2016}\natexlab{b}.
\newblock \showarticletitle{Generalization and exploration via randomized value functions}. In \bibinfo{booktitle}{\emph{International Conference on Machine Learning}}.
\newblock


\bibitem[\protect\citeauthoryear{Papoudakis, Christianos, Rahman, and Albrecht}{Papoudakis et~al\mbox{.}}{2019}]%
        {papoudakis2019nonstationarity}
\bibfield{author}{\bibinfo{person}{Georgios Papoudakis}, \bibinfo{person}{Filippos Christianos}, \bibinfo{person}{Arrasy Rahman}, {and} \bibinfo{person}{Stefano~V. Albrecht}.} \bibinfo{year}{2019}\natexlab{}.
\newblock \showarticletitle{Dealing with Non-Stationarity in Multi-Agent Deep Reinforcement Learning}.
\newblock \bibinfo{journal}{\emph{arXiv preprint 1906.04737}} (\bibinfo{year}{2019}).
\newblock


\bibitem[\protect\citeauthoryear{Papoudakis, Christianos, Sch\"afer, and Albrecht}{Papoudakis et~al\mbox{.}}{2021}]%
        {papoudakis2021benchmarking}
\bibfield{author}{\bibinfo{person}{Georgios Papoudakis}, \bibinfo{person}{Filippos Christianos}, \bibinfo{person}{Lukas Sch\"afer}, {and} \bibinfo{person}{Stefano~V. Albrecht}.} \bibinfo{year}{2021}\natexlab{}.
\newblock \showarticletitle{Benchmarking Multi-Agent Deep Reinforcement Learning Algorithms in Cooperative Tasks}. In \bibinfo{booktitle}{\emph{Advances in Neural Information Processing Systems, Track on Datasets and Benchmarks}}.
\newblock


\bibitem[\protect\citeauthoryear{Rashid, Samvelyan, de~Witt, Farquhar, Foerster, and Whiteson}{Rashid et~al\mbox{.}}{2020}]%
        {rashid2020monotonic}
\bibfield{author}{\bibinfo{person}{Tabish Rashid}, \bibinfo{person}{Mikayel Samvelyan}, \bibinfo{person}{Christian~Schroeder de Witt}, \bibinfo{person}{Gregory Farquhar}, \bibinfo{person}{Jakob Foerster}, {and} \bibinfo{person}{Shimon Whiteson}.} \bibinfo{year}{2020}\natexlab{}.
\newblock \showarticletitle{Monotonic value function factorisation for deep multi-agent reinforcement learning}.
\newblock \bibinfo{journal}{\emph{Journal of Machine Learning Research}} \bibinfo{volume}{21}, \bibinfo{number}{1} (\bibinfo{year}{2020}).
\newblock


\bibitem[\protect\citeauthoryear{Sch\"afer, Christianos, Hanna, and Albrecht}{Sch\"afer et~al\mbox{.}}{2022}]%
        {schaefer2022derl}
\bibfield{author}{\bibinfo{person}{Lukas Sch\"afer}, \bibinfo{person}{Filippos Christianos}, \bibinfo{person}{Josiah~P. Hanna}, {and} \bibinfo{person}{Stefano~V. Albrecht}.} \bibinfo{year}{2022}\natexlab{}.
\newblock \showarticletitle{Decoupled Reinforcement Learning to Stabilise Intrinsically-Motivated Exploration}. In \bibinfo{booktitle}{\emph{International Conference on Autonomous Agents and Multiagent Systems}}.
\newblock


\bibitem[\protect\citeauthoryear{Scott}{Scott}{2010}]%
        {scott2010modern}
\bibfield{author}{\bibinfo{person}{Steven~L. Scott}.} \bibinfo{year}{2010}\natexlab{}.
\newblock \showarticletitle{A modern {Bayesian} look at the multi-armed bandit}.
\newblock \bibinfo{journal}{\emph{Applied Stochastic Models in Business and Industry}} \bibinfo{volume}{26}, \bibinfo{number}{6} (\bibinfo{year}{2010}).
\newblock


\bibitem[\protect\citeauthoryear{Sessa, Kamgarpour, and Krause}{Sessa et~al\mbox{.}}{2022}]%
        {sessa2022efficient}
\bibfield{author}{\bibinfo{person}{Pier~Giuseppe Sessa}, \bibinfo{person}{Maryam Kamgarpour}, {and} \bibinfo{person}{Andreas Krause}.} \bibinfo{year}{2022}\natexlab{}.
\newblock \showarticletitle{Efficient Model-based Multi-agent Reinforcement Learning via Optimistic Equilibrium Computation}. In \bibinfo{booktitle}{\emph{International Conference on Machine Learning}}.
\newblock


\bibitem[\protect\citeauthoryear{Shen and How}{Shen and How}{2023}]%
        {shen2021implicit}
\bibfield{author}{\bibinfo{person}{Macheng Shen} {and} \bibinfo{person}{Jonathan~P. How}.} \bibinfo{year}{2023}\natexlab{}.
\newblock \showarticletitle{Implicit Ensemble Training for Efficient and Robust Multiagent Reinforcement Learning}.
\newblock \bibinfo{journal}{\emph{Transactions on Machine Learning Research}} (\bibinfo{year}{2023}).
\newblock


\bibitem[\protect\citeauthoryear{Sunehag, Lever, Gruslys, Czarnecki, Zambaldi, Jaderberg, Lanctot, Sonnerat, Leibo, Tuyls, and Graepel}{Sunehag et~al\mbox{.}}{2018}]%
        {sunehag2018value}
\bibfield{author}{\bibinfo{person}{Peter Sunehag}, \bibinfo{person}{Guy Lever}, \bibinfo{person}{Audrunas Gruslys}, \bibinfo{person}{Wojciech~M. Czarnecki}, \bibinfo{person}{Vin{\'i}cius~Flores Zambaldi}, \bibinfo{person}{Max Jaderberg}, \bibinfo{person}{Marc Lanctot}, \bibinfo{person}{Nicolas Sonnerat}, \bibinfo{person}{Joel~Z. Leibo}, \bibinfo{person}{Karl Tuyls}, {and} \bibinfo{person}{Thore Graepel}.} \bibinfo{year}{2018}\natexlab{}.
\newblock \showarticletitle{Value-Decomposition networks for cooperative multi-agent learning}. In \bibinfo{booktitle}{\emph{International Conference on Autonomous Agents and Multi-Agent Systems}}.
\newblock


\bibitem[\protect\citeauthoryear{Tan}{Tan}{1993}]%
        {tan1993multi}
\bibfield{author}{\bibinfo{person}{Ming Tan}.} \bibinfo{year}{1993}\natexlab{}.
\newblock \showarticletitle{Multi-agent reinforcement learning: Independent vs. cooperative agents}. In \bibinfo{booktitle}{\emph{International Conference on Machine Learning}}.
\newblock


\bibitem[\protect\citeauthoryear{Thompson}{Thompson}{1933}]%
        {thompson1933likelihood}
\bibfield{author}{\bibinfo{person}{William~R. Thompson}.} \bibinfo{year}{1933}\natexlab{}.
\newblock \showarticletitle{On the likelihood that one unknown probability exceeds another in view of the evidence of two samples}.
\newblock \bibinfo{journal}{\emph{Biometrika}} \bibinfo{volume}{25}, \bibinfo{number}{3/4} (\bibinfo{year}{1933}), \bibinfo{pages}{285--294}.
\newblock


\bibitem[\protect\citeauthoryear{Wang, Wang, Wu, and Zhang}{Wang et~al\mbox{.}}{2020}]%
        {wang2019influence}
\bibfield{author}{\bibinfo{person}{Tonghan Wang}, \bibinfo{person}{Jianhao Wang}, \bibinfo{person}{Yi Wu}, {and} \bibinfo{person}{Chongjie Zhang}.} \bibinfo{year}{2020}\natexlab{}.
\newblock \showarticletitle{Influence-based multi-agent exploration}. In \bibinfo{booktitle}{\emph{International Conference on Learning Representations}}.
\newblock


\bibitem[\protect\citeauthoryear{Yu, Velu, Vinitsky, Wang, Bayen, and Wu}{Yu et~al\mbox{.}}{2022}]%
        {yu2022surprising}
\bibfield{author}{\bibinfo{person}{Chao Yu}, \bibinfo{person}{Akash Velu}, \bibinfo{person}{Eugene Vinitsky}, \bibinfo{person}{Yu Wang}, \bibinfo{person}{Alexandre Bayen}, {and} \bibinfo{person}{Yi Wu}.} \bibinfo{year}{2022}\natexlab{}.
\newblock \showarticletitle{The Surprising Effectiveness of {PPO} in Cooperative Multi-Agent Games}. In \bibinfo{booktitle}{\emph{Advances in Neural Information Processing Systems, Track on Datasets and Benchmarks}}.
\newblock


\bibitem[\protect\citeauthoryear{Zheng, Chen, Wang, He, Hu, Chen, Fan, Gao, and Zhang}{Zheng et~al\mbox{.}}{2021}]%
        {zheng2021episodic}
\bibfield{author}{\bibinfo{person}{Lulu Zheng}, \bibinfo{person}{Jiarui Chen}, \bibinfo{person}{Jianhao Wang}, \bibinfo{person}{Jiamin He}, \bibinfo{person}{Yujing Hu}, \bibinfo{person}{Yingfeng Chen}, \bibinfo{person}{Changjie Fan}, \bibinfo{person}{Yang Gao}, {and} \bibinfo{person}{Chongjie Zhang}.} \bibinfo{year}{2021}\natexlab{}.
\newblock \showarticletitle{Episodic multi-agent reinforcement learning with curiosity-driven exploration}. In \bibinfo{booktitle}{\emph{Advances in Neural Information Processing Systems}}.
\newblock


\bibitem[\protect\citeauthoryear{Zhou, Li, and Zhu}{Zhou et~al\mbox{.}}{2020}]%
        {zhou2019posterior}
\bibfield{author}{\bibinfo{person}{Yichi Zhou}, \bibinfo{person}{Jialian Li}, {and} \bibinfo{person}{Jun Zhu}.} \bibinfo{year}{2020}\natexlab{}.
\newblock \showarticletitle{Posterior sampling for multi-agent reinforcement learning: Solving extensive games with imperfect information}. In \bibinfo{booktitle}{\emph{International Conference on Learning Representations}}.
\newblock


\end{thebibliography}

\newpage
\appendix
\onecolumn

\section{Pseudocode of EMAX}
\label{app:sec:emax:pseudocode}

In this section, we provide pseudocode for training of IDQN and QMIX with EMAX in \Cref{alg:dqn-emax} and \Cref{alg:qmix-emax}, respectively. We note that the pseudocode for VDN with EMAX is analogous to the pseudocode for QMIX with EMAX without the mixing network and computing value estimates and targets as defined in \Cref{eq:emax:ensemble_vdn_estimate,eq:emax:ensemble_vdn_target}, respectively. Lastly, we provide the pseudocode for evaluations with any EMAX algorithm in \Cref{alg:emax-eval}.

\begin{algorithm}[h!]
    \caption{Training IDQN with EMAX}
    \label{alg:dqn-emax}
    \begin{algorithmic}
        \State {\bfseries Initialise:} $\{\theta_i^k\}_{k=1}^K$ of value functions $\{Q^k_i\}_{k=1}^K$ for each agent $i\in\Ag$
        \State {\bfseries Initialise:} empty episodic replay buffer $\data \gets \emptyset$
        \For{each episode}
            \State Obtain initial state $\st^0 \sim \instdist$ and joint observation $\job^0$
            \For{each step $t=0, \cdots, T$}
                \For{each agent $i\in\Ag$}
                    \State Select action $\ac^t_i \sim \pi_i^{\text{expl}}(\his^t_i; \theta_i)$ (\Cref{eq:emax:expl_policy})
                \EndFor
                \State Apply joint action $\jac^t = (\ac^t_1, \ldots, \ac^t_N)$
                \State Receive next state $\st^{t+1} \sim \Stf(\st^t, \jac^t)$, rewards $\rew_i^t = \Rew_i(\st^t, \jac^t, \st^{t+1})$ for each agent $i\in\Ag$, and joint observation $\job^{t+1} \sim \Obf(\st^t, \jac^t)$
            \EndFor
            \State Sample bootstrap masks $\{m^k\}_{k=1}^K$ from $\text{Bernoulli}(p)$
            \State Store episode $(\st^{0:T}, \jhis^{0:T}, \jac^{0:T}, \jrew^{0:T}, \{m_k\}_{k=1}^K)$ in $\mathcal{D}$
            \For{each agent $i \in \Ag$}
                \For{each model in the ensemble $k = 1, \ldots, K$}
                    \State Sample batch of episodes $\batch$ from $\data$ with $m_k = 1$
                    \State Update $\theta_i^k$ by minimising $\loss(\theta_i^k)$~\seehere{eq:emax:imeanq_loss} averaged over each time step $t$ in $\batch$
                \EndFor
            \EndFor
        \EndFor
    \end{algorithmic}
\end{algorithm}

\begin{algorithm}[h!]
    \caption{Training QMIX with EMAX}
    \label{alg:qmix-emax}
    \begin{algorithmic}
        \State {\bfseries Initialise:} parameters $\{\theta_i^k\}_{k=1}^K$ of value functions $\{Q^k_i\}_{k=1}^K$ for each agent $i\in\Ag$
        \State {\bfseries Initialise:} parameters $\theta_{\text{mix}}$ and $\widebar{\theta}_{\text{mix}}$ of the main and target mixing networks
        \State {\bfseries Initialise:} empty episodic replay buffer $\data \gets \emptyset$
        \For{each episode}
            \State Obtain initial state $\st^0 \sim \instdist$ and joint observation $\job^0$
            \For{each step $t=0, \cdots, T$}
                \For{each agent $i\in\Ag$}
                    \State Select action $\ac^t_i \sim \pi_i^{\text{expl}}(\his^t_i; \theta_i)$ (\Cref{eq:emax:expl_policy})
                \EndFor
                \State Apply joint action $\jac^t = (\ac^t_1, \ldots, \ac^t_N)$
                \State Receive next state $\st^{t+1} \sim \Stf(\st^t, \jac^t)$, reward $\rew^t = \Rew(\st^t, \jac^t, \st^{t+1})$, and joint observation $\job^{t+1} \sim \Obf(\st^t, \jac^t)$
            \EndFor
            \State Sample bootstrap masks $\{m^k\}_{k=1}^K$ from $\text{Bernoulli}(p)$
            \State Store episode $(\st^{0:T}, \his^{0:T}, \jac^{0:T}, \rew^{0:T}, \{m_k\}_{k=1}^K)$ in $\mathcal{D}$
            \For{each model in the ensemble $k = 1, \ldots, K$}
                \State Sample batch of episodes $\batch$ from $\data$ with $m_k = 1$
                \State Update $\theta^k$ and $\theta_{\text{mix}}$ by minimising $\loss(\theta^k)$~\seehere{eq:vd_loss,eq:emax:ensemble-qmix-estimate} averaged over each time step $t$ in $\batch$
            \EndFor
            \State In a set interval, update target mixing network $\widebar{\theta}_{\text{mix}} \leftarrow \theta_{\text{mix}}$
        \EndFor
    \end{algorithmic}
\end{algorithm}

\begin{algorithm}[h!]
    \caption{Evaluating with EMAX}
    \label{alg:emax-eval}
    \begin{algorithmic}
        \Require Trained ensemble of value functions $\{Q^k_i\}_{k=1}^K$ for each agent $i\in\Ag$
        \State Obtain initial state $\st^0 \sim \instdist$ and joint observation $\job^0$
        \For{each step $t=0, \cdots, T$}
            \For{each agent $i\in\Ag$}
                \State Select action $\ac^t_i \sim \pi_i^{\text{eval}}(\his^t_i; \theta_i)$ (\Cref{eq:emax:eval_policy})
            \EndFor
            \State Apply joint action $\jac^t = (\ac^t_1, \ldots, \ac^t_N)$
            \State Receive next state $\st^{t+1} \sim \Stf(\st^t, \jac^t)$, rewards $\rew_i^t = \Rew_i(\st^t, \jac^t, \st^{t+1})$ for each agent $i\in\Ag$, and joint observation $\job^{t+1} \sim \Obf(\st^t, \jac^t)$
        \EndFor
    \end{algorithmic}
\end{algorithm}

\newpage
\section{Bootstrapped Sampling for EMAX Training}
\label{app:sec:emax:bootstrapped_sampling}

For the boostrapped sampling process, we follow the methodology of \citet{osband2016deep} to draw bootstrapped samples for each model as subsets of the entire training experiences to train on. More specifically, we draw a Bernoulli mask $\{m_k\}_{k=1}^K$ for each model in the ensemble whenever an episode is added to the episodic replay buffer. This mask determines whether the $k$-th model within the ensemble is trained on this episode ($m_k = 1$) or not ($m_k = 0$). Each mask is drawn from a Bernoulli distribution with probability $p$ of being $1$ and $1-p$ of being $0$, i.e. $m_k \sim \text{Bernoulli}(p)$. For $p=1$, each episode would be used to train each model in the ensemble so all models in the ensemble would receive the same training data. In contrast for a small $p$, the training data is likely to be diverse across models in the ensemble but each model would also only be trained on a small subset of the episodes which might sacrifice learning efficiency. In our experiments, we adopt to use $p=0.9$ but similar to prior work \citep{osband2016deep} we have not found the choice of $p$ to significantly affect the performance of our algorithm.

\section{Relation to Standard Upper-Confidence Bound (UCB) Exploration}
\label{app:sec:emax:ucb_relation}
The classical upper-confidence bound (UCB) algorithm~\citep{auer2002using} algorithm was proposed for bandit problems and defines an exploratory policy
\begin{equation}
    \pol(t) \in \argmax_{\ac\in\Ac} Q(\ac, t) + \beta \sqrt{\frac{2 \log(t)}{n(\ac, t)}}
    \label{eq:ucb_policy}
\end{equation}
with $Q(\ac, t)$ denoting the sample mean of action $\ac$ at timestep $t$, $\beta$ is a weighting coefficient, and $n(\ac, t)$ denotes the number of times $\ac$ has been chosen until timestep $t$. The uncertainty term of UCB is derived from upper bound of the confidence interval over the value of actions. In this way, the UCB policy selects actions that have a high value estimate, as measured by $Q(\ac, t)$, and with high uncertainty as measured by the uncertainty term. This uncertainty term decreases as the action $\ac$ is being selected more often and increases over time to ensure that exploration does not collapse too early.

The uncertainty term of UCB strongly correlates with the epistemic uncertainty of value functions that can be represented through the standard deviation across the ensemble, as done by prior work~\citep{osband2016deep,lee2021sunrise,liang2022reducing}. A core contribution of our work is the insight that within multi-agent systems, an additional source of uncertainty arises from the action selection of other agents. In EMAX, we propose to leverage this additional source of uncertainty to guide the exploration towards states and actions with the potential for interactions with other agents. Sufficiently exploring these states and actions is often essential to learn optimal policies that solve problems that require coordination across multiple agents, as indicated by the empirical benefits observed with EMAX in \Cref{sec:emax:experiments_results}.

\section{Additional Evaluation Details}
\label{app:sec:emax:eval_details}

\subsection{Environment Details}
\label{app:sec:environments}

\looseness=-1
\textbf{Level-Based Foraging:} The level-based foraging (LBF) environment~\citep{albrecht2013game,papoudakis2021benchmarking} contains diverse tasks in which agents and food are randomly scattered in a gridworld. Agents observe the location of themselves as well as all other agents and food in the gridworld, and are able to choose between discrete actions $\Ac = \{\text{do nothing}, \text{move up}, \text{move down}, \text{move left}, \text{move right}, \text{pick-up}\}$. Agents and food are assigned levels and agents can only pick-up food if the level of all agents standing next to the food and choosing the pick-up action together is greater or equal to the level of the food. Agents only receive rewards for successful collection of food. Episodes terminate after all food has been collected or after at most $50$ timesteps. Each episode randomises the level and starting locations of agents and food. Tasks vary in the size of the gridworld, the number of agents and food, and the level assignment. 

\textbf{Boulder-Push:} In the boulder-push environment (BPUSH)~\citep{christianos2022pareto}, agents need to navigate a gridworld to move a boulder to a target location. Agents observe the location of the boulder, all other agents, and the direction the boulder needs to be pushed in. The action space of all agents consists of the same discrete actions $\Ac = \{\text{move up}, \text{move down}, \text{move left}, \text{move right}\}$. Agents only receive rewards of $0.1$ per agent for successfully pushing the boulder forward in its target direction, which requires cooperation of all agents, and a reward of $1$ per agent for the boulder reaching its target location. Unsuccessful pushing of the boulder by some but not all agents leads to a penalty reward of $-0.01$. Episodes terminate after the boulder reached its target location or after at most $50$ timesteps. BPUSH tasks considered in this work vary in the size of the gridworld and the number of agents varying between two and four.

\looseness=-1
\textbf{Multi-Robot Warehouse:} The multi-robot warehouse environment (RWARE)~\citep{christianos2020shared,papoudakis2021benchmarking} represents gridworld warehouses with blocks of shelves. Agents need to navigate the warehouse and collect currently requested items. Agents only observe nearby agents and shelves immediately next to their location, and choose discrete actions $\Ac = \{\text{turn left}, \text{turn right}, \text{move forward}, \text{load/ unload shelf}\}$. Agents are only rewarded for successful deliveries of requested shelves, which require long sequences of actions, with a reward of $1$, thus rewards are very sparse making RWARE tasks hard exploration problems. At each timestep, the total number of requested shelves is equal to the number of agents and once requested shelves, a currently unrequested shelf is uniformly at random sampled and added to the list of requested shelves. Episodes terminate after $500$ timesteps. While agents are capable of delivering shelves without interaction with other agents, agents need to cooperate to avoid blocking each others path in narrow parts of the warehouse and trying to move to and load identical requested shelves. To maximise episodic returns, agents need to learn to avoid such conflicts with other agents which requires them to learn conventions and cooperate. It is worth highlighting that no value-based algorithm achieved non-zero rewards in this environment within four million timesteps of training in prior evaluations~\citep{papoudakis2021benchmarking}.

\looseness=-1
\textbf{Multi-Agent Particle Environment:} In the multi-agent particle environment (MPE)~\citep{mordatch2018emergence,lowe2017multi}, agents navigate continuous two-dimensional, fully-observable environments. In all tasks, agents observe the relative position and velocity of all agents, as well as the relative positions of landmarks in the environment. Agents choose between five discrete actions consisting of doing nothing and movement in all four cardinal directions. We evaluate agents in three diverse tasks within MPE which all require cooperation between all agents with densely rewarded objectives. (1) Predator-prey in which three agents control predators in an environment with three landmarks, representing obstacles, and a faster, pre-trained\footnote{Pre-trained agents are obtained from the EPyMARL codebase~\citep{papoudakis2021benchmarking}. They were obtained by training all agents (including adversaries) with the MADDPG algorithm for 25,000 episodes.} prey. (2) Spread in which three agents need to cover three landmarks while avoiding collisions with each other. (3) Adversary in which two agents are in an environment with an pre-trained adversary and two landmarks. At the beginning of each episode, one of the two landmarks is randomly determined as the goal landmark for the agents (agents observe this goal landmark but the adversary has no information about it). Both agents are rewarded for one of them being close to the goal landmark but they are negatively rewarded for the adversary agent moving close to the goal location. Therefore, agents are incentivised to cover both landmarks in order to maximise their rewards since then they are receiving rewards for covering the goal landmark but also preventing the adversary agent from identifying which landmark is the goal landmark.

\subsection{Computational Resources}
\label{app:sec:emax:comp_resources}

All experiments were conducted on (1) desktop computers with two Nvidia RTX 2080 Ti GPUs, Intel i9-9900X @ 3.50GHz CPU, 62GB RAM, running Ubuntu 20.04, (2) two server machines with four Nvidia V100 GPUs, Intel Xeon Platinum 8160 @ 2.10GHz CPU, 503GB RAM, running CentOS Linux 7 OS, and (3) one server machine with Nvidia RTX A4500 GPUs, an AMD EPYC 7763 CPU with 64 cores at up to 3.6GHz, 1TB RAM, running Ubuntu 22.04. The speedtest for varying ensemble sizes reported in \Cref{tab:emax:ensemble_size_speedtest} has been conducted on the desktop computer (1). 

\subsection{Hyperparameter Optimisation}
\label{app:sec:emax:deep_hyperparams}
For IDQN, VDN, QMIX and extensions with EMAX, we conduct a gridsearch to identify best hyperparameters in one selected task within each environment by evaluating each algorithm configuration for three runs and selecting the hyperparameter configuration which led to highest average evaluation returns throughout training. We largely based our configurations on the reported hyperparameters from \citet{papoudakis2021benchmarking} with minimal hyperparameter tuning. \Cref{app:tab:emax:hyperparams_search} shows the hyperparameters that are searched over for each algorithm, and \Cref{app:tab:emax:hyperparams_search_tasks} shows the tasks and number of training time steps used for the hyperparameter optimisation in each environment. \Cref{app:tab:emax:deep_hyperparams} shows the identified best-performing hyperparameters for each algorithm in all four environments.

Our implementation of IDQN, VDN, QMIX, and EMAX are based on the EPyMARL codebase\footnote{Available at \url{https://github.com/uoe-agents/epymarl}.}. For the baseline of MAVEN, CDS, and EMC, we migrated the provided codebase from the authors\footnote{Available at \url{https://github.com/AnujMahajanOxf/MAVEN}, \url{https://github.com/lich14/CDS} and \url{https://github.com/kikojay/EMC}.}
into EPyMARL to support all environments. For MAVEN, CDS, and EMC, we use the hyperparameters identified for QMIX for each environment with the algorithm-specific hyperparameters provided by the authors. For IPPO and MAPPO, we use the best identified hyperparameters reported in \citet{papoudakis2021benchmarking}.

\begin{table}[ht!]
    \centering
    \caption{Hyperparameters that are searched over in one representative task of each environment.}
    \label{app:tab:emax:hyperparams_search}
    \begin{tabular}{l l c}
        \toprule
        Algorithm & Hyperparameter  & Values\\
        \midrule
        \multirow{3}{*}{Every algorithm} & Network architecture    & FC, GRU\\
                                         & Network size            & 64, 128\\
                                         & Reward standardisation  & False, True\\ \midrule
        Baselines & $\epsilon$ decay steps  & \num{50000}, \num{200000}\\ \midrule
        EMAX algorithms & UCB uncertainty coefficient $\beta$ & 0.1, 0.3, 1\\
        \bottomrule
    \end{tabular}
\end{table}

\begin{table}[ht!]
    \centering
    \caption{Tasks and number of training time steps used for hyperparameter optimisation in each environment.}
    \label{app:tab:emax:hyperparams_search_tasks}
    \begin{tabular}{l l c}
        \toprule
        Environment & Task  & Time steps\\
        \midrule
        LBF & Foraging-10x10-4p-3f-coop & 4M\\
        BPUSH & BPUSH $12\times12$ 2ag & 7.5M\\
        RWARE & RWARE $11\times10$ 4ag & 5M\\
        MPE & MPE spread & 1M\\
        \bottomrule
    \end{tabular}
\end{table}

\begin{table}[ht!]
    \centering
    \caption{Hyperparameters for IDQN, VDN, QMIX and extensions with EMAX for all environments.}
    \label{app:tab:emax:deep_hyperparams}
    \begin{tabular}{l l c c c c}
        \toprule
        Algorithm & Hyperparameter  & LBF & BPUSH & RWARE & MPE \\
        \midrule
        \multirow{14}{*}{Shared} & $\gamma$ & \multicolumn{4}{c}{0.99}\\
                                 & Activation function     & \multicolumn{4}{c}{ReLU}\\
                                 & Parameter sharing       & \multicolumn{4}{c}{True}\\
                                 & Optimiser               & \multicolumn{4}{c}{Adam}\\
                                 & Maximum gradient norm   & \multicolumn{4}{c}{5}\\
                                 & Minimum $\epsilon$      & \multicolumn{4}{c}{0.05}\\
                                 & Evaluation $\epsilon$   & \multicolumn{4}{c}{0.05}\\
                                 & Learning rate           & \multicolumn{4}{c}{$e^{-4}$}\\
                                 & Target update frequency & \multicolumn{4}{c}{200}\\
                                 & Replay buffer capacity (episodes) & \multicolumn{4}{c}{\num{5000}}\\
                                 & Batch size (episodes)   & \multicolumn{4}{c}{32}\\ \midrule
        \multirow{4}{*}{IDQN} & Network architecture    & GRU & GRU & GRU & FC\\
                              & Network size            & 128 & 128 & 128 & 128\\
                              & Reward standardisation  & True & True & True & True\\
                              & $\epsilon$ decay steps  & \num{50000} & \num{50000} & \num{50000} & \num{50000} \\ \midrule
        \multirow{4}{*}{VDN} & Network architecture    & GRU & GRU & GRU & FC\\
                             & Network size            & 128 & 128 & 128 & 128\\
                             & Reward standardisation  & True & True & True & True\\
                             & $\epsilon$ decay steps  & \num{200000} & \num{200000} & \num{50000} & \num{50000} \\ \midrule
        \multirow{6}{*}{QMIX} & Mixing embedding size   & \multicolumn{4}{c}{32}\\
                              & Hypernetwork embedding size & \multicolumn{4}{c}{64} \\
                              & Network architecture    & FC & FC & FC & FC \\
                              & Network size            & 128 & 128 & 128 & 128\\
                              & Reward standardisation  & True & True & True & True\\
                              & $\epsilon$ decay steps  & \num{200000} & \num{200000} & \num{50000} & \num{50000} \\ \midrule
        \multirow{4}{*}{IDQN-EMAX} & Network architecture    & GRU & GRU & GRU & GRU\\
                                   & Network size            & 128 & 128 & 128 & 128\\
                                   & Reward standardisation  & True & True & True & True\\
                                   & UCB uncertainty coefficient $\beta$ & 1 & 1 & 1 & 1\\ \midrule
        \multirow{4}{*}{VDN-EMAX} & Network architecture    & GRU & GRU & GRU & GRU\\
                                  & Network size            & 128 & 128 & 128 & 128\\
                                  & Reward standardisation  & True & True & True & True\\
                                  & UCB uncertainty coefficient $\beta$ & 0.1 & 0.1 & 0.3 & 0.1\\ \midrule
        \multirow{4}{*}{QMIX-EMAX} & Network architecture    & GRU & GRU & GRU & GRU\\
                                   & Network size            & 128 & 128 & 128 & 128\\
                                   & Reward standardisation  & True & True & True & True\\
                                   & UCB uncertainty coefficient $\beta$ & 0.3 & 0.3 & 0.3 & 0.3\\
        \bottomrule
    \end{tabular}
\end{table}

\clearpage

\section{Individual Task Evaluation Returns in Mixed-Objective Tasks}
\label{app:sec:emax:ind_deep_results}
\begin{figure*}[h!]
    \centering
    \includegraphics[width=.4\linewidth]{images/results/ind_rewards_eval/legend.pdf}

    \begin{subfigure}{.3\linewidth}
        \centering
        \includegraphics[width=\linewidth]{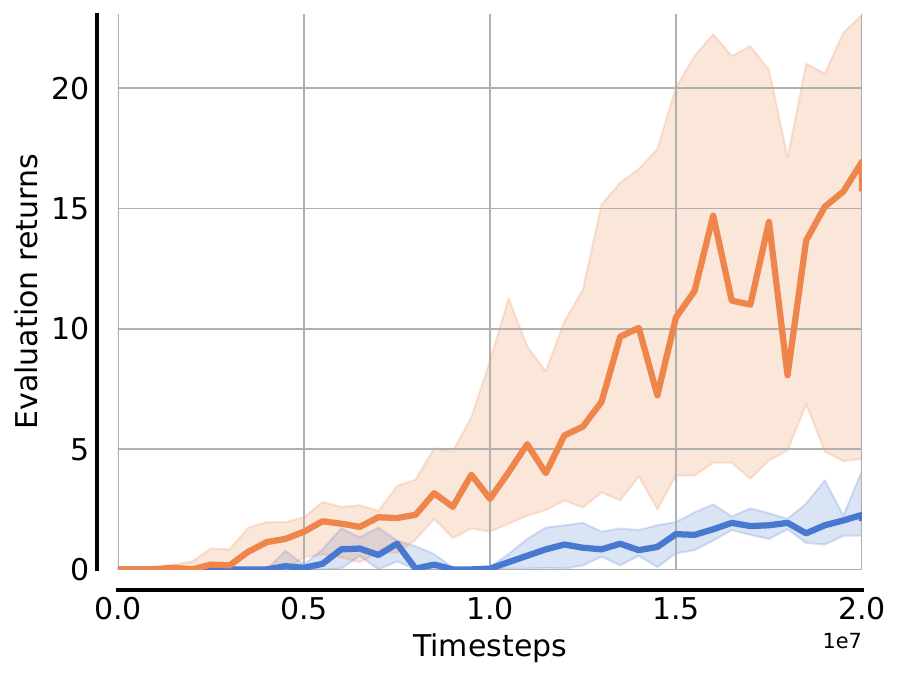}
        \caption{RWARE 11x10 2ag}
    \end{subfigure}
    \begin{subfigure}{.3\linewidth}
        \centering
        \includegraphics[width=\linewidth]{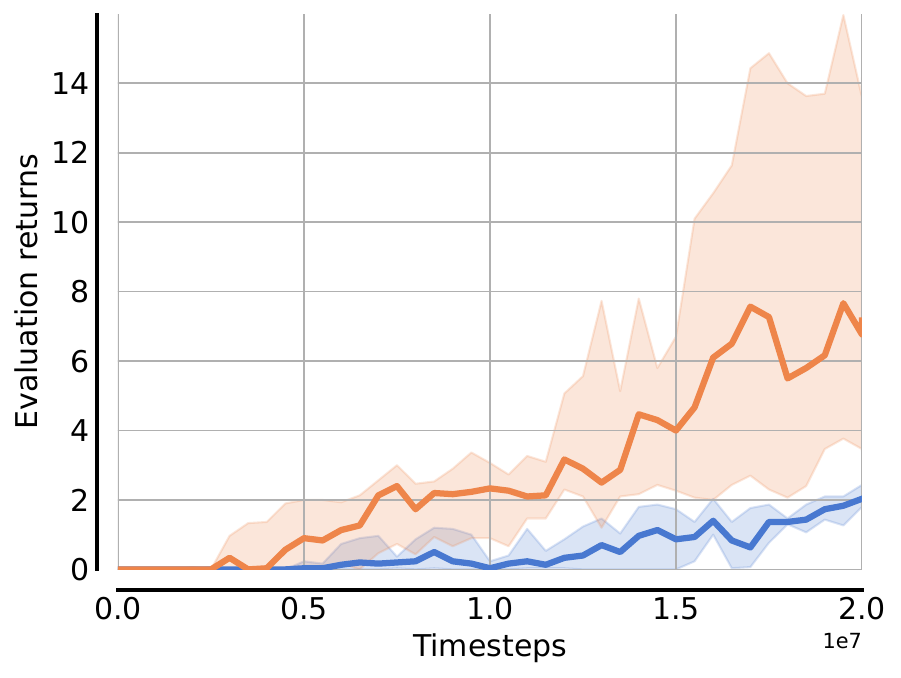}
        \caption{RWARE 11x10 2ag hard}
    \end{subfigure}
    \begin{subfigure}{.3\linewidth}
        \centering
        \includegraphics[width=\linewidth]{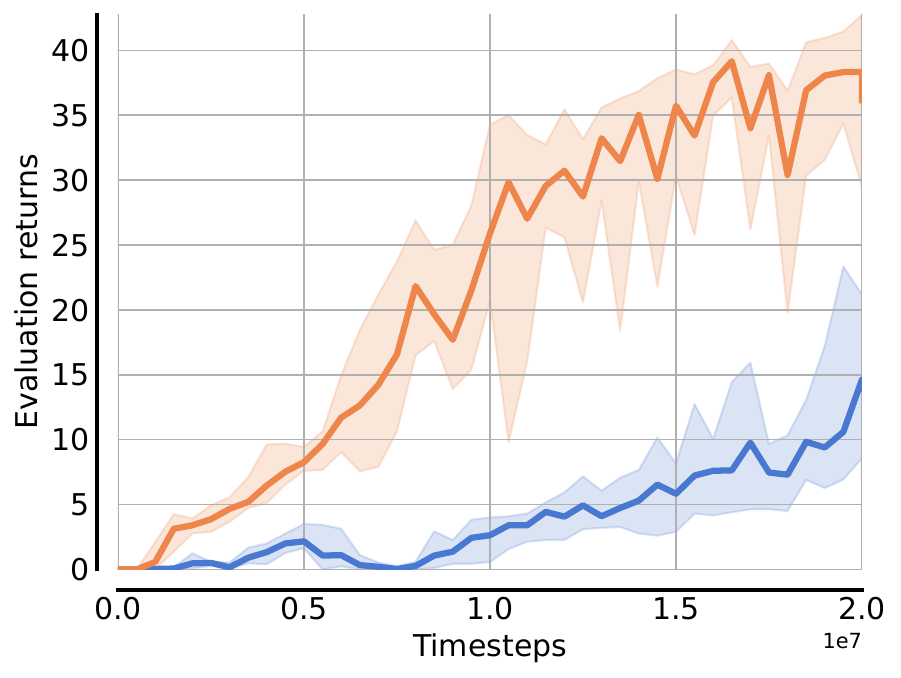}
        \caption{RWARE 11x10 4ag}
    \end{subfigure}
    
    \begin{subfigure}{.3\linewidth}
        \centering
        \includegraphics[width=\linewidth]{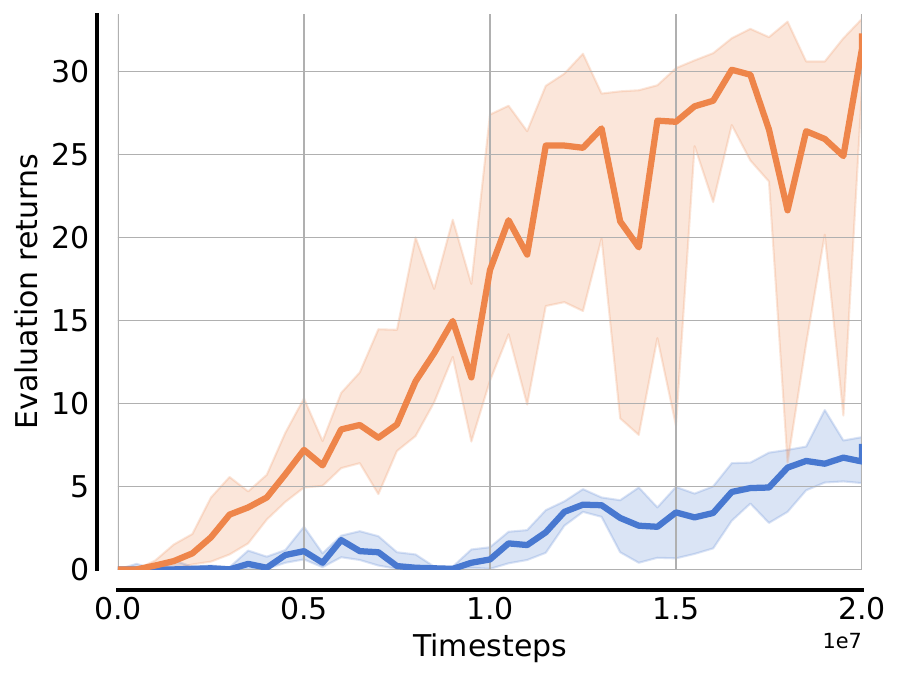}
        \caption{RWARE 11x10 4ag hard}
    \end{subfigure}
    \begin{subfigure}{.3\linewidth}
        \centering
        \includegraphics[width=\linewidth]{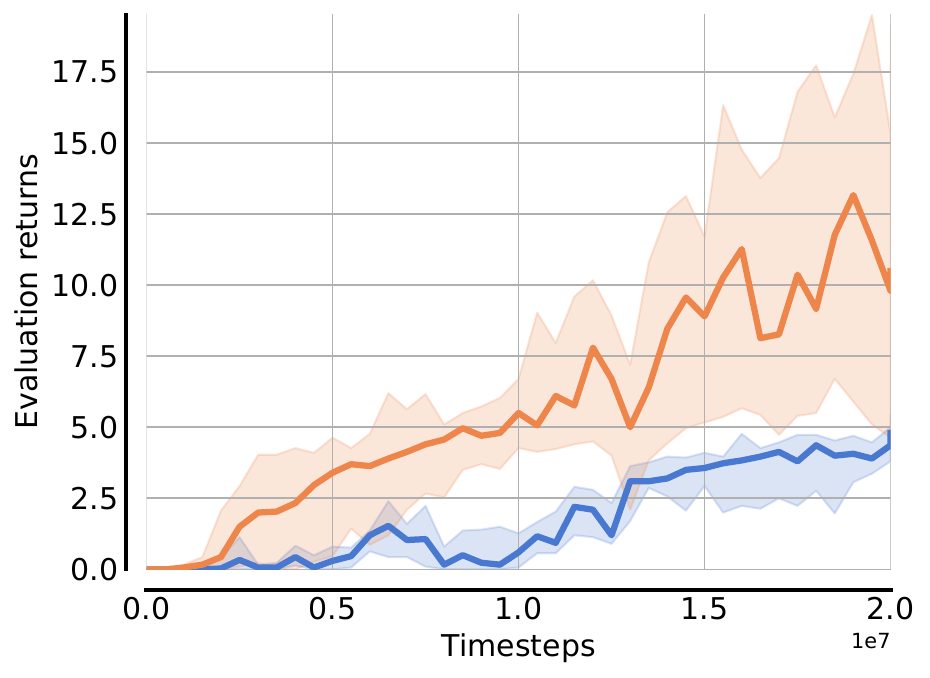}
        \caption{RWARE 20x10 4ag}
    \end{subfigure}
    \begin{subfigure}{.3\linewidth}
        \centering
        \includegraphics[width=\linewidth]{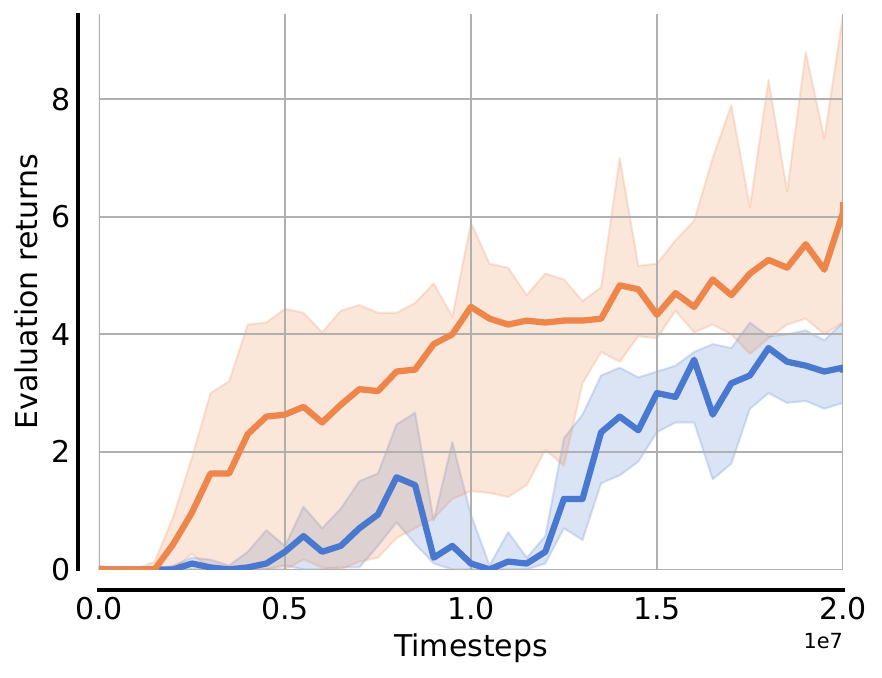}
        \caption{RWARE 20x16 4ag}
    \end{subfigure}

    \begin{subfigure}{.3 \linewidth}
        \centering
        \includegraphics[width=\linewidth]{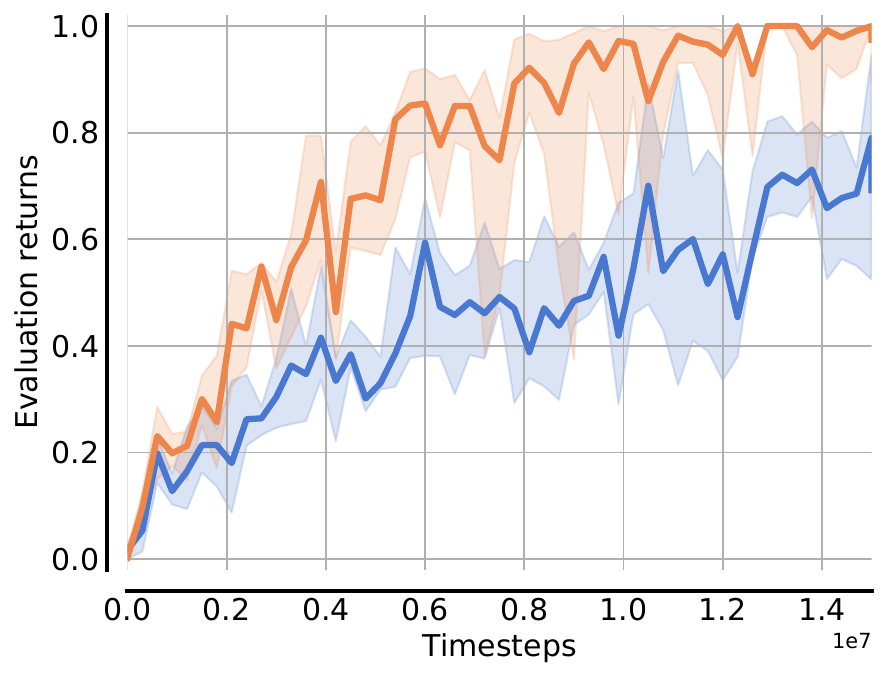}
        \caption{LBF 10x10-3p-5f}
    \end{subfigure}
    \begin{subfigure}{.3\linewidth}
        \centering
        \includegraphics[width=\linewidth]{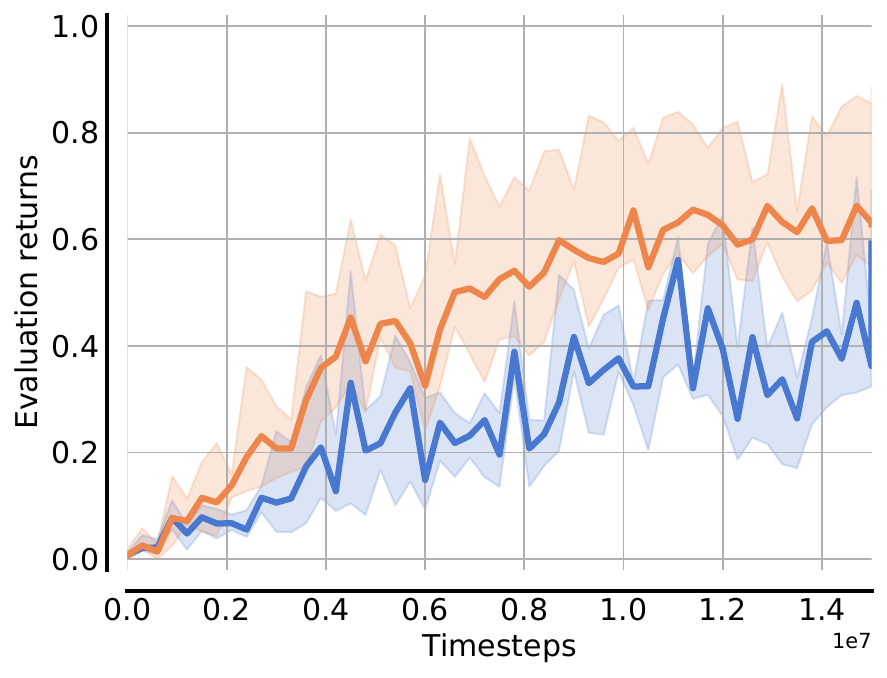}
        \caption{LBF 15x15-3p-5f}
    \end{subfigure}
    \begin{subfigure}{.3\linewidth}
        \centering
        \includegraphics[width=\linewidth]{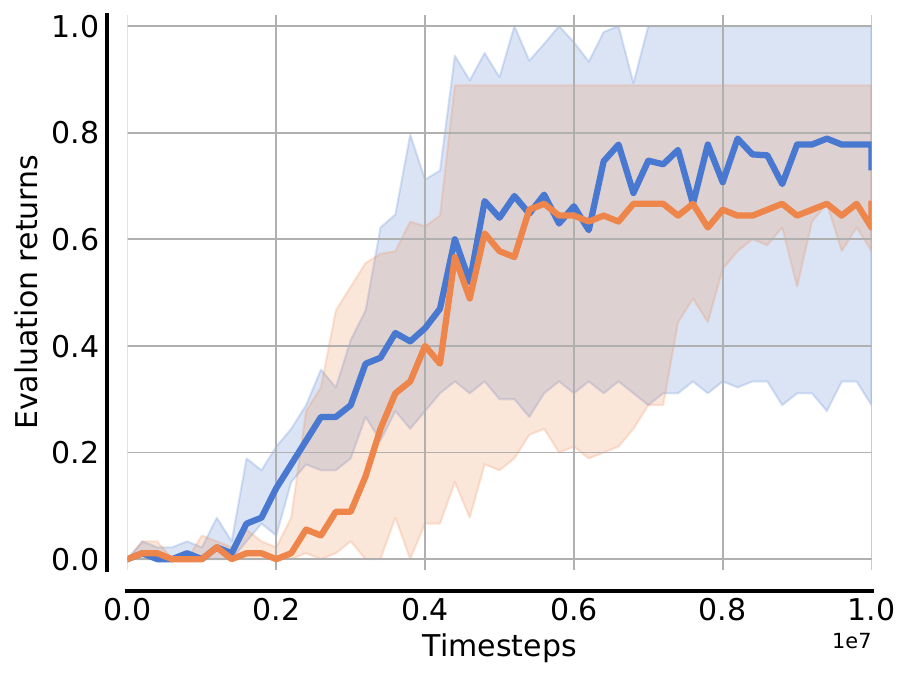}
        \caption{LBF 10x10-4p-3f-coop}
    \end{subfigure}

    \begin{subfigure}{.3\linewidth}
        \centering
        \includegraphics[width=\linewidth]{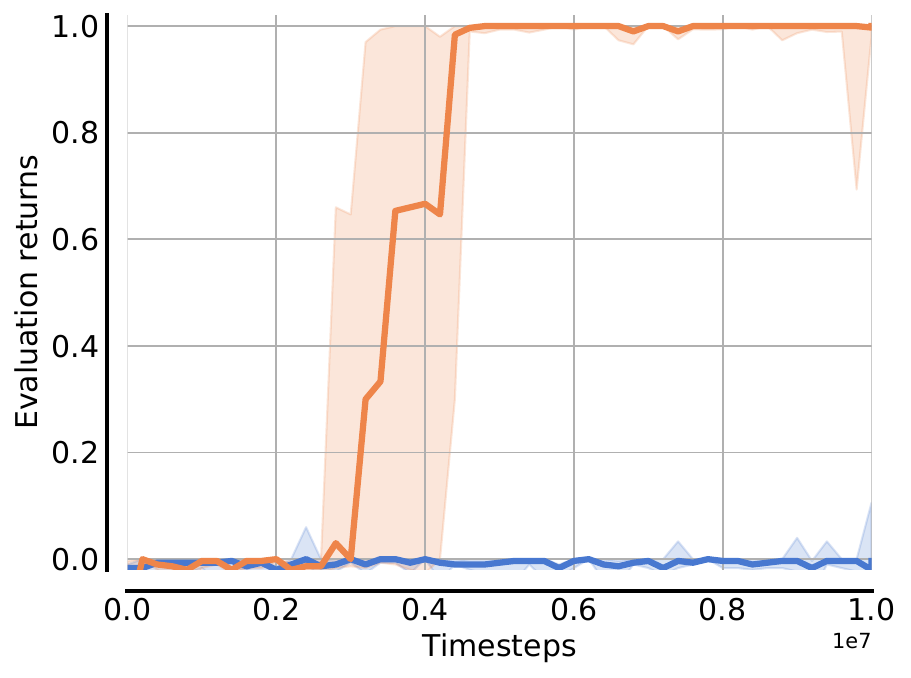}
        \caption{LBF 5x5-2p-1f-coop-pen}
    \end{subfigure}
    \begin{subfigure}{.3\linewidth}
        \centering
        \includegraphics[width=\linewidth]{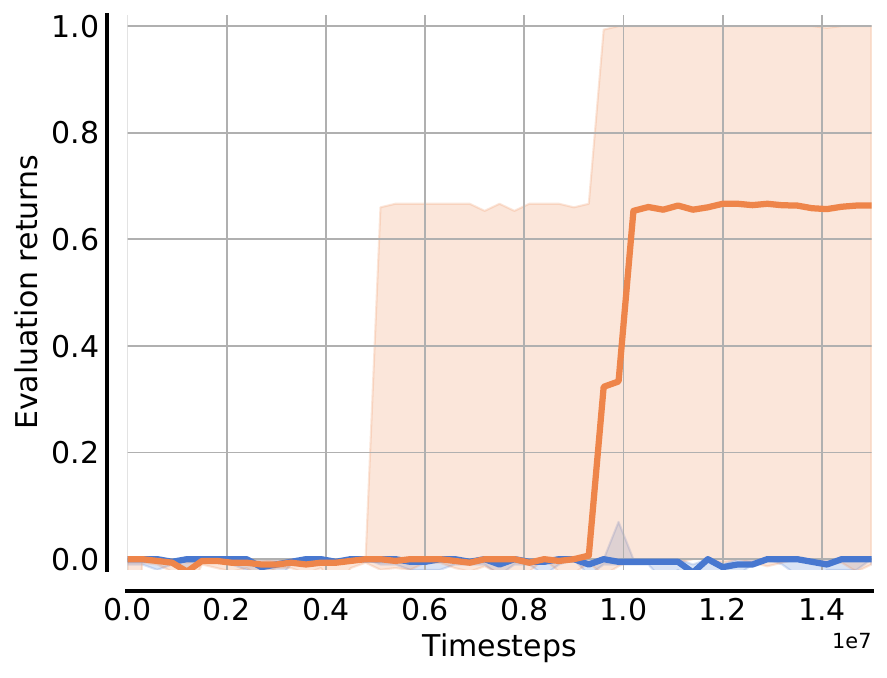}
        \caption{LBF 8x8-2p-1f-coop-pen}
    \end{subfigure}

    \caption{Average evaluation returns and 95\% confidence intervals for IDQN with and without EMAX in mixed-objective LBF and RWARE tasks.}
\end{figure*}

\clearpage
\section{Normalised Environment Returns in Cooperative Tasks}
\label{app:sec:emax:deep_norm_eval_returns}

\begin{figure*}[h!]
    \centering
    \includegraphics[width=\linewidth]{images/results/legend.pdf}

    \begin{subfigure}{.49\linewidth}
        \centering
        \includegraphics[width=\linewidth]{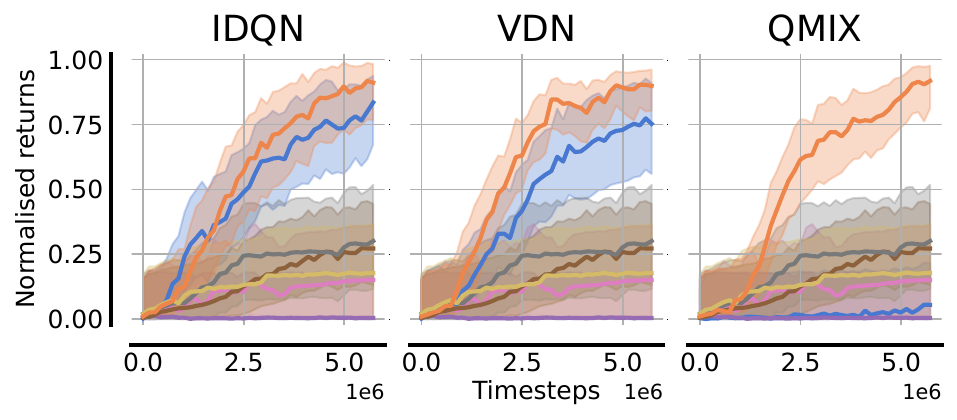}
        \caption{LBF}
        \label{fig:emax:lbf_norm}
    \end{subfigure}
    \begin{subfigure}{.49\linewidth}
        \centering
        \includegraphics[width=\linewidth]{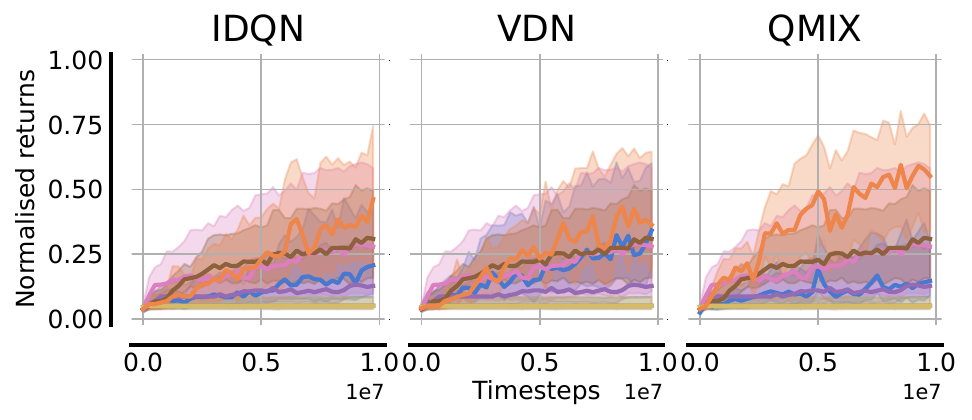}
        \caption{BPUSH}
        \label{fig:emax:bpush_norm}
    \end{subfigure}
    \begin{subfigure}{.49\linewidth}
        \centering
        \includegraphics[width=\linewidth]{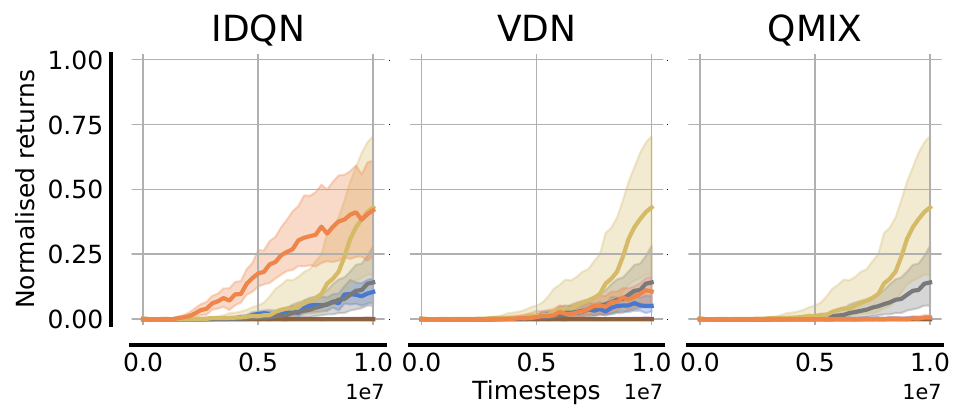}
        \caption{RWARE}
        \label{fig:emax:rware_norm}
    \end{subfigure}
    \begin{subfigure}{.49\linewidth}
        \centering
        \includegraphics[width=\linewidth]{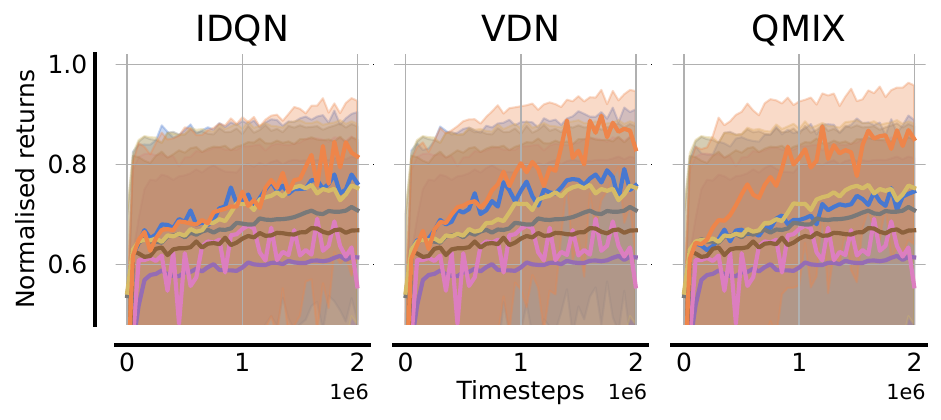}
        \caption{MPE}
        \label{fig:emax:mpe_norm}
    \end{subfigure}
    \caption{Interquartile mean and 95\% confidence intervals of normalised evaluation returns for all algorithms in each environment.}
    \label{fig:emax:emax-norm}
\end{figure*}

\clearpage
\section{Individual Task Evaluation Returns in Cooperative Tasks}
\label{app:sec:emax:deep_ind_eval_returns}

\begin{figure*}[h!]
    \centering
    \includegraphics[width=\linewidth]{images/results/legend.pdf}

    \begin{subfigure}{.49\linewidth}
        \centering
        \includegraphics[width=\linewidth]{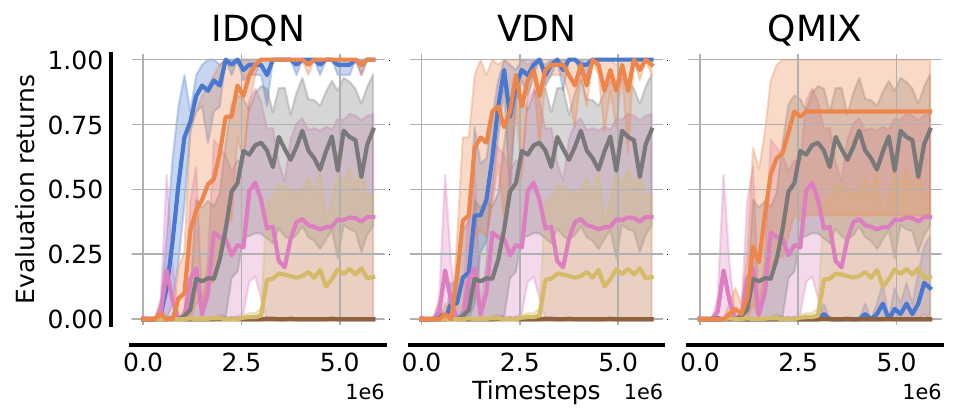}
        \caption{LBF 10x10-4p-1f-coop}
        \label{app:fig:emax:lbf_10x10-4p-1f-coop}
    \end{subfigure}
    \begin{subfigure}{.49\linewidth}
        \centering
        \includegraphics[width=\linewidth]{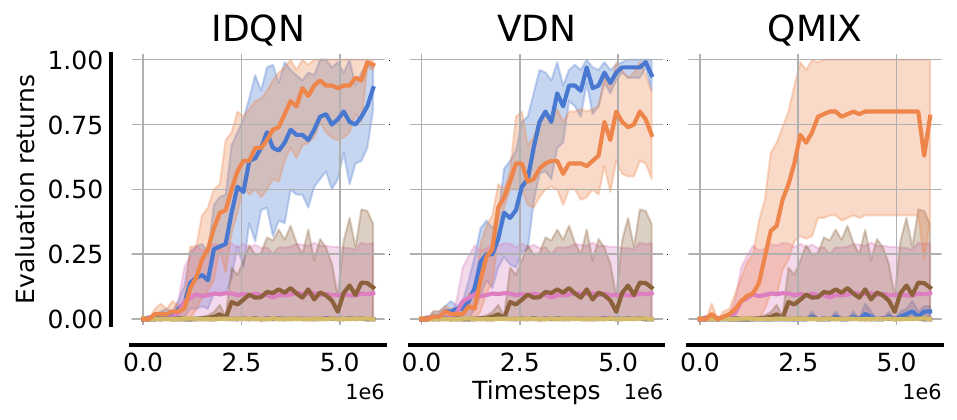}
        \caption{LBF 10x10-4p-2f-coop}
        \label{app:fig:emax:lbf_10x10-4p-2f-coop}
    \end{subfigure}

    \begin{subfigure}{.49\linewidth}
        \centering
        \includegraphics[width=\linewidth]{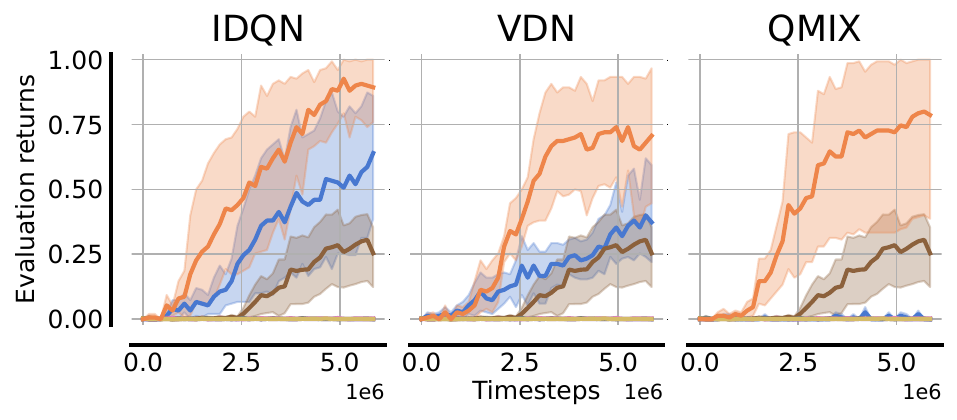}
        \caption{LBF 10x10-4p-3f-coop}
        \label{app:fig:emax:lbf_10x10-4p-3f-coop}
    \end{subfigure}
    \begin{subfigure}{.49\linewidth}
        \centering
        \includegraphics[width=\linewidth]{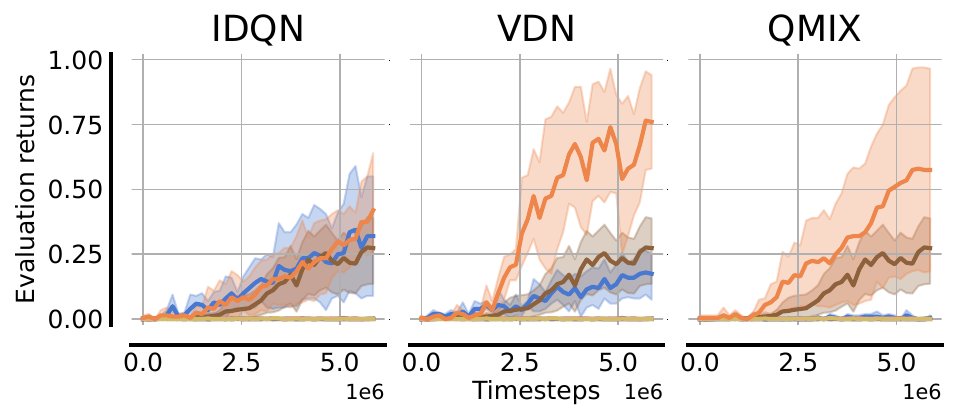}
        \caption{LBF 10x10-4p-4f-coop}
        \label{app:fig:emax:lbf_10x10-4p-4f-coop}
    \end{subfigure}

    \begin{subfigure}{.49\linewidth}
        \centering
        \includegraphics[width=\linewidth]{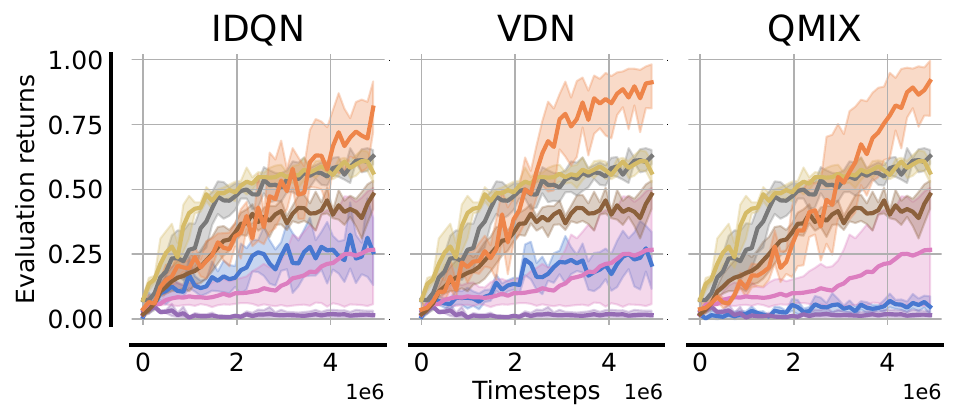}
        \caption{LBF 10x10-3p-5f}
        \label{app:fig:emax:lbf_10x10-3p-5f}
    \end{subfigure}
    \begin{subfigure}{.49\linewidth}
        \centering
        \includegraphics[width=\linewidth]{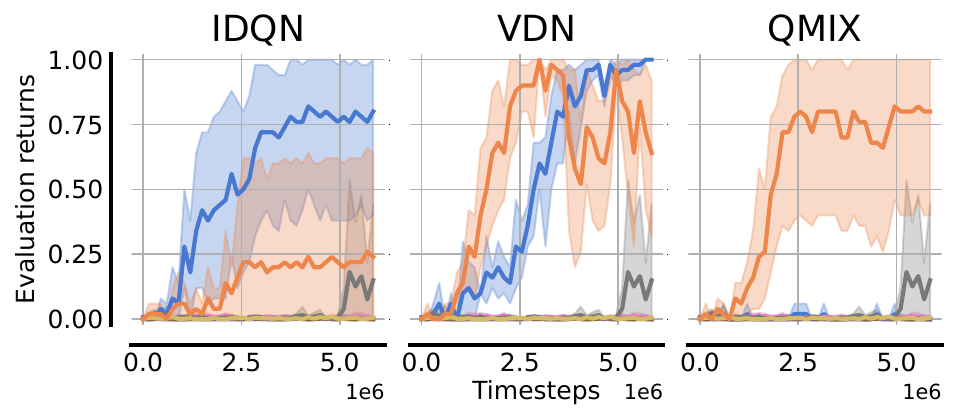}
        \caption{LBF 15x15-8p-1f-coop}
        \label{app:fig:emax:lbf_15x15-8p-1f-coop}
    \end{subfigure}

    \begin{subfigure}{.49\linewidth}
        \centering
        \includegraphics[width=\linewidth]{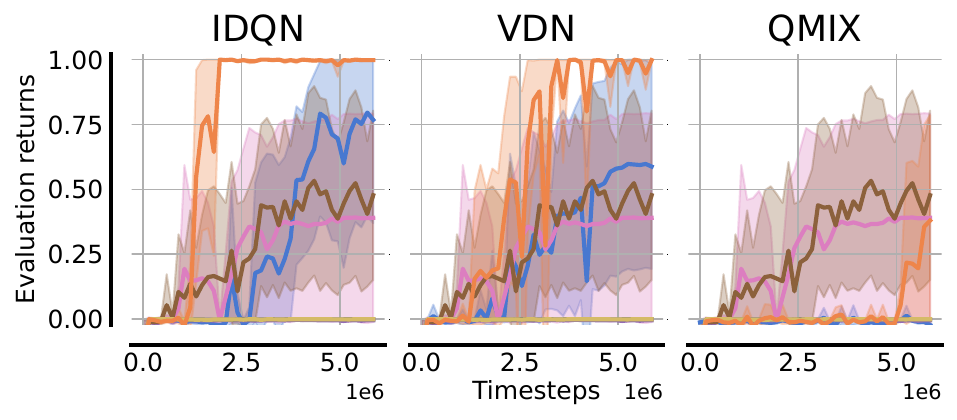}
        \caption{LBF 5x5-2p-1f-coop-pen}
        \label{app:fig:emax:lbf_5x5-2p-1f-coop-pen}
    \end{subfigure}
    \begin{subfigure}{.49\linewidth}
        \centering
        \includegraphics[width=\linewidth]{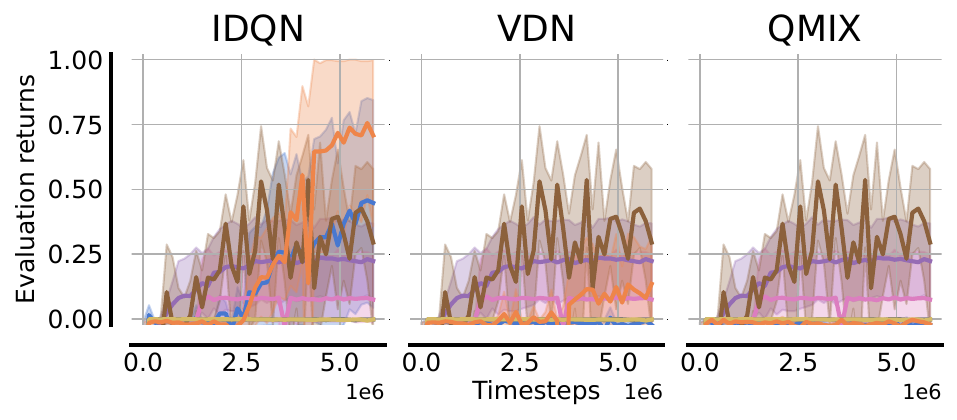}
        \caption{LBF 5x5-2p-2f-coop-pen}
        \label{app:fig:emax:lbf_5x5-2p-2f-coop-pen}
    \end{subfigure}
    \caption{Average evaluation returns and 95\% confidence intervals for all algorithms in LBF tasks.}
    \label{app:fig:emax:lbf_individual_returns}
\end{figure*}

\begin{figure*}[h!]
    \centering
    \includegraphics[width=\linewidth]{images/results/legend.pdf}

    \begin{subfigure}{.49\linewidth}
        \centering
        \includegraphics[width=\linewidth]{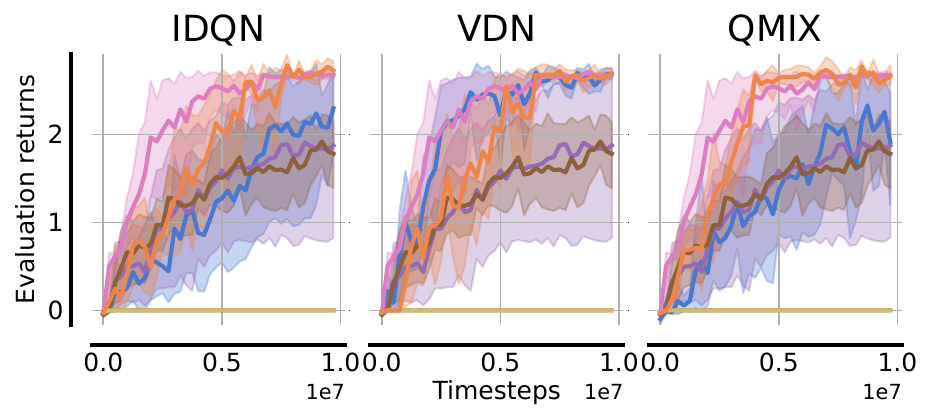}
        \caption{BPUSH 8x8 2ag}
        \label{app:fig:emax:bpush_small-2ag-easy}
    \end{subfigure}
    \begin{subfigure}{.49\linewidth}
        \centering
        \includegraphics[width=\linewidth]{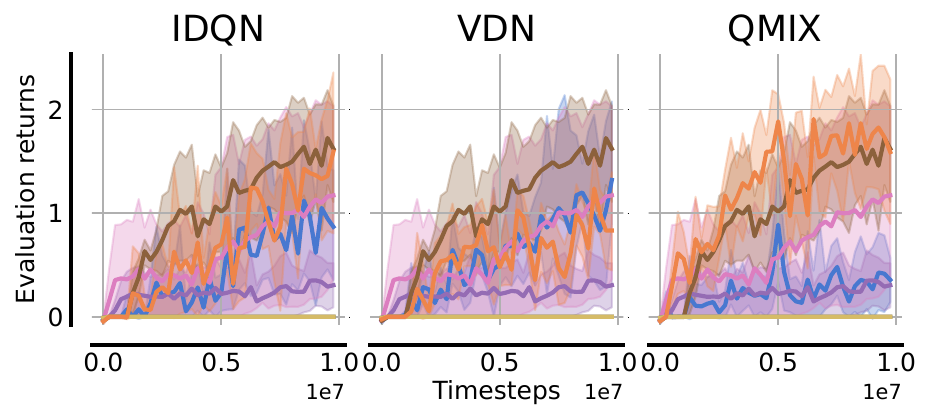}
        \caption{BPUSH 12x12 2ag}
        \label{app:fig:emax:bpush_medium-2ag-easy}
    \end{subfigure}

    \begin{subfigure}{.49\linewidth}
        \centering
        \includegraphics[width=\linewidth]{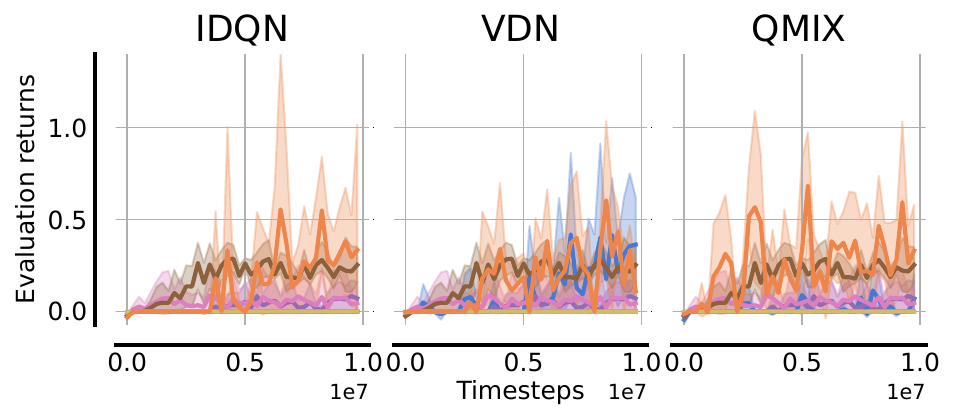}
        \caption{BPUSH 20x20 2ag}
        \label{app:fig:emax:bpush_large-2ag-easy}
    \end{subfigure}
    \begin{subfigure}{.49\linewidth}
        \centering
        \includegraphics[width=\linewidth]{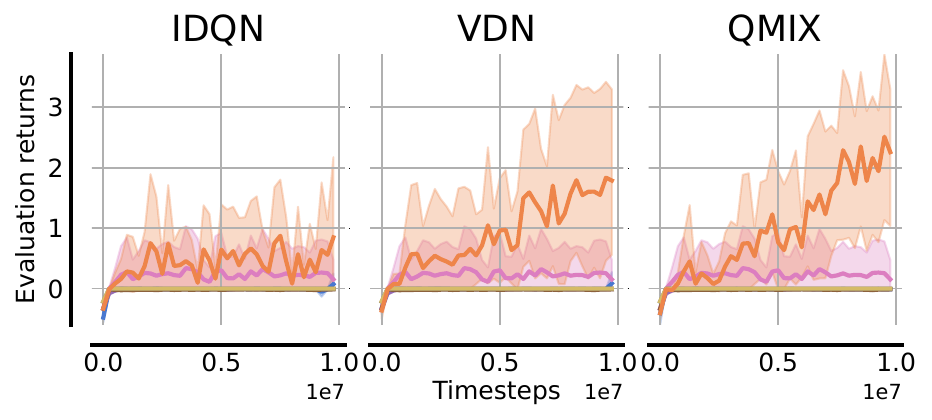}
        \caption{BPUSH 5x5 4ag}
        \label{app:fig:emax:bpush_tiny-4ag-easy}
    \end{subfigure}
    \caption{Average evaluation returns and 95\% confidence intervals for all algorithms in BPUSH tasks.}
    \label{app:fig:emax:bpush_individual_returns}
\end{figure*}

\begin{figure*}[h!]

    \begin{subfigure}{.49\linewidth}
        \centering
        \includegraphics[width=\linewidth]{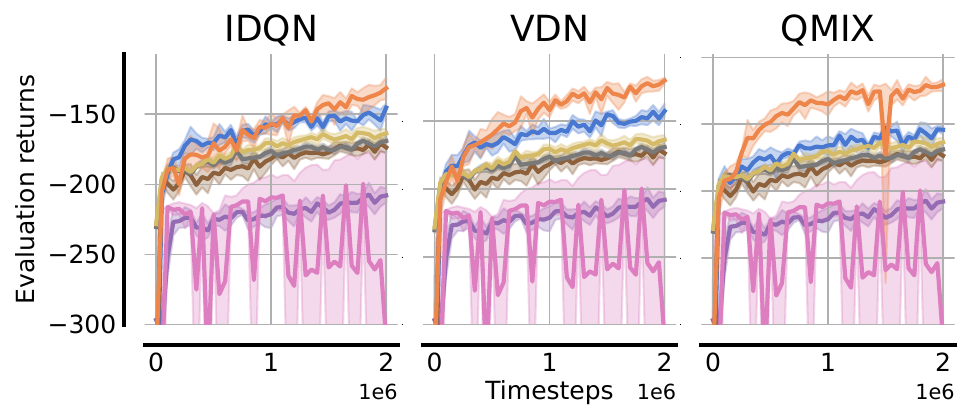}
        \caption{MPE spread}
        \label{app:fig:emax:mpe-spread}
    \end{subfigure}
    \begin{subfigure}{.49\linewidth}
        \centering
        \includegraphics[width=\linewidth]{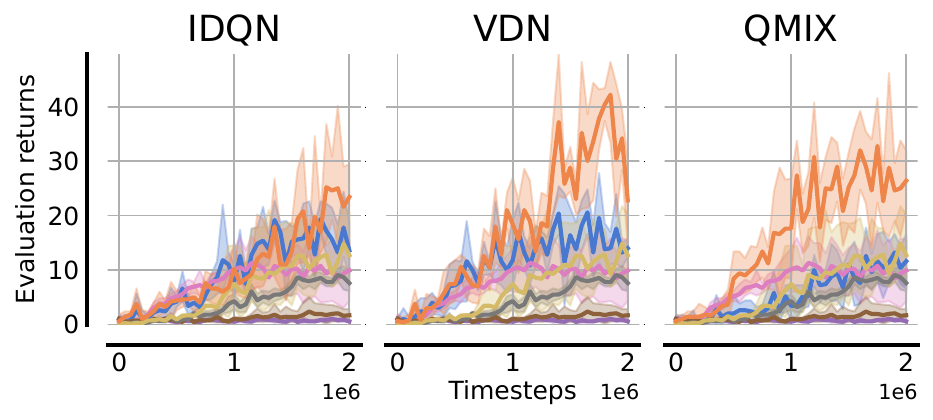}
        \caption{MPE predator-prey}
        \label{app:fig:emax:mpe-tag}
    \end{subfigure}

    \begin{subfigure}{.49\linewidth}
        \centering
        \includegraphics[width=\linewidth]{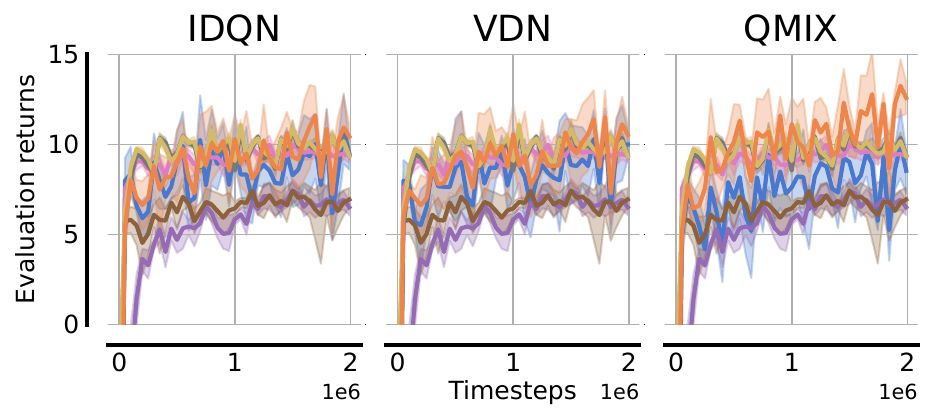}
        \caption{MPE adversary}
        \label{app:fig:emax:mpe-adversary}
    \end{subfigure}
    \caption{Average evaluation returns and 95\% confidence intervals for all algorithms in MPE tasks.}
    \label{app:fig:emax:mpe_individual_returns}
\end{figure*}

\begin{figure*}[h!]
    \centering
    \includegraphics[width=\linewidth]{images/results/legend.pdf}

    \begin{subfigure}{.49\linewidth}
        \centering
        \includegraphics[width=\linewidth]{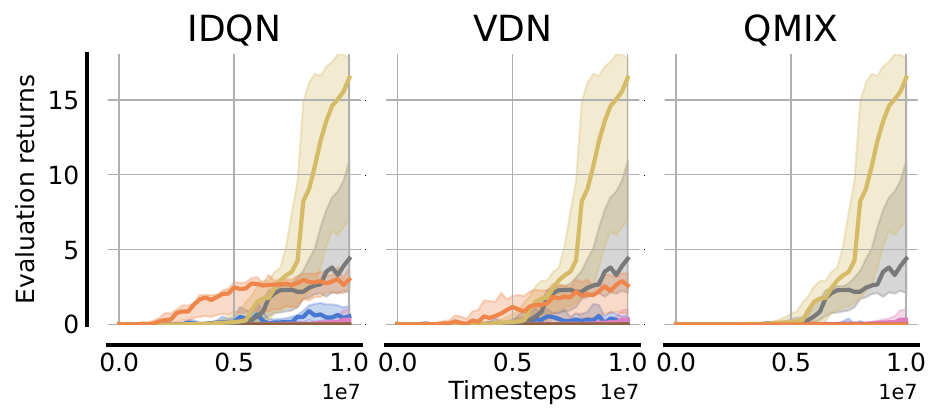}
        \caption{RWARE 11x10 2ag}
        \label{app:fig:emax:rware_tiny-2ag}
    \end{subfigure}
    \begin{subfigure}{.49\linewidth}
        \centering
        \includegraphics[width=\linewidth]{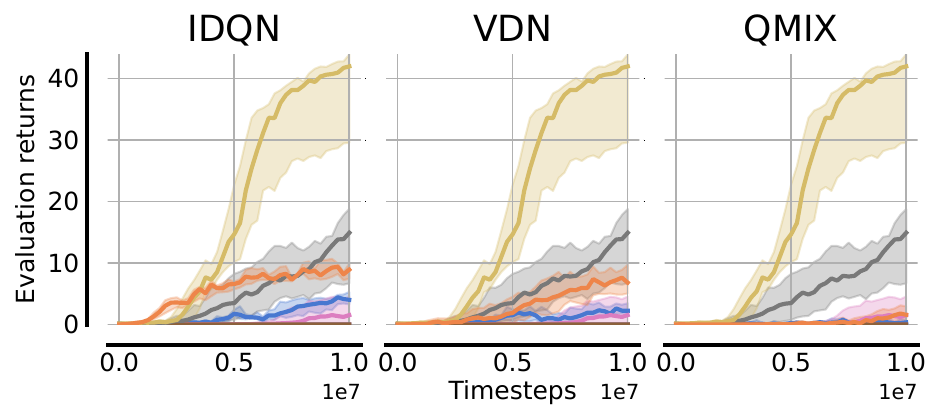}
        \caption{RWARE 11x10 4ag}
        \label{app:fig:emax:rware_tiny-4ag}
    \end{subfigure}

    \begin{subfigure}{.49\linewidth}
        \centering
        \includegraphics[width=\linewidth]{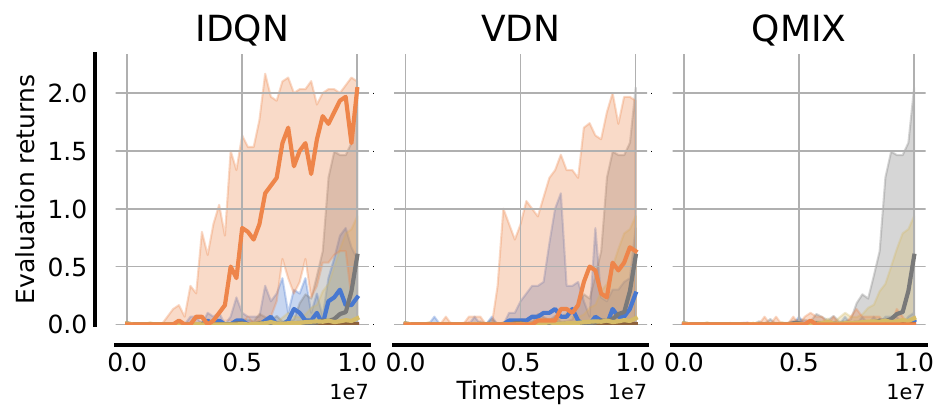}
        \caption{RWARE 20x10 2ag}
        \label{app:fig:emax:rware_small-2ag}
    \end{subfigure}
    \begin{subfigure}{.49\linewidth}
        \centering
        \includegraphics[width=\linewidth]{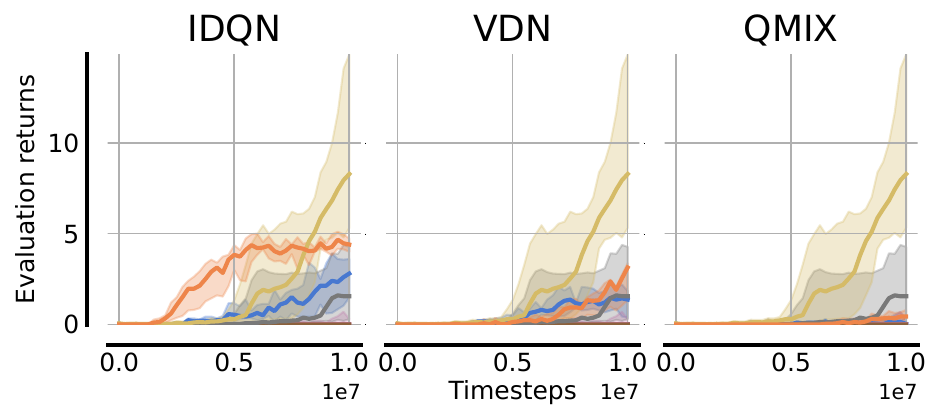}
        \caption{RWARE 20x10 4ag}
        \label{app:fig:emax:rware_small-4ag}
    \end{subfigure}
    
    \begin{subfigure}{.49\linewidth}
        \centering
        \includegraphics[width=\linewidth]{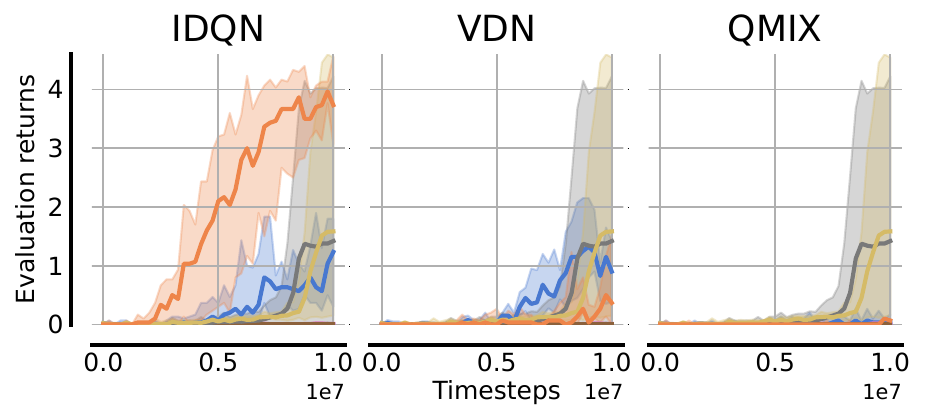}
        \caption{RWARE 20x16 4ag}
        \label{app:fig:emax:rware_medium-4ag}
    \end{subfigure}
    \begin{subfigure}{.49\linewidth}
        \centering
        \includegraphics[width=\linewidth]{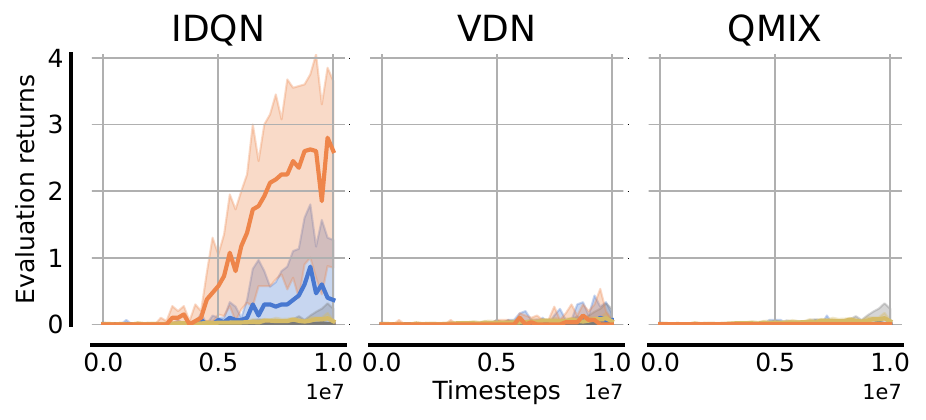}
        \caption{RWARE 29x16 4ag}
        \label{app:fig:emax:rware_large-4ag}
    \end{subfigure}
    \caption{Average evaluation returns and 95\% confidence intervals for all algorithms in RWARE tasks.}
    \label{app:fig:emax:rware_individual_returns}
\end{figure*}

\clearpage
\section{Network Size Analysis}
\label{app:sec:emax:network_size_analysis}

\subsection{Ensemble Size Comparison}
\label{app:sec:emax:ensemble_size_cost}
The computational cost of training an ensemble of models scales with the ensemble size $K$. To illustrate the additional cost, we investigate the training speed of EMAX for varying $K$. \Cref{tab:emax:ensemble_size_speedtest} shows the average time to train IDQN, VDN, QMIX, and their corresponding EMAX extensions with $K \in \{2, 5, 8\}$ for \num{10000} time steps in the LBF 10x10-3p-3f task. These times were averaged across ten runs. We can see that training an ensemble of $K=5$ value functions, as applied in our evaluation, increases the training time by less than 100\%. While this cost is significant, the increase in computational cost is notably less than a linear increase due to parallelisation on modern hardware, and we believe that it is justified in cases where the significant improvements of EMAX in both sample efficiency and stability are of importance. 

\begin{table}[h]
    \centering
    \begin{tabular}{l c c c c}
        \toprule
        \textbf{Algorithm}  & Baseline & $K=2$ & $K=5$ & $K=8$ \\ \midrule
        IDQN    & 16.80 & 21.29 (+27\%) & 33.04 (+97\%) & 48.06 (+186\%)\\
        VDN     & 16.92 & 21.56 (+27\%) & 33.25 (+97\%) & 48.16 (+185\%)\\
        QMIX    & 17.70 & 22.53 (+27\%) & 33.71 (+90\%) & 48.66 (+175\%)\\
        \bottomrule
    \end{tabular}
    \caption[Average training time for algorithms with and without EMAX, and for varying ensemble sizes.]{Average training time (in seconds) for vanilla and EMAX algorithms with varying ensemble sizes $K$ to complete \num{10000} time steps of training in the LBF 10x10-3p-3f task. Relative increase to the training time of the baseline algorithm ($K=1$) is given in parenthesis. Times are averaged across ten runs.}
    \label{tab:emax:ensemble_size_speedtest}
\end{table}

\subsection{Comparison with Baselines Using Larger Networks}
\label{app:sec:emax:large_network_baselines}
To further investigate whether the performance benefits of EMAX arise solely due to its larger number of learnable parameters compared to the vanilla algorithms, we evaluate all vanilla algorithms for larger network sizes. We keep the overall architecture of networks identical, so all value function networks consist of one hidden layer projecting the input observations to a hidden size of $d^h$, followed by a gated recurrent unit (GRU)~\citep{cho2014learning} with identical hidden dimensionality, and a final linear layer projecting the hidden output state of the GRU to action-values for each action with the dimensionality of the action space of an individual agent $i$, i.e.\ $|\Ac_i|$. We evaluate the vanilla algorithms with hidden sizes of $d^h \in \{128, 256, 512\}$ and compare their performance to EMAX with $K=5$ models in the ensemble and hidden size of $128$. The number of total parameters resulting from these models for one LBF and one RWARE task are shown in \Cref{tab:emax:model_size_comparison}. As we can see, EMAX with $K=5$ (and hidden size of $128$) has exactly five times more parameters in the model compared to the baseline with one model with hidden size of $128$. The baseline model with hidden size of $256$ is comparable to the model size of EMAX while the baseline model with hidden size of $512$ is roughly three times larger than the ensemble of EMAX.

\begin{table}[h!]
    \centering
    \begin{tabular}{l c c c c c c}
        \toprule
        \textbf{Task}  & $|o|$ & $|\Ac_i|$ & Base (128) & Base (256) & Base (512) & EMAX ($K=5$) \\ \midrule
        LBF 10x10-4p-3f-coop & 25 & 6 & \num{103174} & \num{402950} & \num{1592326} & \num{515870}\\
        RWARE $11\times20$ 4ag & 95 & 5 & \num{112005} & \num{420613} & \num{1627653} & \num{560025}\\
        \bottomrule
    \end{tabular}
    \caption[Comparison of network sizes of baseline algorithms and EMAX for varying model sizes and tasks with varying observation dimensionality.]{Observation dimensionality and the resulting number of parameters within the main value function networks for baseline algorithms with hidden sizes of $128$, $256$, and $512$, as well as for EMAX with $K=5$ models in the ensemble and hidden size of $128$ for one LBF and RWARE task.}
    \label{tab:emax:model_size_comparison}
\end{table}

\begin{figure*}[t]
    \centering
    \includegraphics[width=\linewidth]{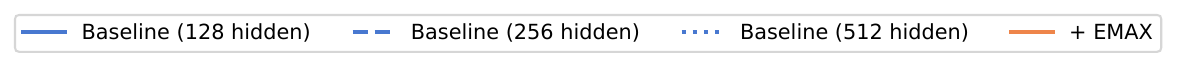}

    \begin{subfigure}{.49\linewidth}
        \centering
        \includegraphics[width=\linewidth]{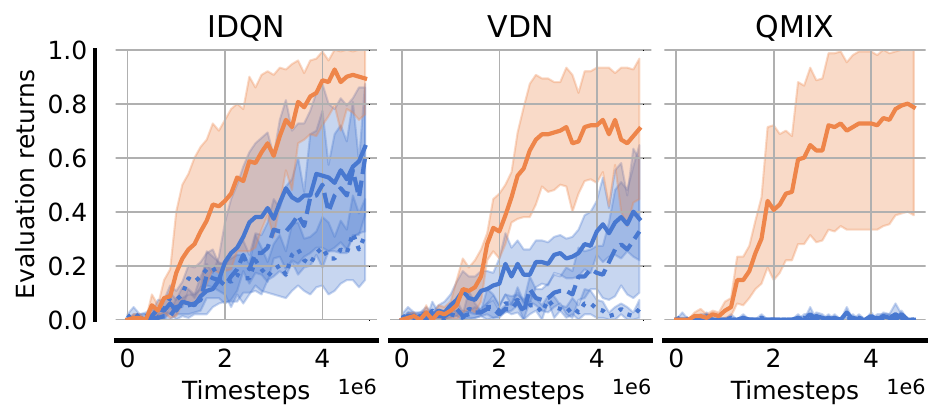}
        \caption{LBF 10x10-4p-3f-coop}
        \label{fig:emax:lbf-10x10-4p-3f-coop_larger_networks}
    \end{subfigure}
    \begin{subfigure}{.49\linewidth}
        \centering
        \includegraphics[width=\linewidth]{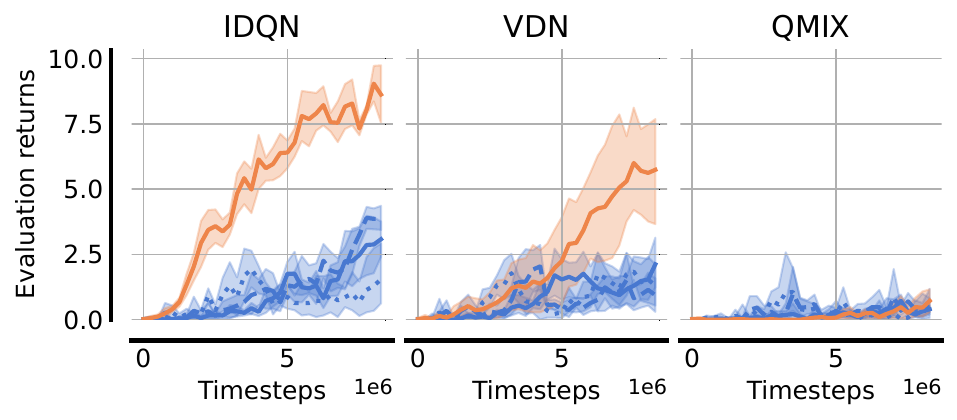}
        \caption{RWARE $11\times20$ 4ag}
        \label{fig:emax:rware-tiny-4ag_larger_networks}
    \end{subfigure}

    \caption[Average evaluation returns and confidence intervals for all vanilla algorithms with varying network sizes and EMAX.]{Mean and 95\% confidence intervals of evaluation returns for all vanilla algorithms with default and larger network sizes, and EMAX extensions.}
    \label{fig:emax:baseline_larger_networks}
\end{figure*}

\Cref{fig:emax:baseline_larger_networks} shows the evaluation returns of all vanilla baseline algorithms for varying model sizes compared to EMAX with $K=5$ models in the ensemble in one LBF and one RWARE task. As we can see, the vanilla algorithms are unable to make effective use of larger networks and reach similar evaluation returns to the original baseline with hidden sizes of $128$ despite four times and fifteen times more parameters in the model. Despite these larger networks, the vanilla algorithms are unable to reach the performance of EMAX with $K=5$ models in the ensemble. This suggests that the ensemble in EMAX and its use in the exploration policy, evaluation policy, and target computation is important to make effective use of the increase in parameters and EMAX does not outperform the baselines due to its larger computational budget.

\section{Sample Efficiency and Wall-Clock Efficiency}
\label{app:sec:emax:efficiency_samples_time}
\begin{figure}[t]
    \centering
    \includegraphics[width=.4\linewidth]{images/results/ind_rewards_eval/legend.pdf}

    \begin{subfigure}{.24\textwidth}
        \includegraphics[width=\textwidth]{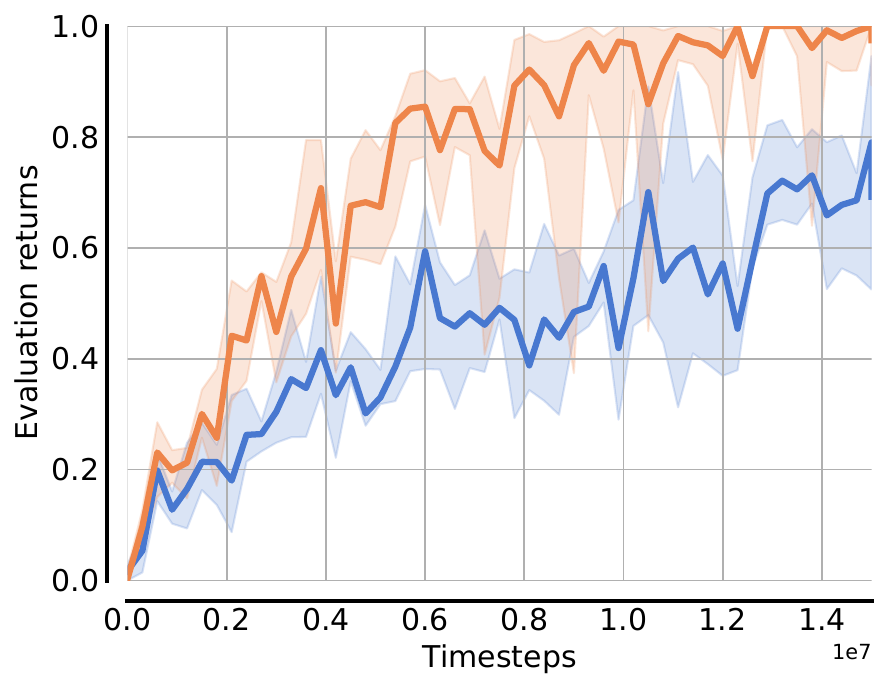}
        \caption{LBF 10x10-3p-5f, samples.}
        \label{fig:emax:lbf-10x10-3p-5f-sample-efficiency}
    \end{subfigure}
    \begin{subfigure}{.24\textwidth}
        \includegraphics[width=\textwidth]{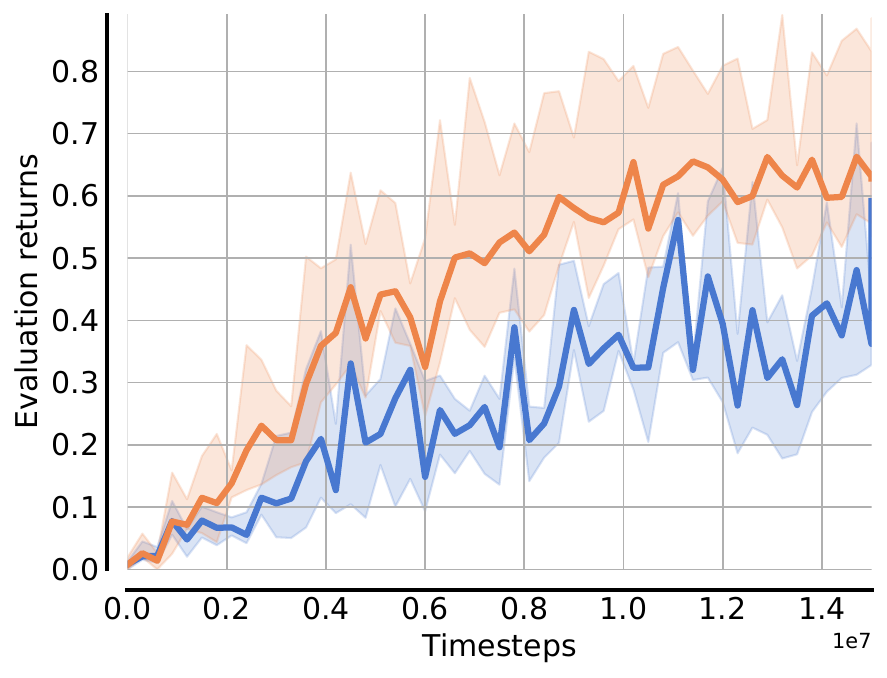}
        \caption{LBF 15x15-3p-5f, samples.}
        \label{fig:emax:lbf-15x15-3p-5f-sample-efficiency}
    \end{subfigure}
    \begin{subfigure}{.24\textwidth}
        \includegraphics[width=\textwidth]{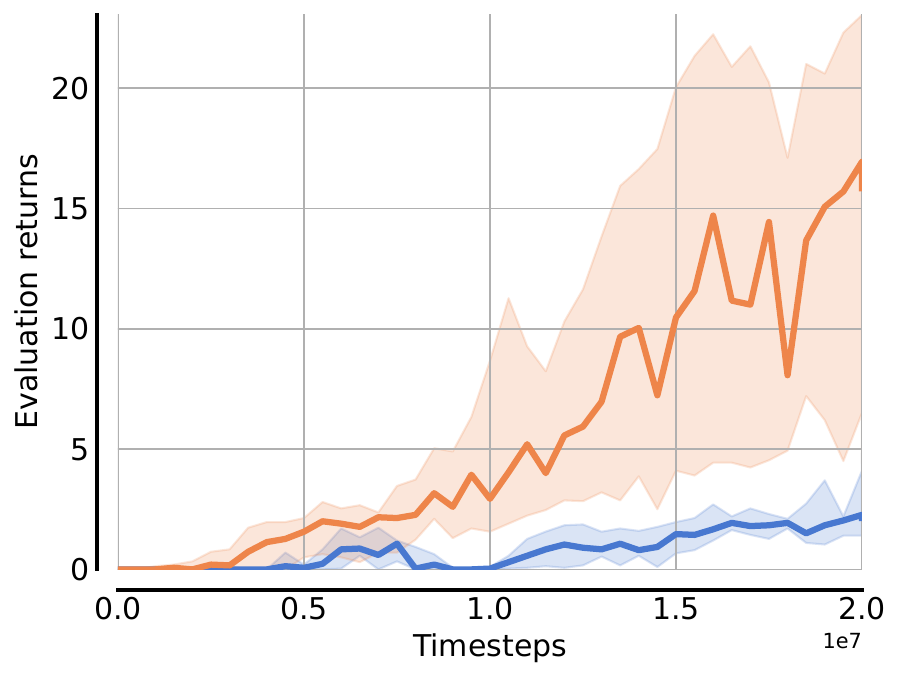}
        \caption{RWARE $11\times10$ 2ag, samples.}
        \label{fig:emax:rware-tiny-2ag-sample-efficiency}
    \end{subfigure}
    \begin{subfigure}{.24\textwidth}
        \includegraphics[width=\textwidth]{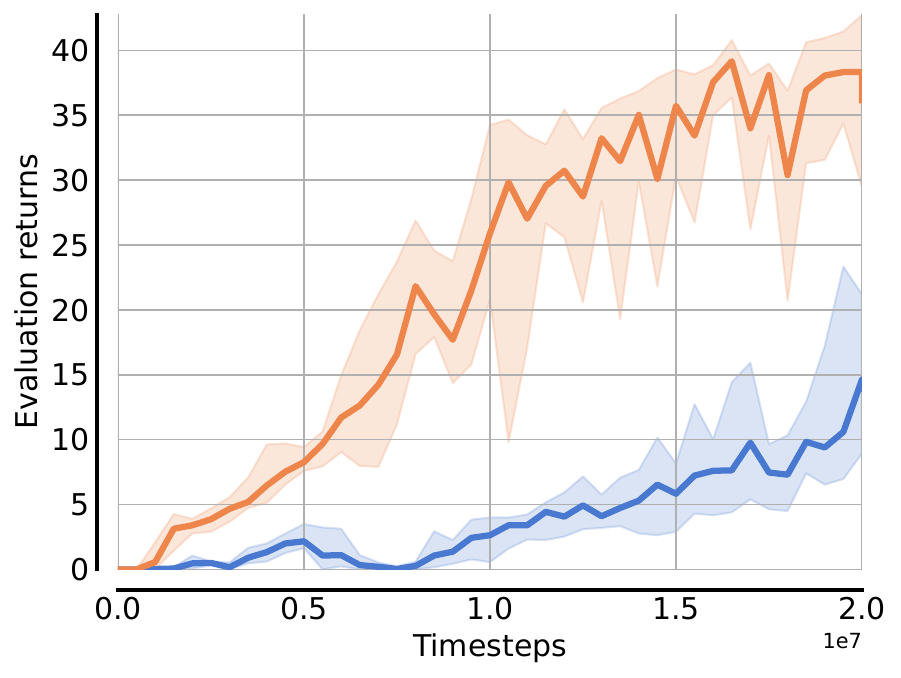}
        \caption{RWARE $11\times10$ 4ag, samples.}
        \label{fig:emax:rware-tiny-4ag-sample-efficiency}
    \end{subfigure}

    \begin{subfigure}{.24\textwidth}
        \includegraphics[width=\textwidth]{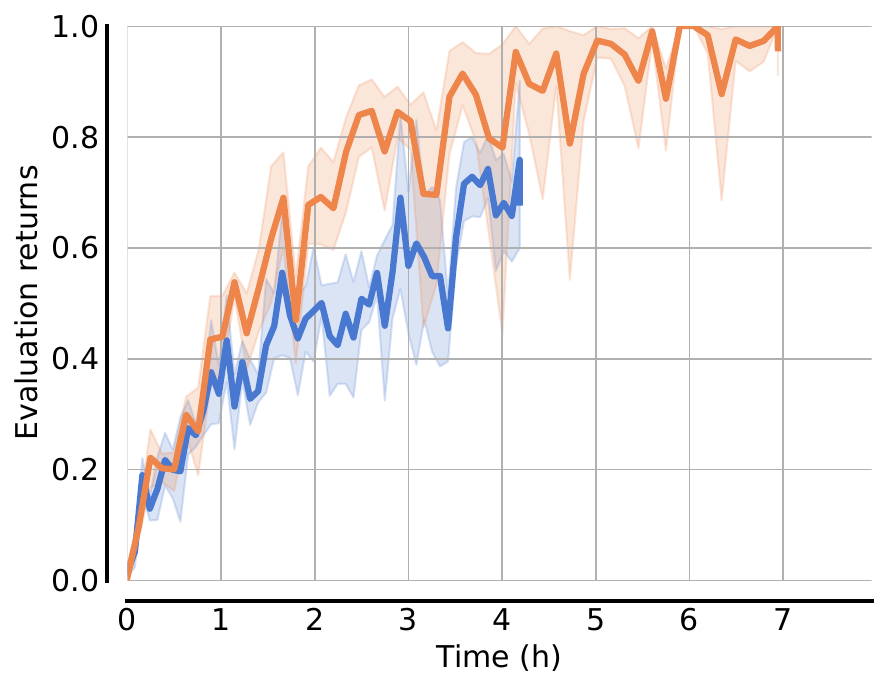}
        \caption{LBF 10x10-3p-5f, time.}
        \label{fig:emax:lbf-10x10-3p-5f-time-efficiency}
    \end{subfigure}
    \begin{subfigure}{.24\textwidth}
        \includegraphics[width=\textwidth]{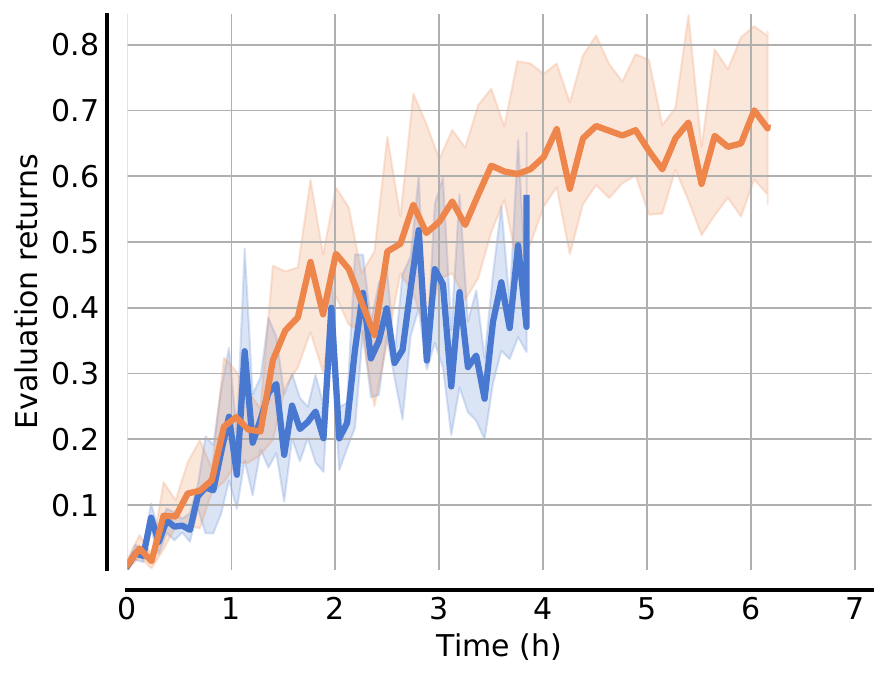}
        \caption{LBF 15x15-3p-5f, time.}
        \label{fig:emax:lbf-15x15-3p-5f-time-efficiency}
    \end{subfigure}
    \begin{subfigure}{.24\textwidth}
        \includegraphics[width=\textwidth]{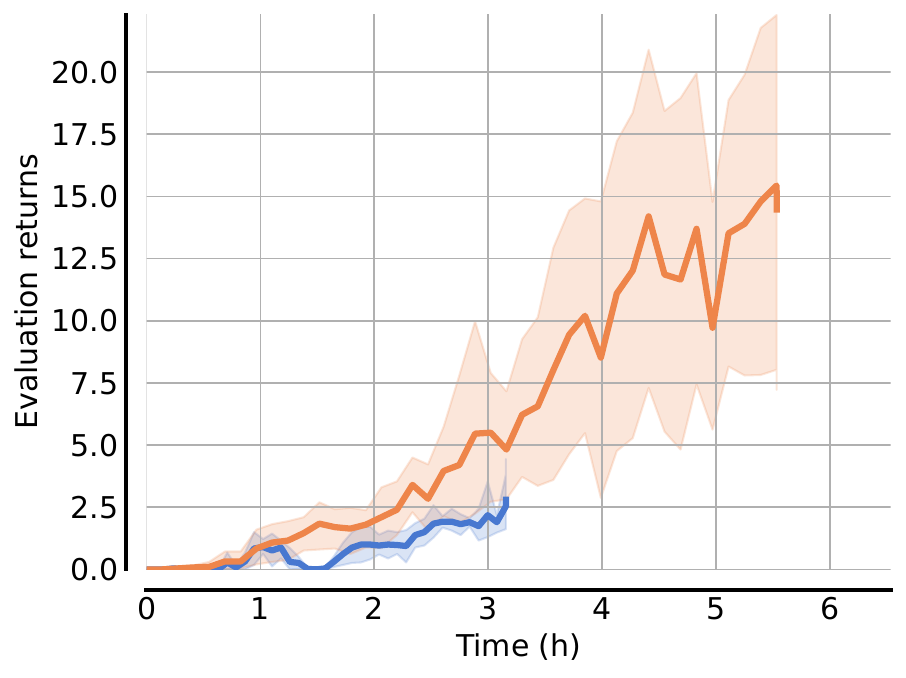}
        \caption{RWARE $11\times10$ 2ag, time.}
        \label{fig:emax:rware-tiny-2ag-time-efficiency}
    \end{subfigure}
    \begin{subfigure}{.24\textwidth}
        \includegraphics[width=\textwidth]{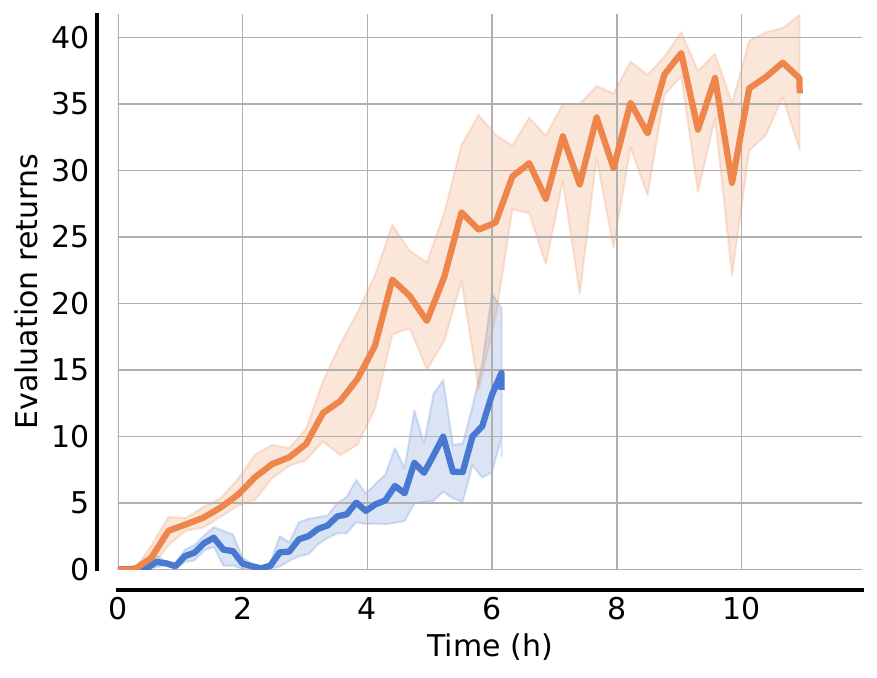}
        \caption{RWARE $11\times10$ 4ag, time.}
        \label{fig:emax:rware-tiny-4ag-time-efficiency}
    \end{subfigure}
    \caption{Sample efficiency (top) and wall-clock (bottom) efficiency of IDQN with and without EMAX in two LBF and RWARE tasks with individual rewards. Plots visualise the mean and 95\% confidence intervals of evaluation returns across five seeds.}
    \label{fig:emax:efficiency_samples_time}
\end{figure}

EMAX is largely motivated by the inefficiency of existing MARL algorithms in regards to their sample efficiency. Based on this motivation, empirical comparisons have been made to compare MARL baselines and EMAX-extended algorithms for a fixed budget of training samples. However, as expected and discussed in \Cref{app:sec:emax:ensemble_size_cost}, training EMAX with ensembles of value functions increases the cost of training. To provide insights in regards to the efficiency of EMAX-extended algorithms in terms of wall-clock time, we present the evaluation returns of IDQN with and without EMAX in two LBF and RWARE tasks with individual rewards in \Cref{fig:emax:efficiency_samples_time}. We train all algorithms for 15M and 20M timesteps in LBF and RWARE, respectively, and results are aggregated across five random seeds. For both RWARE tasks, we find that EMAX significantly improves the learning efficiency of IDQN even given an identical training budget in terms of wall-clock time. For LBF tasks, improvements are less pronounced but IDQN-EMAX still perform slightly better or comparable to base IDQN given identical training time.

\clearpage

\section{Uncertainty $\beta$ Sensitivity Analysis}
\label{app:sec:emax:beta_analysis}
\begin{figure}[h!]
    \centering
    \includegraphics[width=.5\textwidth]{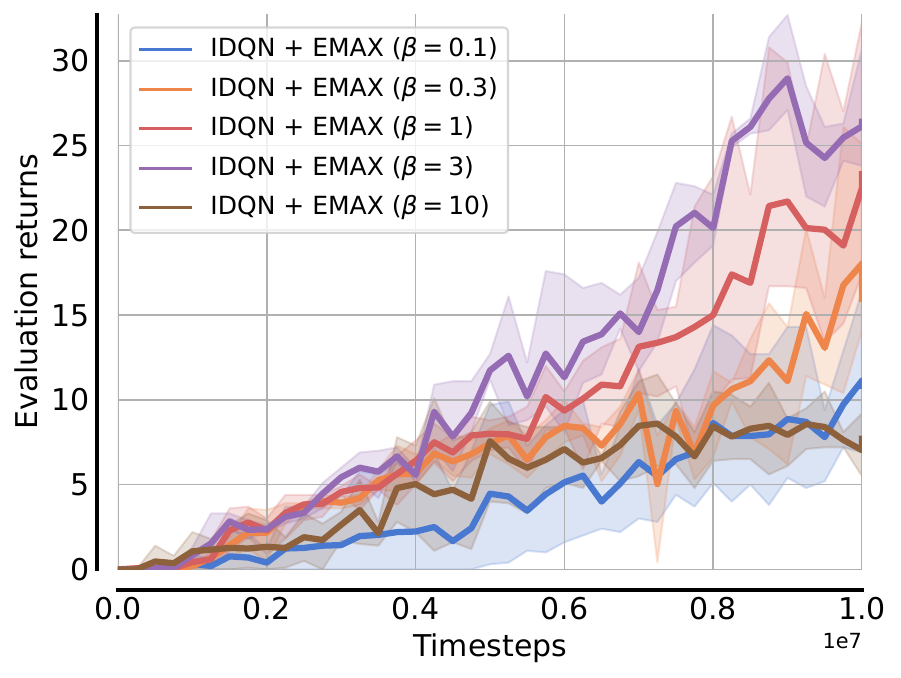}
    \caption{Evaluation returns of IDQN-EMAX for varying $\beta$ values in the RWARE $11\times10$ 4ag task with individual rewards. Visualised are the mean and 95\% confidence intervals of evaluation returns across three seeds.}
    \label{fig:emax:beta_sensitivity}
\end{figure}

EMAX introduces two additional hyperparameters to the underlying MARL algorithm: (1) $K$ to determine the number of value functions in the ensemble, and (2) $\beta$ to determine the weight of the exploration policy's uncertainty term \seehere{eq:emax:expl_policy}. In \Cref{fig:emax:rware-tiny-4ag_ensemble}, we already provide insights on how values of $K$ impact the performance of EMAX. In \Cref{fig:emax:beta_sensitivity}, we provide similar analysis for varying values of $\beta$ for the IDQN-EMAX algorithm trained for 10 timesteps in the RWARE $11\times10$ 4ag task with individual rewards. We consider $\beta \in \{0.1, 0.3, 1, 3, 10\}$ and visualise mean and 95\% confidence intervals of evaluation returns across three random seeds. While we find that the performance of IDQN-EMAX varies notably for different $\beta$, we remark that IDQN-EMAX outperforms the baseline IDQN algorithm for all values of $\beta$ when trained for a comparable number of timesteps.

\end{document}